\documentclass[%
 reprint,
nofootinbib,
 amsmath,amssymb,
 aps,
]{revtex4-2}

\usepackage{graphicx}
\usepackage{dcolumn}
\usepackage{bm}

\usepackage{amssymb,amsmath,verbatim,mathtools,needspace,enumitem,etoolbox,graphicx,physics,microtype,afterpage,xspace,tabularx,lmodern,multirow}

\usepackage{booktabs}
\usepackage{gensymb}
\usepackage{soul}
\usepackage{multirow}
\usepackage{subcaption}
\usepackage{ragged2e}
\usepackage{bbm}
\usepackage[dvipsnames, usenames]{xcolor}
\definecolor{linkcolor}{rgb}{0.0,0.3,0.5}
\usepackage[unicode, colorlinks=true, linkcolor=linkcolor, citecolor=linkcolor, filecolor=linkcolor, urlcolor=linkcolor, linktocpage, breaklinks]{hyperref}
\usepackage{orcidlink}
\usepackage[all]{hypcap}
\usepackage{cancel}
\usepackage{empheq}
\usepackage[T1]{fontenc}
\usepackage[utf8]{inputenc}
\usepackage[usenames,dvipsnames]{xcolor}
\usepackage{csquotes}
\usepackage{mathrsfs}
\hypersetup{colorlinks=true,citecolor=Periwinkle,linkcolor=Periwinkle,urlcolor=Periwinkle}

\definecolor{romared}{RGB}{142,0,28}

\newcommand{\be}{\begin{equation}}
\newcommand{\ee}{\end{equation}}

\def\be{\begin{equation}}
\def\ee{\end{equation}}
\newcommand{\beq}{\begin{eqnarray}}
\newcommand{\eeq}{\end{eqnarray}}
\usepackage{makecell}
\usepackage{soul}

\newcolumntype{Y}{>{\centering\arraybackslash}X}

\makeatletter
\newcommand*{\rom}[1]{\expandafter\@slowromancap\romannumeral #1@}
\makeatother

\makeatletter
\let\jnl@style=\rm
\def\ref@jnl#1{{\jnl@style#1}}

\def\aj{\ref@jnl{AJ}}                   
\def\actaa{\ref@jnl{Acta Astron.}}      
\def\araa{\ref@jnl{ARA\&A}}             
\def\apj{\ref@jnl{ApJ}}                 
\def\apjl{\ref@jnl{ApJ}}                
\def\apjs{\ref@jnl{ApJS}}               
\def\ao{\ref@jnl{Appl.~Opt.}}           
\def\apss{\ref@jnl{Ap\&SS}}             
\def\aap{\ref@jnl{A\&A}}                
\def\aapr{\ref@jnl{A\&A~Rev.}}          
\def\aaps{\ref@jnl{A\&AS}}              
\def\azh{\ref@jnl{AZh}}                 
\def\baas{\ref@jnl{BAAS}}               
\def\bac{\ref@jnl{Bull. astr. Inst. Czechosl.}}
\def\caa{\ref@jnl{Chinese Astron. Astrophys.}}
\def\cjaa{\ref@jnl{Chinese J. Astron. Astrophys.}}
\def\icarus{\ref@jnl{Icarus}}           
\def\jcap{\ref@jnl{J. Cosmology Astropart. Phys.}}
\def\jrasc{\ref@jnl{JRASC}}             
\def\memras{\ref@jnl{MmRAS}}            
\def\mnras{\ref@jnl{MNRAS}}             
\def\na{\ref@jnl{New A}}                
\def\nar{\ref@jnl{New A Rev.}}          
\def\pra{\ref@jnl{Phys.~Rev.~A}}        
\def\prb{\ref@jnl{Phys.~Rev.~B}}        
\def\prc{\ref@jnl{Phys.~Rev.~C}}        
\def\prd{\ref@jnl{Phys.~Rev.~D}}        
\def\pre{\ref@jnl{Phys.~Rev.~E}}        
\def\prl{\ref@jnl{Phys.~Rev.~Lett.}}    
\def\pasa{\ref@jnl{PASA}}               
\def\pasp{\ref@jnl{PASP}}               
\def\pasj{\ref@jnl{PASJ}}               
\def\rmxaa{\ref@jnl{Rev. Mexicana Astron. Astrofis.}}%
\def\qjras{\ref@jnl{QJRAS}}             
\def\skytel{\ref@jnl{S\&T}}             
\def\solphys{\ref@jnl{Sol.~Phys.}}      
\def\sovast{\ref@jnl{Soviet~Ast.}}      
\def\ssr{\ref@jnl{Space~Sci.~Rev.}}     
\def\zap{\ref@jnl{ZAp}}                 
\def\nat{\ref@jnl{Nature}}              
\def\iaucirc{\ref@jnl{IAU~Circ.}}       
\def\aplett{\ref@jnl{Astrophys.~Lett.}} 
\def\apspr{\ref@jnl{Astrophys.~Space~Phys.~Res.}}
\def\bain{\ref@jnl{Bull.~Astron.~Inst.~Netherlands}} 
\def\fcp{\ref@jnl{Fund.~Cosmic~Phys.}}  
\def\gca{\ref@jnl{Geochim.~Cosmochim.~Acta}}   
\def\grl{\ref@jnl{Geophys.~Res.~Lett.}} 
\def\jcp{\ref@jnl{J.~Chem.~Phys.}}      
\def\jgr{\ref@jnl{J.~Geophys.~Res.}}    
\def\jqsrt{\ref@jnl{J.~Quant.~Spec.~Radiat.~Transf.}}
\def\memsai{\ref@jnl{Mem.~Soc.~Astron.~Italiana}}
\def\nphysa{\ref@jnl{Nucl.~Phys.~A}}   
\def\physrep{\ref@jnl{Phys.~Rep.}}   
\def\physscr{\ref@jnl{Phys.~Scr}}   
\def\planss{\ref@jnl{Planet.~Space~Sci.}}   
\def\procspie{\ref@jnl{Proc.~SPIE}}   

\makeatother

\makeatletter
\def\etal@author{author}
\newcount\etal@count
\def\etal@first\bibinfo#1#2{\bibinfo{#1}{#2}\ \textit{et~al.}\etal@skip}
\def\etal@skip#1\bibinfo#2#3{\advance\etal@count by -1
  \ifnum\etal@count>1 \expandafter\etal@skip\fi}
\def\bibfield#1#2{\def\etal@tmp{#1}%
  \ifx\etal@tmp\etal@author
    \etal@count=0
    \setbox0=\hbox{\def\bibinfo##1##2{\def\etal@tmp{##1}%
      \ifx\etal@tmp\etal@author\global\advance\etal@count by 1 \fi}#2}%
    \ifnum\etal@count>10 \etal@first#2\else #2\fi
  \else #2\fi}
\makeatother

\usepackage{etoolbox} 
\makeatletter
\patchcmd{\@outputpage@head}{\@ifnum{\@mpcol+\@ne}{\@disablepaircolumn}{}}{}{}{}
\makeatother

\begin{document}


\title{\textsc{SPLIT}: Loosely-coherent inference of long-inspiral gravitational-wave sources \\ for robustness against modeling and data outliers}

\author{Shubham Kejriwal\,\orcidlink{0009-0004-5838-1886}}
 \email{shubhamkejriwal@u.nus.edu}
 \affiliation{Department of Physics, National University of Singapore, 21 Lower Kent Ridge Rd, Singapore 117551}


\begin{abstract}
The Laser Interferometer Space Antenna (LISA) will detect gravitational waves (GWs) from dozens of extreme- and intermediate-mass-ratio inspirals (EMRIs/IMRIs). These sources will stay in-band for months to years, offering extraordinary scientific potential. However, their fully phase-coherent analysis standard in current pipelines imposes stringent waveform accuracy requirements; failing to model the signal over such long durations can result in significant systematic biases. To address this, we formulate a robust Bayesian inference framework that segments the data into blocks, infers each block coherently, and recombines the results using loosely-coherent phase information. By restricting full phase-tracking to much shorter block durations, this approach prevents significant accumulation of phase errors. We implement this methodology in \textsc{SPLIT} (Segmented Posteriors for Long-Inspiral Templates), a GPU-accelerated Python package. Applying \textsc{SPLIT} to an environment-rich injection, we demonstrate that while a fully-coherent vacuum-GR analysis incurs a maximum 1D systematic bias of $\approx 4.8\sigma$ from the truth, the shorter integration window of our loosely-coherent approach restricts such biases to $\lesssim 0.5\sigma$. Overall, despite a fractional loss of optimal signal-to-noise ratio, the substantial improvement in parameter accuracy offered by our method presents a highly advantageous trade-off for LISA and other future GW detectors.
\end{abstract}

\maketitle


\section{Introduction}\label{sec:introduction}

The Laser Interferometer Space Antenna (LISA) is an upcoming space-based observatory that will unlock the gravitational-wave (GW) universe in the milli-Hz band~\cite{2024arXiv240207571C}. Alongside other exciting sources, LISA will detect dozens of extreme- and intermediate-mass-ratio inspirals (EMRIs/IMRIs) ~\cite{2017PhRvD..95j3012B,2026arXiv260317072S,2024arXiv240207571C}. These sources are characterized by the mass ratio between the secondary, typically a lighter stellar-origin black hole, orbiting a much heavier massive black hole (MBH) in its strong-gravity regime~\cite{2024arXiv240207571C}. The detection and characterization of EMRIs/IMRIs is one of the main science objectives of the LISA mission~\cite{2024arXiv240207571C}. The large number of strong-field orbits completed by the secondary in the LISA band allows high precision measurements of their modeled parameters~\cite{2017PhRvD..95j3012B}, promising stringent constraints on black hole properties like mass and spin~\cite{2009CQGra..26i4034G,2010PhRvD..81j4014G,2017PhRvD..95j3012B,2020PhRvD.102l4054B,2026arXiv260115198S}, deviations from general relativity (GR)~\cite{2013LRR....16....7G,2017PhRvD..96h4039C,2026PhRvD.113b3036S}, astrophysical environments~\cite{2011PhRvD..84b4032K,2023PhRvX..13b1035S,2025PhRvD.111h4006D,2026arXiv260317072S}, and cosmology~\cite{2021arXiv210602053L,2023LRR....26....5A}.

However, realizing this immense scientific potential in a robust Bayesian inference framework presents a formidable data analysis challenge. Firstly, waveform templates used in the analysis should be sufficiently accurate, introducing at most $\sim 1$ radian dephasing with respect to the signal at the true parameters over the entire observed inspiral lasting between months and years~\cite{2007PhRvD..76j4018C,2008PhRvD..78l4020L,2024PhRvD.109l4048B,2025PhRvD.111h2010K,2025PhRvD.112j4023C,2025LRR....28....9L}. Secondly, studies have shown that secular perturbations induced by astrophysical environments or modified gravity effects can introduce severe systematic biases during inference over the full inspiral if such effects are not modeled correctly~\cite{2014PhRvD..89j4059B,2023PhRvX..13b1035S,Kejriwal:2023djc,2026PhRvD.113b3036S}. Last but not the least, instrumental artifacts such as data gaps and transient noise glitches in realistic LISA data may also introduce biases in the analysis~\cite{2025CQGra..42f5018C,2025arXiv251216322B,2025PhRvD.111l4053B,2026arXiv260317072S}. 

Fundamentally, the challenges outlined above originate from enforcing strict phase coherence over the long duration of the inspiral in the LISA band. Thus, a natural mitigation strategy is to segment the data stream into shorter, contiguous blocks, each analyzed independently using fully-coherent methods and combine the results incoherently. This \textit{semi-coherent} strategy offers various advantages: first, the stringent 1 radian dephasing threshold need only be satisfied over the much shorter block duration, relaxing waveform accuracy requirements on theoretical models; second, unmodeled secular perturbations cannot accumulate significantly within sufficiently short duration blocks, thereby improving local consistency with vacuum-GR templates; and finally, transient noise glitches and data gaps can be isolated to a small subset of all blocks, insulating the inference of the broader inspiral trajectory from such localized effects.

However, fully incoherent combination of blocks discards more information than necessary, hindering the inference prospects of EMRIs and IMRIs. First, when the block duration is much shorter than the radiation-reaction timescale of the orbit, the local orbital frequencies become quasi-monochromatic, leading to strong degeneracies in the parameter space, e.g., between the orbital separation and the phase parameters (Sec.~\ref{sec:studenttprior}). Second, and more fundamentally, the stringent constraints on astrophysics and GR these sources promise are accumulated by tracking the phase coherently over the full inspiral; by completely severing phase coherence across blocks, we significantly degrade the measurability and void the long baseline advantage.

In this paper, we propose an inference framework that segments the data, but where a controlled degree of phase coherence is encoded between adjacent segments, and is implemented at the level of the Bayesian prior (Sec.~\ref{sec:studenttprior}). In the two extreme cases, requiring exact phase consistency recovers a fully-coherent analysis while requiring no consistency recovers the semi-coherent one. Our framework bridges the two, and is thus one of \textit{loosely-coherent} inference, following the nomenclature commonly used in the continuous-wave (CW) literature~\cite{2010CQGra..27t5017D}.

It is useful at this point to distinguish between the three stages that together constitute the inverse problem in GW data analysis~\cite{2022PhRvD.106j4051C}: \textit{detection}, establishing the statistically significant presence of signal in noisy data; \textit{identification}, mapping a detected signal sufficiently accurately onto the parameters of an approximate forward model; and \textit{inference}, estimating the Bayesian posterior of the true source parameters. The first two are collectively referred to as search. Importantly, while search can proceed with phenomenological or approximate waveform templates, inference relies on the most accurate model available. To the best of our knowledge, semi- and loosely-coherent methods in the GW literature have been developed primarily for search, and are reviewed below. In contrast, the framework presented in this work mainly focuses on their application to the inference stage, with the aim to robustly recover the signal parameters even in the presence of noise artifacts or unmodeled physical effects.

Semi-coherent and loosely-coherent methods have long been used in searches for CW signals in ground-based detector networks~\cite{2014ApJ...785..119A,
2017ApJ...839...12A,2023LRR....26....3R,2023APh...15302880W,2025ApJ...983...99A}. CWs originate from rapidly spinning neutron stars with non-axisymmetric mass deformations or fluid instabilities~\cite{1979PhRvD..20..351Z,2023LRR....26....3R,2023APh...15302880W}. Some of these sources undergo sudden, unpredictable spin transitions which induce random frequency jumps in the signal. Consequently, a fully-coherent search statistic aiming to map the evolution between frequency bins will quickly lose phase coherence after such \textit{spin-wandering}. Instead, semi- and loosely-coherent approaches are routinely applied in search algorithms~\cite{1998PhRvD..57.2101B,2010CQGra..27t5017D,2011PhRvD..84b3007P,2015PhRvD..91j2003L,2016PhRvD..93l3009S,2017PhRvD..96f3004A,2021PhRvD.104d2003M,2025PhRvD.111h3016C,2025PhRvD.111j4002T,2026arXiv260314168T}, including hierarchical MCMC follow-up of search candidates~\cite{2018PhRvD..97j3020A,2021PhRvD.104h4012T,2024PhRvD.110l4049M} and robustness of search statistics against noise transients~\cite{2014PhRvD..89f4023K,2016PhRvD..93h4024K}. We particularly highlight the hidden Markov model (HMM) formulation used with the Viterbi algorithm for CW search~\cite{2016PhRvD..93l3009S,2021PhRvD.104d2003M}, in which the signal frequency is treated as a hidden state evolving across contiguous segments under a transition probability that accommodates unpredictable spin-wandering. Our Markovian prior~\eqref{eq:markovprior} is similar in spirit, but acts on the continuous set of trajectory parameters, and informs the Bayesian posterior distribution instead of returning the single most probable path. See Sec.~\ref{sec:studenttprior} for more details.

In the context of LISA, semi-coherent methods have similarly been applied to search of stellar-mass black hole binaries~\cite{2023PhRvD.108h4014B,2024PhRvD.110j3026B} and more recently for single-harmonic search of EMRIs in the time-frequency domain~\cite{2026PhRvD.113b4061S}. Ref.~\cite{2026PhRvD.113b4061S} in particular, after performing semi-coherent search, further identifies the EMRI parameters through a staged procedure similar to Bayesian search strategies adopted in the CW literature~\cite{2018PhRvD..97j3020A,2021PhRvD.104h4012T,2024PhRvD.110l4049M}: once the search has localized a candidate and narrowed the region of interest, the semi-coherent detection statistic is adopted as the likelihood in an MCMC that resolves the posterior over the global source parameters. Common to such approaches, and to  of the semi-coherent search statistic itself~\cite{2011PhRvD..84b3007P}, is that the local segment-level parameters are either maximized, or marginalized under explicit priors. In this work, in contrast, the segmentation follows through to the inference of the block-level parameters themselves, carrying the robustness advantage enabled by the shorter integration window into the recovered posteriors, at the cost of a sampling dimensionality that grows linearly with $N_{\rm blocks}$~\eqref{eq:inferencedim}. We further note that Ref.~\cite{2026PhRvD.113b4061S} demonstrates the effectiveness of their semi-coherent pipeline on vacuum-GR injections in stationary Gaussian noise, whereas the framework proposed in this paper is motivated to mitigate the systematic biases induced by noise artifacts and modeling outliers.

In this paper, we formulate a general loosely-coherent Bayesian inference framework for long-inspiral signals, and demonstrate its utility on EMRI/IMRI injections. We propose a natural decomposition of the EMRI parameter space into static and evolving parameters, where the latter vary on the radiation-reaction timescale while the former remain constant throughout the observation window (Sec.~\ref{sec:generalformalism}). With this decomposition, we formulate the loosely-coherent inference posterior (Eq.~\eqref{eq:jointposteriorprod}) and prescribe useful explicit forms of the constituent likelihood and prior functions: to down-weight the impact of data outliers in each inference block, we adapt a heavy-tailed Student-t likelihood from previous works~\cite{2011PhRvD..84l2004R,2011CQGra..28a5010R} (Sec.~\ref{sec:studenttlikelihood}); and to accommodate modeling outliers across blocks, we construct the Markovian Student-t prior, whose transition probabilities loosely link adjacent blocks, allowing departures from the predicted vacuum-GR trajectory while penalizing excessive discontinuity in its evolution (Sec.~\ref{sec:studenttprior}). Since the optimal signal-to-noise ratio (SNR)~\cite{1992PhRvD..46.5236F} is distributed across the blocks, the block-wise posteriors are broadened in comparison to a fully-coherent analysis. We quantify this loss of SNR, and derive how the resulting posterior widths, and hence the severity of the recovered systematic biases, scale with the number of blocks (Sec.~\ref{sec:snrloss}). Furthermore, we qualitatively verify this scaling with the help of an example in Sec.~\ref{sec:results_example3_Nblocks3}.

To manage the cost of inference over the large parameter space dimensionality, we further propose a sequential Gibbs sampling strategy for efficient posterior recovery (Sec.~\ref{sec:results_setup}). We implement these methods in a new Python package called \textsc{SPLIT} (Segmented Posteriors for Long-Inspiral Templates)\footnote{\url{https://github.com/perturber/SPLIT}}~\cite{kejriwal_2026_20290209}. The package has a highly configurable interface and the capability to parallelize over multiple Graphics Processing Units (GPUs). This allows us to perform loosely-coherent inference for a variety of EMRI/IMRI injections, and compare the results against fully-coherent baselines (Sec.~\ref{sec:results_example1},~\ref{sec:results_example2}, and~\ref{sec:results_example3}).

The remainder of this paper is organized as follows: in Sec.~\ref{sec:methods}, we present the loosely-coherent inference framework and quantify the impact of the fractional loss of optimal SNR in each block; in Sec.~\ref{sec:results}, we apply this framework using the \textsc{SPLIT} package to three distinct injection examples: (i) a vacuum-GR EMRI signal with no detector noise injection (Sec.~\ref{sec:results_example1}), (ii) a vacuum-GR EMRI signal with a zero-mean Gaussian noise injection (Sec.~\ref{sec:results_example2}), and (iii) an IMRI signal where the secondary is perturbed by its astrophysical environment (Sec.~\ref{sec:results_example3}); finally, in Sec.~\ref{sec:discussion}, we summarize our findings, comment on the practical utility of the loosely-coherent method, and discuss future directions.

\section{Methods}\label{sec:methods}

\subsection{General formalism}~\label{sec:generalformalism}

We segment the discrete time-series data $\vec{d}$ into $N_{\rm blocks}$ contiguous blocks $\{\vec{d}_i\}_{i=0}^{N_{\rm blocks}-1}$. Furthermore, we decompose the parameter set $\vec{\theta}_i$ describing the template $\vec{h}(\vec{\theta}_i)$ for the $i{\rm th}$ block as $\vec{\theta}_i := \{\vec{\vartheta},\vec{\varphi}_i\}$; here, $\vec{\vartheta}$ is generically the set of ``static'' parameters, i.e., those that remain constant in time throughout the signal duration (e.g., the BH masses, spins, etc.),\footnote{Formally, the BH masses and spins also change over time due to GW radiation. However, this occurs at a much slower rate compared to the evolution of the source's orbital parameters during the inspiral, and we therefore neglect their evolution.} and the remaining parameter set $\vec{\varphi}_i := \vec{\theta}_i \setminus \vec{\vartheta}$ contains model parameters that vary over time (e.g., the orbital separation, eccentricity, etc.).\footnote{Here $\setminus$ denotes the set difference.}

With this decomposition, we can write the joint posterior probability density function (pdf) of $\left(\{\vec\varphi_i\}, \vec\vartheta\right)$ from Bayes theorem as
\begin{align}
    p(\{\vec\varphi_i\},\vec\vartheta|\{\vec{d}_i\}) &= \frac{p(\{\vec{d}_i\}|\{\vec\varphi_i\},\vec\vartheta)p(\{\vec\varphi_i\},\vec\vartheta)}{p(\{\vec{d}_i\})}\\
    &= \frac{p(\{\vec{d}_i\}|\{\vec\varphi_i\},\vec\vartheta)p(\{\vec\varphi_i\}|\vec\vartheta)p(\vec\vartheta)}{p(\{\vec{d}_i\})}~\label{eq:jointposterior}
\end{align}
Here $p(\{\vec{d}_i\}|\{\vec\varphi_i\},\vec\vartheta)$ is the joint likelihood and $ p(\{\vec\varphi_i\},\vec\vartheta)$ the joint prior of $\{\vec\varphi_i\},\vec\vartheta$. In the second line, we have decomposed the joint prior in terms of the conditional and marginal pdfs. The denominator denotes the evidence for the data, explicitly written as 
\begin{align}
    p(\{\vec{d}_i\}) := \int{\rm d}\vec{\vartheta}~{\rm d}\{\vec\varphi_i\}~p(\{\vec{d}_i\}|\{\vec\varphi_i\},\vec{\vartheta})p(\{\vec\varphi_i\},\vec{\vartheta}).
\end{align}
To proceed, we assume a locally stationary noise process, i.e., that the noise is stationary in each block (but not necessarily across the entire observation window). Furthermore, we assume that the noise in block $i$ is independent of the noise in block $j \neq i$. Formally, the detector noise can be finitely correlated across the entire observation window. However, by setting appropriate correlation thresholds, we can assign sufficiently accurate ``decorrelation times'' such that the block-wise noise independence approximation holds reasonably well. We concretize this argument in Appendix~\ref{app:blocindependence} for realistic detector noise with the help of upper limits on $N_{\rm blocks}$ and appropriate time windowing functions. Additionally, we validate this assumption in Example~\rom{2} (Sec.~\ref{sec:results_example2}).

Under these conditions, the likelihood in Eq.~\eqref{eq:jointposterior} can be rewritten as
\begin{align}
    p(\{\vec{d}_i\}|\{\vec{\varphi}_i\},\vec{\vartheta}) = \prod_i p(\vec{d}_i|\vec{\varphi}_i,\vec{\vartheta}).~\label{eq:localstationary}
\end{align}
Substituting Eq.~\eqref{eq:localstationary}
in~\eqref{eq:jointposterior}, we can write the joint posterior pdf as
\begin{align}
    p(\{\vec\varphi_i\},\vec\vartheta|\{\vec{d}_i\}) = \frac{\left(\prod_i p(\vec{d}_i|\vec{\varphi}_i,\vec{\vartheta})\right)p(\{\vec{\varphi}_i\}|\vec{\vartheta})p(\vec\vartheta)}{p(\{\vec{d}_i\})}.~\label{eq:jointposteriorprod}
\end{align}

\subsection{A Student-t likelihood for robustness against transient features}~\label{sec:studenttlikelihood}

We now focus on the explicit form of the single-block likelihood, $p(\vec{d}_i|\vec\varphi_i,\vec\vartheta)$. To motivate our framework, we first write the typical GW likelihood and its noise assumptions, highlight its limitations under realistic detector noise, and then formulate the more robust Student-t likelihoood used in this work. We first omit the block index notation for clarity but restore it when defining the explicit likelihood used in this work.

Typically, GW data analysis assumes that the noise process can be approximated as a stationary zero-mean Gaussian process. Note that this is a stronger condition than necessary to write the likelihood~\eqref{eq:localstationary} and therefore everything discussed in Sec.~\ref{sec:generalformalism} still holds. Then, the likelihood $p(\vec{d}|\vec\theta)$ for a set of model parameters $\vec\theta$ under the data $\vec{d}$ is the Whittle likelihood~\cite{10.1111/j.2517-6161.1957.tb00242.x}
\begin{align}
    p(\vec{d}|\vec\theta) \propto \exp\left(-\frac{1}{2}\left.\langle\vec{d}-\vec{h}(\vec\theta)\right|\vec{d}-\vec{h}(\vec\theta)\rangle\right)~\label{eq:gaussianlikelihood}
\end{align}
where $\langle \vec{a} | \vec{b} \rangle$ is the noise-weighted inner-product~\cite{1992PhRvD..46.5236F,1994PhRvD..49.2658C}
\begin{align}
    \langle\vec{a}|\vec{b}\rangle &:= 4\mathfrak{R}\left(\int_{f_{\rm min}}^{f_{\rm max}} {\rm d}f \frac{\tilde a(f)\tilde b^*(f)}{S_n(f)}\right)\\
    &\approx \frac{4}{T}\mathfrak{R}\left(\sum_{f_k = f_{\rm min}}^{f_{\rm max}}\frac{\tilde a(f_k)\tilde b^*(f_k)}{S_n(f_k)}\right).
\end{align}
Here overhead-tilde represents the Fourier transform, $^*$ represents the complex conjugate, and $S_n(f)$ is the one-sided noise power spectral density (PSD). $f_{\rm min}$, $f_{\rm max} \in (0,\infty)$ are the minimum and maximum detector frequencies, respectively. In the second line, we make a discrete approximation with $T$ the observation time.

In a realistic setup, in contrast, non-stationary transient features such as instrumental glitches may be present in the data, which voids the stationary, Gaussian noise assumption intrinsic to Eq.~\eqref{eq:gaussianlikelihood}. Due to the quadratic dependence of the Gaussian exponent~\eqref{eq:gaussianlikelihood}, such outliers can incur excessive penalties. Consequently, if we employ the standard Gaussian likelihood for block-level inference, an unmodeled transient feature in a single block can dominate the total likelihood product~\eqref{eq:localstationary}, suppressing the total signal evidence. 

Various implementations of robust likelihood functions designed to accommodate such departures from Gaussianity have been proposed in the GW literature~\cite{2011PhRvD..84l2004R,2011CQGra..28a5010R,2023PhRvD.108j3005S,2025PhRvD.111b2005K,2026PhRvD.114b4004S}.\footnote{Similar marginalization strategies have also been proposed for robust CW search statistics, see e.g.,~\cite{2016PhRvD..93h4024K,2014PhRvD..89f4023K}.} The common strategy is to replace the Gaussian with a heavier-tailed distribution, so that a large outlier at a single frequency is down-weighted instead of incurring the drastic quadratic penalty of Eq.~\eqref{eq:gaussianlikelihood}. Here, we adopt the Student-t distribution approach~\cite{2011PhRvD..84l2004R,2011CQGra..28a5010R}, which follows from relaxing the assumption that the noise PSD is known exactly. Instead, the noise variance at a given frequency is assigned an inverse-gamma prior and the Gaussian likelihood is marginalized over it~\cite{2011PhRvD..84l2004R,2011CQGra..28a5010R,bretthorst2013bayesian}. Explicitly, the Student-t likelihood can be written as
\begin{align}
    &p(\vec{d}|\vec\theta) \propto \nonumber\\
    &\exp\left\{{-\sum_{f_k=f_{\rm min}}^{f_{\rm max}}\frac{\nu_k+2}{2}}\log\left[1+\frac{\langle\left.\vec{d}-\vec{h}(\vec\theta)\right|\vec{d}-\vec{h}(\vec\theta)\rangle_k}{\nu_k}\right]\right\}~\label{eq:studenttgeneral}
\end{align}
where we define the individual-bin inner product,
\begin{align}
    \langle\left.\vec{a}\right|\vec{b}\rangle_k := \frac{4}{T}\mathfrak{R}\left(\frac{\tilde{a}(f_k)\tilde{b}^*(f_k)}{S_n(f_k)}\right).
\end{align}
In Eq.~\eqref{eq:studenttgeneral}, $\nu_k > 0$ is the degrees-of-freedom parameter. Unlike a Gaussian likelihood, which induces a quadratic decay inside the exponent, the Student-t decays logarithmically, i.e., with a heavier tail, and where the tail's weightedness is inversely proportional to $\nu_k$ such that we recover the Gaussian likelihood when $\nu_k \to \infty$. This heavy-tailed likelihood ensures that data segments with large outliers do not dominate the total likelihood product~\eqref{eq:localstationary}. In the following, we assume that the degrees-of-freedom parameter is constant across all bins, such that $\nu_k \equiv \nu$. This reflects the \textit{a priori} assumption that all frequency bins are expected to have outliers of the same statistical significance.

Other strategies for relaxing the assumptions underlying Eq.~\eqref{eq:gaussianlikelihood} are available. One such approach is to adopt the more general family of hyperbolic functions as the likelihood~\cite{2023PhRvD.108j3005S,2025PhRvD.111b2005K,2026PhRvD.114b4004S}, of which the Student-t function is a limiting case. Another is to infer the noise spectrum directly rather than marginalizing over it, e.g., by modeling the residuals between the data and the template as an autoregressive process and jointly sampling over the noise model's parameters, as proposed in Ref.~\cite{2024EPJC...84.1023M}. We adopt the Student-t distribution here for its simple, analytic, and single-hyperparameter form, which sufficiently demonstrates the main ideas of the loosely-coherent framework. We emphasize, however, that the construction of Sec.~\ref{sec:generalformalism} is agnostic to this choice and that future renditions can readily substitute an alternative robust likelihood as appropriate to the application at hand.

We can apply the Student-t likelihood to our formalism by assigning an independent $\nu_i$ to each $i{\rm th}$ block, and correspondingly writing the block-level likelihood as
\begin{widetext}
\begin{align}
    p(\vec{d}_i|\vec\varphi_i,\vec\vartheta) \propto \exp\left\{-\frac{\nu_{i}+2}{2}\sum_{f_k=f^i_{\rm min}}^{f^i_{\rm max}}\log\left[1+\frac{\langle\left.\vec{d}_i-\vec{h}(\vec\varphi_i,\vec\vartheta)\right|\vec{d}_i-\vec{h}(\vec\varphi_i,\vec\vartheta)\rangle_k}{\nu_{i}}\right]\right\}~\label{eq:studenttblocklikelihood}.
\end{align}
\end{widetext}
Clearly, the Student-t likelihood for each $i{\rm th}$ block~\eqref{eq:studenttblocklikelihood} is independent of all other blocks $j\neq i$ (conditioned on $\vec\vartheta$) such that Eq.~\eqref{eq:localstationary} holds.

In this work, we make the further simplifying assumption that the degrees of freedom parameter is the same across all blocks: $\nu_i \equiv \nu_{\rm like}$. More realistically, one may assign smaller values of $\nu_i$ to the likelihood in later blocks to account for, e.g., instrumental degradation over time, which may in turn lead to an increased rate of noise outliers~\cite{2026arXiv260317072S}. Additionally, $\nu_{\rm like}$ (or the set $\{\nu_i\}$) may be treated as a free hyperparameter during inference, with its prior informed by the expected rate at which such outliers occur. However, to reduce the inference space dimensionality, we treat $\nu_{\rm like}$ as a fixed model parameter in the remainder of this work.

\subsection{A Markovian Student-t prior for robust phase tracking}~\label{sec:studenttprior}

Next, we describe the conditional prior on the evolving parameters, $p(\{\vec\varphi_i\}|\vec\vartheta)$ in Eq.~\eqref{eq:jointposteriorprod}. A simple form could be a block-independent prior: $p(\{\vec\varphi_i\}|\vec\vartheta_i) = \prod_i p(\vec\varphi_i|\vec\vartheta_i)$, allowing the sampler to freely adjust the evolving parameters in each block; this is the fully phase-incoherent (semi-coherent) limit discussed in Sec.~\ref{sec:introduction}. However, in cases where the time duration in individual blocks is much shorter than the radiation-reaction timescale of the orbit, the local frequencies become quasi-monochromatic, which can lead to strong degeneracies in the posterior space of the evolving parameters. To break these degeneracies, we propose a first-order discrete Markov chain prior,
\begin{align}
    p(\{\vec\varphi_i\}|\vec\vartheta) := \pi(\vec\varphi_0|\vec\vartheta)\prod_{i=1}^{N_{\rm blocks}-1} \pi(\vec\varphi_i|\vec\varphi_{i-1},\vec\vartheta).~\label{eq:markovprior}
\end{align}
Here, $\pi(\vec\varphi_0|\vec\vartheta)$ is the prior of the evolving parameters in the first block, equivalently a Markovian initial condition, and $\pi(\vec\varphi_i|\vec\varphi_{i-1},\vec\vartheta)$ is the Markov chain transition probability. We note that our implementation is similar in spirit to the HMM/Viterbi search algorithm formulated in the CW context~\cite{2016PhRvD..93l3009S,2021PhRvD.104d2003M}. In fact, the block-wise independent likelihood formulated in Eqs.~\eqref{eq:localstationary},~\eqref{eq:studenttblocklikelihood} is effectively the \textit{emission} probability of the data in the $i$th block, such that the joint posterior~\eqref{eq:jointposteriorprod} constitutes a Markovian state-space model conditioned on the static parameters $\vec\vartheta$. However, our implementation differs in two main ways: the \textit{hidden} state is the continuous set of trajectory parameters rather than a discrete set of frequency bins, and the objective is a Bayesian posterior distribution instead of the single most probable path through the state space.

If we were to \textit{a priori} assume that the data represent purely vacuum-GR trajectories, one could model the transition probability deterministically, $\pi(\vec\varphi_i|\vec\varphi_{i-1},\vec\vartheta) = \delta(\vec\varphi_i-f(\vec\varphi_{i-1},\vec\vartheta;\Delta t_i))$, where $\delta$ is the Dirac-delta function and $\Delta t := t_i - t_{i-1}$. Additionally, $f \equiv f(\vec\varphi,\vec\vartheta;t)$ is a smooth, invertible function describing the forward-in-time, vacuum-GR trajectory evolution of $\vec\varphi$ evaluated at time $t$; With this choice, we fall back to a fully-coherent framework, where the evolving parameter set of the $i$th block is deterministic given those in the $i-1$ block. Realistically, however, we can allow for robustness against \textit{modeling} uncertainties from, e.g., secular perturbative effects (astrophysical environments, modified gravity effects, etc.) by modeling the transition probability as another multivariate Student-t distribution,
\begin{align}
    \pi(\vec\varphi_i|\vec\varphi_{i-1},\vec\vartheta) \propto \left(1 + \frac{\Delta\vec\varphi_i^T\Sigma_{\rm prior}^{-1}\Delta\vec\varphi_i}{\nu_{\rm prior}} \right)^{-(\nu_{\rm prior}+D_{\rm evol})/2}.~\label{eq:priortransition}
\end{align}
Here $\Delta\vec\varphi_i := \vec\varphi_i - f(\vec\varphi_{i-1},\vec\vartheta;\Delta t_i)$ represents the deviation of the $i$th block's parameters from the deterministic vacuum-GR prediction (modeled by $f$), $\nu_{\rm prior}$ is the degrees of freedom parameter, and $D_{\rm evol}$ is the dimensionality of $\vec\varphi_i$. $\Sigma_{\rm prior}$ is a $D_{\rm evol}\times D_{\rm evol}$ matrix which quantifies our \textit{a priori} knowledge of modeling errors. In the limit $\Sigma_{\rm prior} \to {\rm diag}(\vec{0})$, the prior is once again fully-coherent. On the other end, $\Sigma_{\rm prior} \to {\rm diag}(\infty)$ represents a fully phase-incoherent (semi-coherent) prior.

Together, the Student-t likelihood~\eqref{eq:studenttblocklikelihood} and the Markovian prior~(\ref{eq:markovprior},~\ref{eq:priortransition}) make the loosely-coherent analysis robust against transient or secular outliers in the data and the model itself, effectively mitigating systematic biases in the inference of the joint posterior~\eqref{eq:jointposteriorprod}. Our choice of hyperparameters $\nu_{\rm like}$ and $(\nu_{\rm prior}, \Sigma_{\rm prior})$ reflects our \textit{a priori} confidence in the data and in the model accuracy.

\subsection{Fractional loss of signal-to-noise ratio and its impact on systematic biases}~\label{sec:snrloss}

While the loosely-coherent framework formulated above offers robustness against data and modeling outliers, it comes at a natural cost: the total SNR of the GW signal is now distributed among the blocks. This in turn leads to a broadening of the block-wise posteriors compared to a fully-coherent analysis and, in the presence of unmodeled effects in the signal, impacts the severity of systematic biases incurred in posterior recovery. In this section, we approximate the relation between $N_{\rm blocks}$ and the fractional loss of optimal SNR in each block and qualitatively comment on the corresponding scaling of posterior widths and systematic biases.

In the GW context, the optimal SNR of a signal $\vec{s}$ observed coherently over the full observation duration is defined as $\rho_{\rm opt} = \langle\vec{s}|\vec{s}\rangle^{1/2}$~\cite{1994PhRvD..49.2658C}. Hence, when segmenting the signal into $N_{\rm blocks}$ blocks $\{\vec{s}_i\}$, the $i$th block retains a \textit{local} SNR of $\rho_i = \langle\vec{s}_i|\vec{s}_i\rangle^{1/2}$. Furthermore, since the blocks are contiguous (and ignoring the small windowing effects formulated in Appendix~\ref{app:blocindependence}), we get $\sum_i \rho_i^2 = \rho_{\rm opt}^2$. Then, under the simplifying assumption that the total optimal SNR is divided equally among all blocks, we can approximately write 
\begin{align}
    \frac{\rho_i}{\rho_{\rm opt}} \approx \frac{1}{\sqrt{N_{\rm blocks}}}.~\label{eq:lossofsnr}
\end{align} 
In words, the SNR in each $i$th block, $\rho_i$, is a factor $\approx \sqrt{N_{\rm blocks}}$ lower than the optimal SNR of the full-duration signal $\rho_{\rm opt}$. We can also qualitatively highlight the impact of loss of optimal SNR~\eqref{eq:lossofsnr} on the 1D marginal posterior variance $\sigma^2$ using the heuristic $\sigma^2 \sim 1/\rho^2$ (see, e.g., Ref.~\cite{2008PhRvD..77d2001V}). Hence, the ratio between the posterior widths of the $j$th parameter in the $i$th block ($\sigma^2_{ij} \sim 1/\rho_i^2$) and in the fully-coherent case ($\sigma^2_{{\rm FC},j} \sim 1/\rho^2_{\rm opt}$) can be written from Eq.~\eqref{eq:lossofsnr} as
\begin{align}
    \frac{\sigma^2_{ij}}{\sigma^2_{{\rm FC},j}} \sim N_{\rm blocks}.~\label{eq:lossofsnrposterior}
\end{align}

Finally, we comment on the impact of $N_{\rm blocks}$ on the severity of systematic biases. Assuming a deviation between the signal and the analysis template, either from an unmodeled physical effect or an inaccurate noise model, the $i$th block incurs a systematic bias $\Delta\vec\theta_i := \vec\theta_i^{\rm truth} - \vec\theta_i^{\rm bf}$ where $\vec\theta_i^{\rm truth}$ is the injected parameter for the $i$th block and $\vec\theta_i^{\rm bf}$ is the recovered best-fit value, and the two are generally not the same (See, e.g., Ref.~\cite{2007PhRvD..76j4018C}). The significance of this bias on the $j$th parameter can be measured by the 1D $z$-score, $z_{ij} = |\Delta\theta_{ij}|/\sigma_{ij}$. Substituting Eq.~\eqref{eq:lossofsnrposterior}, we have
\begin{align}
    \frac{z_{ij}}{z_{{\rm FC},j}} \sim \frac{1}{\sqrt{N_{\rm blocks}}}\frac{\Delta\theta_{ij}}{\Delta\theta_{{\rm FC},j}}.~\label{eq:zscorescaling}
\end{align}
For a secularly accumulating deviation, the second factor on the right side is bounded above by unity; a perturbation which integrates over the full inspiral is only partially accumulated over the much shorter block duration. Thus, the $z$-scores are suppressed twice: first by the number of segments $N_{\rm blocks}$ which broadens the posterior widths and second by the ratio of systematic biases which reduces the bias itself.

In practice, blocks that correspond to the earlier (later) inspiral phase of the binary's evolution will have $\sigma^2_{ij}/\sigma^2_{\rm FC} \gtrsim N_{\rm blocks}$ ($\lesssim N_{\rm blocks}$). In addition, the heavy-tailed Student-t likelihood and prior pdfs may lead to a broader posterior spread in some blocks in the presence of actual data outliers or unmodeled physics.

\subsection{Joint posterior estimate at a fixed initial time}~\label{sec:t0_backprop}

In some instances, particularly when comparing the loosely-coherent posteriors against fully-coherent ones (Appendix~\ref{app:plots}), we may be interested in the joint posterior on the parameter set $(\vec\varphi_0, \vec\vartheta)$ where $\vec\varphi_0$ is the evolving set evaluated at a fixed initial time $t_0$. Explicitly, the evolving parameter set $\vec\varphi(t)$ can be written as
\begin{align}
    \vec{\varphi}(t) \equiv f(\vec{\varphi}_0,\vec\vartheta;t)~\label{eq:varphievolution}
\end{align}
where $f$ is the forward-in-time, vacuum-GR trajectory evolution function defined in Sec.~\ref{sec:studenttprior}. Additionally, for $t_i$ the start time of the $i$th block, we define the shorthand $\vec\varphi_i := \vec\varphi(t_i)$. Then, if we set the start of the first block as the initial time, we have $\vec\varphi_{i=0} \equiv \vec\varphi_0$, and the evolution of all $\{\vec\varphi_{i>0}\}$ is given by Eq.~\eqref{eq:varphievolution}. Furthermore, the joint posterior on $(\vec\varphi_0,\vec\vartheta)$ from all $i$ independent blocks can be expressed as the push-forward of the joint posterior in Eq.~\eqref{eq:jointposteriorprod},
\begin{align}
    p(\{\vec\varphi_{0,i}\},\vec\vartheta|\{\vec{d}_i\}) = &p(\{\vec\varphi_i = f(\vec\varphi_{0,i},\vec\vartheta;t_i)\},\vec\vartheta|\{\vec{d}_i\})\times \nonumber\\
    &\prod_i \left|{\rm det}(J_i)\right|.
\end{align}
The notation $\{\vec\varphi_{0,i}\}$ is deliberate, to emphasize that estimates on $\vec\varphi_0$ are obtained independently from each $i{\rm th}$ block. Here, the determinant of the Jacobian matrix $J_i$ is explicitly, 
\begin{align}
    |{\rm det}(J_i)| &:= \left|\begin{matrix}
        \frac{\partial \vec\varphi_i}{\partial \vec\varphi_{0,i}} & \frac{\partial \vec\varphi_i}{\partial \vec\vartheta}\\
        \frac{\partial \vec\vartheta}{\partial \vec\varphi_{0,i}} & \frac{\partial \vec\vartheta}{\partial \vec\vartheta}
    \end{matrix}\right| \nonumber\\
    &= \left|\begin{matrix}
        \frac{\partial \vec\varphi_i}{\partial \vec\varphi_{0,i}} & \frac{\partial \vec\varphi_i}{\partial \vec\vartheta}\\
        0 & I_{D_{\rm static}}
    \end{matrix}\right| \nonumber \\
    &= \left|\frac{\partial\vec\varphi_i}{\partial\vec\varphi_{0,i}}\right|.
\end{align} 
In the second line, we note that $\vec\vartheta$ is independent of $\vec\varphi_{0,i}$ such that the bottom left block of the Jacobian is zero, and that $\partial\vec\vartheta/\partial\vec\vartheta = I_{D_{\rm static}}$ where $D_{\rm static}$ is the dimensionality of $\vec\vartheta$. 

In practice, given $m$ samples $\{\hat\varphi_i^k,\hat\vartheta^k\}_{k=1}^m$ from the marginal posterior $p(\vec\varphi_i,\vec\vartheta|\{\vec{d}_i\}) = \int_{j\neq i}{\rm d}\{\vec\varphi_j\}p(\{\vec\varphi_i\},\vec\vartheta|\{\vec{d}_i\})$, valid samples from the joint posterior of $(\{\vec\varphi_{0,i}\},\vec\vartheta)$ are simply obtained as,
\begin{align}
    (\hat\varphi^{(k)}_{0,i},\hat\vartheta^{(k)}) = f^{-1}(\hat\varphi^{(k)}_i,\hat\vartheta^{(k)};t_i),\hat\vartheta^{(k)}
\end{align}
where $f^{-1}(\vec\varphi,\vec\vartheta;t)$ is the inverse of $f$ describing the \textit{backwards-in-time}, vacuum-GR trajectory evolution of $\vec\varphi$ at time $-t$, and $k$ is the sample index. See, e.g., Ref.~\cite[Sec. 2.1]{CaseBerg:01}.

\section{Results}\label{sec:results}

We now apply the loosely-coherent framework constructed in Sec.~\ref{sec:methods} to various data injection examples and compare its results against fully-coherent baselines. In all following examples, the analysis model comprises a vacuum-GR EMRI/IMRI template. In addition, we consider three distinct injection examples: (i) a vacuum-GR signal with a zero-noise injection to check for the pipeline's consistency (Sec.~\ref{sec:results_example1}); (ii) a vacuum-GR signal with a stationary Gaussian noise injection to validate the block-independence assumption (Sec.~\ref{sec:results_example2}); and (iii) an environment-rich IMRI signal with an accretion effect and a zero-noise injection to check for robustness against secular perturbations (Sec.~\ref{sec:results_example3}). We additionally repeat the analysis in Example~\rom{3} with a different number of blocks to test the scaling derived in Sec.~\ref{sec:snrloss} (Sec.~\ref{sec:results_example3_Nblocks3}). The setup is described in more detail in Sec.~\ref{sec:results_setup} and the respective example subsections.

Our methods are implemented in a Python package which we call \textsc{SPLIT} (Segmented Posteriors for Long-Inspiral Templates)\footnote{\url{https://github.com/perturber/SPLIT}}~\cite{kejriwal_2026_20290209}. The package provides ample control over EMRI/IMRI model parameters and the analysis configuration. It employs the \textsc{FastEMRIWaveforms} (FEW) package~\cite{2021PhRvL.126e1102C,2021PhRvD.104f4047K,2023arXiv230712585S,2025PhRvD.112j4023C} for rapidly generating the injected signal and the analysis templates, and the \textsc{Eryn} ensemble sampler~\cite{2023MNRAS.526.4814K,michael_katz_2025_17162828,2013PASP..125..306F} for performing Markov Chain Monte Carlo (MCMC) sampling of the posterior~\eqref{eq:jointposteriorprod}. The static and evolving parameter decomposition in \textsc{SPLIT} is naturally handled through \textsc{Eryn}'s hierarchical \textit{branch} and \textit{leaf} structure~\cite{2023MNRAS.526.4814K}. We define two branches: one corresponding to the evolving parameter set with $N_{\rm blocks}$ independent leaves and another for the static parameter set with exactly 1 leaf. \textsc{SPLIT} takes advantage of GPU-based acceleration native to both \textsc{FEW} and \textsc{Eryn}. If available, \textsc{SPLIT} can also parallelize over multiple GPUs to manage the cost of loosely-coherent inference (see also Sec.~\ref{sec:discussion_future} for a discussion). Ultimately, we hope that \textsc{SPLIT}'s modular framework and accessible interface will allow for widespread adoption by the GW community for the analysis of long-inspiral sources in the LISA band, as well as other future detectors~\cite{2020JCAP...03..050M,2021arXiv210909882E}.

\subsection{General setup}~\label{sec:results_setup}

We first describe the general setup of our analysis which is consistent across all examples. 

\paragraph{Signal and noise formulation.} We invoke the long-wavelength approximation (LWA)~\cite{2019CQGra..36j5011R} to extract LISA's one-sided noise PSD, $S_n(f)$. We exclude the galactic binary confusion foreground from our analysis for simplicity.  Furthermore, to save computation costs and avoid aliasing, we pass all injected signals and recovery templates through a band-pass filter with $(f_{\rm min}, f_{\rm max}) = (10^{-4},10^{-1})$ Hz, corresponding to the range in which the detector is the most sensitive~\cite{2019CQGra..36j5011R}. In addition, the data in each block is passed through a Tukey window with shape parameter $\alpha = 0.02$. Roughly $\alpha/2$ fraction of the total signal duration is suppressed at the ends of each block to suffice the block-wise noise-independence condition assumed in the likelihood's definition~\eqref{eq:localstationary}. See Appendix~\ref{app:blocindependence} for more details.

\paragraph{Inference setup.} While the physical parameters vary by example, the total dimensionality $D_{\rm infer}$ of the inferred parameter space can be generically written as 
\begin{align}
    D_{\rm infer} = N_{\rm blocks} \times D_{\rm evol} + D_{\rm static}~\label{eq:inferencedim}
\end{align}
where $D_{\rm evol}$ is the dimensionality of the inferred evolving parameters and $D_{\rm static}$ is that of the static parameters. For loosely-coherent inference, we always choose $N_{\rm blocks}=5$ except in the reanalysis of Example~\rom{3} where we assume $N_{\rm blocks} = 3$ (Sec.~\ref{sec:results_example3_Nblocks3}; see also Sec.~\ref{sec:discussion_future} for a discussion on the choice of $N_{\rm blocks}$). Fully-coherent inference is toggled in \textsc{SPLIT} by simply setting $N_{\rm blocks} = 1$. In all examples, we evaluate the Student-t likelihood~\eqref{eq:studenttblocklikelihood} with degrees-of-freedom parameter $\nu_{\rm like} = 5$. The Markovian Student-t transition probabilities in the prior~\eqref{eq:priortransition} are configured with $\nu_{\rm prior} = 10$. Our choice of $\nu_{\rm like}$ and $\nu_{\rm prior}$ is based on typically assumed values; distributions with $\nu \gg 10$ approach the Gaussian limit while $\nu \leq 2$ have excessively heavy-tails and undefined mean and/or variance. A detailed analysis quantifying the impact of different choices of $\nu_{\rm like}$ and $\nu_{\rm prior}$ is beyond the scope of this study, and should be explored in a future work.

\paragraph{Fully-coherent setup.} For fully-coherent inference, we always initialize 15 MCMC walkers across 3 parallel-tempering steps. We use \textsc{Eryn}'s default \texttt{Stretch} and \texttt{Gaussian} moves for proposing new samples. The \texttt{Stretch} move, originally proposed in Ref.~\cite{2013PASP..125..306F}, helps with convergence in highly-correlated parameter spaces, such as that of EMRIs and IMRIs~\cite{2021PhRvD.104f4047K}. The \texttt{Gaussian} move proposes a new point from a simple Gaussian kernel with a fixed kernel size. See Ref.~\cite{2023MNRAS.526.4814K} for more details. For convergence, we require that the Gelman-Rubin statistic, which compares the variance within and between all walker chains~\cite{1992StaSc...7..457G}, is $\hat{R} \leq 1.05$. In addition, we require the total number of samples per chain $N \geq 50\tau$ where $\tau$ is the maximum of the integrated autocorrelation times~\cite{2010CAMCS...5...65G} across all inferred parameters~\cite{2013PASP..125..306F}. The latter condition ensures that we have an adequate number of independent samples from the posterior.

\paragraph{Loosely-coherent setup.} The number of MCMC walkers in the loosely-coherent case vary by example and are described in the respective example sections. There are 5 parallel-tempering steps in all examples, and the convergence criteria remain identical to the fully-coherent case. Crucially, in the loosely-coherent framework, the choice of proposal moves is critical to how quickly---if at all---the chains converge. Because the total dimensionality $D_{\rm infer}$ scales linearly with $N_{\rm blocks}$~\eqref{eq:inferencedim}, the expanded inference space leads to a drastically lower global acceptance rate when using standard MCMC moves.\footnote{We note that this cost scaling is inherent to the inference stage; in the application of semi- and loosely-coherent methods, e.g., to CW \textit{search}, the associated computational cost in fact reduces. This is because CW search typically relies on template banks, and a broadened semi-coherent search statistic relaxes the requirement on template bank density. See Sec.~\ref{sec:discussion_future} for more details.} To circumvent this bottleneck, we introduce custom moves in the \textsc{SPLIT} pipeline built on a block-wise sequential Gibbs sampling strategy. The core idea is to avoid full-dimensional jumps by sequentially updating the evolving and static parameters, holding all others fixed. Specifically, a single proposal cycle steps through the sequence $\vec\varphi_0 \to \vec\varphi_1 \to \ldots \to \vec\varphi_{N_{\rm blocks}-1}\to \vec\vartheta$, after which the sequence repeats. The dimensionality of each step in the sequence is independent of $N_{\rm blocks}$, enabling healthy acceptance rates. To ensure robust exploration of the posteriors in each block, we combine three distinct Markov chain moves in \textsc{SPLIT}: a stretch move, a Gaussian move with a fixed covariance kernel, and a Gaussian move with an adaptive covariance kernel in which the kernel covariance updates proportionately with the sample covariance at the current iteration after every few steps.

\subsection{Example \rom{1}: vacuum-GR EMRI with zero-noise injection}~\label{sec:results_example1}

\begin{table}[tbp]
    \centering
    \caption{\justifying Summary of the injected \texttt{Pn5AAK} EMRI parameters for Example \rom{1} (Sec.~\ref{sec:results_example1}). The total signal duration is $T=1.0$ year. The evolving parameters are reported at the initial time $t=0$.}
    \label{tab:example1_params}
    \renewcommand{\arraystretch}{1.2}
    \begin{tabular}{@{}lll@{}}
        \toprule
        \textbf{Parameter (units)} & \textbf{Symbol} & \textbf{Value} \\
        \midrule
        \multicolumn{3}{@{}l}{\textbf{Intrinsic static parameters}} \\
        \midrule
        Primary mass ($M_\odot$) & $m_1$ & $10^6$  \\
        Secondary mass ($M_\odot$) & $m_2$ & $10$ \\
        Primary spin (--) & $a$ & $0.8$ \\
        \midrule
        \multicolumn{3}{@{}l}{\textbf{Intrinsic evolving parameters (at $\boldsymbol{t=0.0}$)}} \\
        \midrule
        Semi-latus rectum ($m_1$)& $p_0$ & $7.447$ \\
        Eccentricity (--) & $e_0$ & $0.2$ \\
        Cosine of inclination (--) & $x_{I0}$ & $1.0$ \\
        Initial azimuthal phase (rad) & $\Phi_{\varphi,0}$ & $1.0$ \\
        Initial polar phase (rad) & $\Phi_{\theta,0}$ & $0.0$ \\
        Initial radial phase (rad) & $\Phi_{r,0}$ & $1.0$ \\
        \midrule
        \multicolumn{3}{@{}l}{\textbf{Extrinsic static parameters}} \\
        \midrule
        Luminosity distance (Gpc) & $d_L$ & $3.251$ \\
        Source polar angle (rad) & $\theta_S$ & $\pi/4$ \\
        Source azimuthal angle (rad) & $\phi_S$ & $\pi/3$ \\
        Spin polar angle (rad) & $\theta_K$ & $\pi/6$ \\
        Spin azimuthal angle (rad) & $\phi_K$ & $\pi/8$ \\
        \bottomrule
    \end{tabular}
\end{table}

We begin with a consistency check of the \textsc{SPLIT} pipeline by injecting a vacuum-GR EMRI into the data with a zero-noise realization $\vec{n}=0$, such that no recovery biases with respect to the truth are expected. This allows for a direct comparison between the inferred posteriors in the loosely- and fully-coherent frameworks. 

For the signal $\vec{s} \equiv \vec{d}$, we inject a $T=1.0$ year long \texttt{Pn5AAK} waveform~\cite{2022PhRvL.128w1101I} from the \textsc{FEW} package, sampled at an interval of $dt = 10.0$ seconds. While \texttt{Pn5AAK} waveforms are less accurate compared to the fully-relativistic adiabatic inspiral waveforms now available in \textsc{FEW}~\cite{2025PhRvD.112j4023C}, the former's fully-analytic trajectories allow for much faster likelihood evaluation during inference. 

The chosen model parameters alongside their descriptions are summarized in Tab.~\ref{tab:example1_params}. Note the decomposition into the static and evolving sets. In particular, the semi-latus rectum $p_0$ is calculated as $p_0 = p_{\rm plunge}(m_1,m_2,a,e_0;T) + 0.1$ where $p_{\rm plunge}$ is the initial semi-latus rectum for which the secondary plunges into the MBH after time $T$. This definition of $p_0$ ensures that the waveform is near-plunge, allowing us to focus on a LISA-relevant EMRI source~\cite{2017PhRvD..95j3012B,2026arXiv260317072S}. We restrict the evolution to equatorial orbits by fixing the cosine of the orbital inclination $(x_{I0})$ to 1.0, and correspondingly setting the initial polar phase angle $(\Phi_{\theta,0})$ to 0.0. In addition, we scale the luminosity distance $d_L$ to ensure that the injection has an optimal SNR $\rho_{\rm opt} := \langle \vec{s} | \vec{s}\rangle$ of exactly 50.0.

We use the same \texttt{Pn5AAK} model as the analysis template $\vec{h}$. We infer a subset of the model parameters: the static parameter set is $\vec\vartheta = (m_1, m_2, a)$ and the evolving parameter set in each $i$th block is $\vec\varphi_i = (p_i, e_i, \Phi_{\phi_i}, \Phi_{r,i})$. From Eq.~\eqref{eq:inferencedim}, this corresponds to an inference space dimensionality of $(D^{\rm FC}_{\rm infer},D^{\rm SC}_{\rm infer}) = (7,23)$ in the fully- and loosely-coherent cases, respectively. All other parameters are fixed to their injection values (Tab.~\ref{tab:example1_params}). The explicit uniform prior intervals used in the fully- and loosely-coherent analyses are available in Tab.~\ref{tab:example1_priors}. Note that the priors on $p_0$ and $e_0$ are broader in the loosely-coherent case to accommodate for their evolution across blocks. To aid convergence, we initialize the MCMC walkers in a small $D_{\rm infer}$-sphere centered at the true parameter values. Finally, for the prior transition probabilities~\eqref{eq:priortransition}, we set $\Sigma_{\rm prior} = {\rm diag}(10^{-4}, 10^{-5}, 0.5, 0.5)$ corresponding to $p_0$, $e_0$, $\Phi_{\phi_0}$ and $\Phi_{r_0}$, respectively. 

\begin{table}[tbp]
    \centering
    \caption{\justifying Summary of the uniform prior bounds for the fully-coherent (FC) and loosely-coherent (LC) inference configurations for Examples~\rom{1} and {2} (Sec.~\ref{sec:results_example1} and~\ref{sec:results_example2}). The bounds on the evolving parameters in subsequent blocks $i>0$ of the loosely-coherent analysis are the same as in the $i=0$ block reported here.}
    \label{tab:example1_priors}
    \renewcommand{\arraystretch}{1.2}
    \begin{tabular}{@{}lll@{}}
        \toprule
        \textbf{Parameter} & \textbf{FC} & \textbf{LC} \\
        \midrule
        \multicolumn{3}{@{}l}{\textbf{Static parameters}} \\
        \midrule
        $m_1$ & $10^6\times[0.999, 1.001]$ & $10^6\times[0.999, 1.001]$ \\
        $m_2$ & $10\times[0.999, 1.001]$ & $10\times[0.999, 1.001]$ \\
        $a$ & $8.0\times[0.999, 1.001]$ & $8.0\times[0.999, 1.001]$ \\
        \midrule
        \multicolumn{3}{@{}l}{\textbf{Evolving parameters}} \\
        \midrule
        $p_0$ & $[7.373, 7.522]$ & $[5.262, 7.522]$ \\
        $e_0$ & $[0.198, 0.202]$ & $[0.097, 0.202]$ \\
        $\Phi_{\varphi,0}$ & $[0, 2\pi]$ & $[0, 2\pi]$ \\
        $\Phi_{r,0}$ & $[0, 2\pi]$ & $[0, 2\pi]$ \\
        \bottomrule
    \end{tabular}
\end{table}

The MCMC walkers are parallelized across 4 NVIDIA H100 GPUs, with 2 worker subprocesses per GPU. The fully-coherent analysis, with its 15 walkers across 3 temperatures, achieved convergence after approximately $10^4$ iterations. The sampling took $\approx 19$ ms per likelihood evaluation corresponding to a total walltime of $\approx 2.4$ hours. In the loosely-coherent analysis, with 50 walkers across 5 temperatures, convergence was achieved after $\approx 9.5 \times 10^4$ iterations (or $\approx 5.2$ days of walltime). The significant jump in walltime required for the loosely-coherent analysis stems from several factors such as a $\sim 2.5-3$ times increase in the number of walkers and temperatures, chain initialization location, the sizes of $N_{\rm blocks}$ and $D_{\rm infer}$, correlations between the static and evolving parameters, etc. Future implementations can mitigate these costs, e.g., by further optimizing the sampling algorithms or by employing simulation-based inference methods. See Sec.~\ref{sec:discussion_future} for a discussion.

\begin{figure}[tbp]
    \centering
    \includegraphics[width=0.98\linewidth]{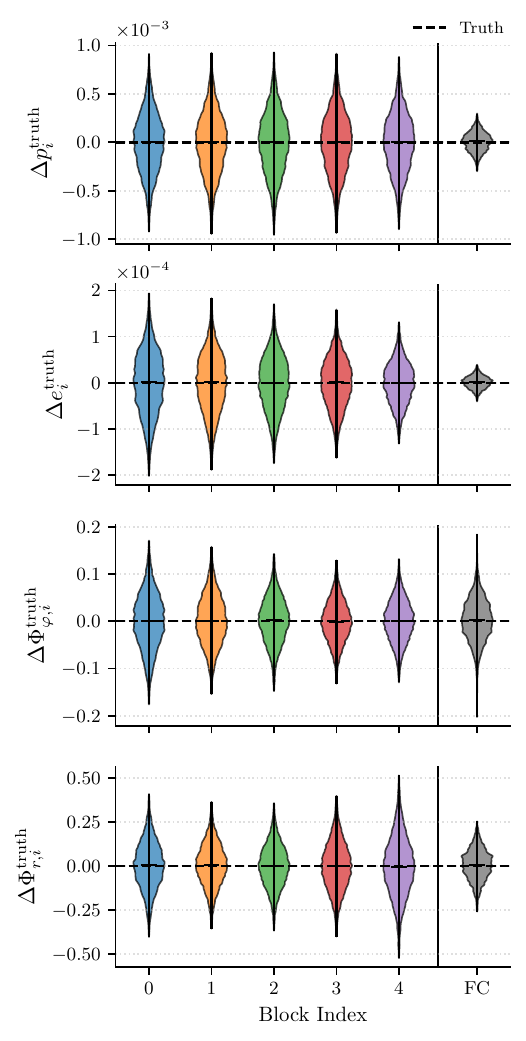}
    \caption{\justifying 1D marginal posterior residuals ($\Delta{\varphi}^{\rm truth}_{ij} = \varphi_{ij}^{\rm truth} - \hat{\varphi}_{ij}$) constructed from the 99.7 percentile evolving parameter samples for the zero-noise vacuum-GR injection (Example \rom{1}, Sec.~\ref{sec:results_example1}). From top to bottom, the panels show the residuals for the semi-latus rectum ($\Delta p_i$), eccentricity ($\Delta e_i$), azimuthal phase ($\Delta \Phi_{\varphi, i}$), and radial phase ($\Delta \Phi_{r, i}$), respectively. In each panel, the first five violin plots display the distributions inferred within blocks $i=0, \dots, 4$ of the loosely-coherent analysis. The rightmost violin plot (FC) corresponds to the residual from the standard fully-coherent analysis. The true injected parameter values lie exactly at zero (horizontal dashed lines).}
    \label{fig:1d_sc_vs_fc_example1}
\end{figure}

To assess the accuracy of the recovered posterior, we compute the residual
$\Delta{\varphi}^{\rm truth}_{ij} = \varphi_{ij}^{\rm truth} - \hat{\varphi}_{ij}$ for the $j$th evolving parameter in the $i$th block, where $\hat{\varphi}_{ij}$ are the 99.7 percentile posterior samples and $\varphi_{ij}^{\rm truth}$ is the true injected value. The resulting 1D residuals are plotted in Fig.~\ref{fig:1d_sc_vs_fc_example1}, together with the residuals from the fully-coherent analysis. As expected for a zero-noise realization, both frameworks recover the true parameters without bias. Because the same SNR is distributed across multiple blocks, the loosely-coherent posteriors are naturally broader than their fully-coherent counterparts as dictated by Eqs.~\eqref{eq:lossofsnr} and~\eqref{eq:lossofsnrposterior}. We obtain similar unbiased recovery for the inferred static parameters $(m_1,m_2,a)$.

The full 7D triangle plot comparing the posterior distributions in both frameworks at the initial time $t=0$ is shown in Fig.~\ref{fig:full_sc_vs_fc_example1} of Appendix~\ref{app:plots}. For this direct comparison, we back-propagate the loosely-coherent samples from all $N_{\rm blocks} > 0$ to $t=0$ using the method described in Sec.~\ref{sec:t0_backprop}. The triangle plot confirms that the loosely-coherent framework retains the strong parameter-space correlations between the static and evolving parameters. 

\subsection{Example \rom{2}: vacuum-GR EMRI with Gaussian noise injection}~\label{sec:results_example2}

In the second example, we validate the block-wise noise independence condition used in the construction of the likelihood~\eqref{eq:localstationary}.

We inject the same \texttt{Pn5AAK} waveform as the signal $\vec{s}$ with identical model parameters as in the previous Example (Sec.~\ref{sec:results_example1}, also see Tab.~\ref{tab:example1_params}). Additionally, for a given one-sided noise PSD function $S_n(f)$, it can be shown from the Wiener-Khinchin theorem~\cite{wiener1930generalized,Khintchine1934} and Parseval's theorem~\cite{2007nras.book.....P} that the variance of the noise at each discrete positive-frequency bin $f_k$ is $\sigma_k^2 = N S_n(f_k) / (4~dt)$, where $N$ is the total number of time samples and $dt$ is the sampling interval. We first construct the frequency-domain noise array $\tilde{n}(f_k)$ by independently drawing its real and imaginary components $\tilde{n}_{k}^{\rm real, imag}$ from a zero-mean Gaussian distribution $\mathcal{N}(0,\sigma_k^2)$ such that
\begin{equation}
    \tilde{n}(f_k) = \hat{n}_{k}^{\rm real} + i\,\hat{n}_{k}^{\rm imag} \,.
\end{equation}
We then apply an inverse fast Fourier transform (iFFT) to $\tilde{n}(f_k)$ to obtain the time-domain noise realization $\vec{n}$, and add it to the EMRI signal to construct the data array: $\vec{d} = \vec{s} + \vec{n}$.

The recovery template, fixed parameters, and uniform prior boundaries (Tab.~\ref{tab:example1_priors}) remain identical to those used in Example~\rom{1} (Sec.~\ref{sec:results_example1}). Again, we achieve convergence after $\approx 10^4$ iterations in the fully-coherent analysis, while the loosely-coherent run converged after $\approx 1.5\times 10^5$ iterations.

We find that both the fully- and loosely-coherent analyses successfully recover the true injected parameters within their respective $\sim1\sigma$ credible intervals, confirming that the noise realization does not introduce any systematic biases into the likelihood~\eqref{eq:localstationary}. In fact, since the loosely-coherent framework accumulates a lower fractional SNR per block~\eqref{eq:lossofsnr}, its resulting posteriors are inherently broader~\eqref{eq:lossofsnrposterior}, and as a result the true parameter values sit more comfortably within its wider credible intervals, as expected from Eq.~\eqref{eq:zscorescaling}. The full 7D joint posteriors at the initial time $t=0$ are presented in Fig.~\ref{fig:full_sc_vs_fc_example2} of Appendix~\ref{app:plots}.

\subsection{Example \rom{3}: environment-rich IMRI injection}~\label{sec:results_example3}

For the third and final example, we increase the astrophysical complexity of the injected signal by introducing an environment-rich source where the secondary is embedded in an accretion disk. While the MBH mass remains $m_1 = 10^6 M_{\odot}$, we increase the secondary mass to $m_2 = 2\times 10^3 M_\odot$. This takes the mass ratio to $q \sim 10^{-3}$, pushing the binary into the IMRI regime~\cite{2024arXiv240207571C,2026arXiv260317072S}. Transitioning to an IMRI allows us to test \textsc{SPLIT}'s robustness against significant secular perturbations with computationally efficient templates as we will describe in more detail shortly. 

The injected signal deviates from a pure vacuum-GR trajectory due to the secondary's interaction with the disk, which we model following the framework of Ref.~\cite{2025PhRvD.111h4006D}. In this model, the disk's surface density and aspect ratio are defined respectively as
\begin{align}
    \Sigma(r) &= \Sigma_0\left(\frac{r}{10m_1}\right)^{-\Sigma_p},\\
    h(r) &= h_0\left(\frac{r}{10m_1}\right)^{(2\Sigma_p-1)/4}.
\end{align}
Here, $\Sigma_0$ and $h_0$ are characteristic values in the inner region of the disk, and $\Sigma_p$ is the power-law scaling that governs the disk profile. We fix $\Sigma_p = -3/2$ in this example which approximates the relevant inner-region of a geometrically-thin $(h \ll r)$, optically-thick Shakura-Sunyaev $\alpha$-disk~\cite{1973A&A....24..337S,2013LRR....16....1A,2011PhRvL.107q1103Y,2014PhRvD..89j4059B}. Ref.~\cite{2025PhRvD.111h4006D} models the secular perturbation on the semi-major axis $a_m$ and orbital eccentricity $e$ of the secondary's orbit around a Schwarzschild MBH as,
\begin{align}
    \frac{t_{\rm gas}}{t_e} &= 0.78(1-e^2)^{1/4}\left({1+\frac{1}{30}\left(\frac{e}{h}\right)^3}\right)^{-1},~\label{eq:eccevolve}\\
    \frac{t_{\rm gas}}{t_{a_m}} &= 2C_{\rm sub}h^2\left(1-e^2\right)\frac{1-\left(\frac{e}{1.25h}\right)^4}{1+(\frac{e}{1.75h})^5},~\label{eq:semimajorevolve}
\end{align}
where $t_e = -e/\dot{e}$ and $t_{a_m} = -a_m/\dot{a}_m$ are the evolution timescales of the orbital eccentricity and semi-major axis, respectively, and $C_{\rm sub} = 2.15 + 0.04\Sigma_p$ is a constant matched against numerical simulations~\cite{2025PhRvD.111h4006D}. Migration torques induced by the disk on the secondary act on the inverse timescale $1/t_{\rm gas} = q(\Sigma r^2/m_1)(\Omega_K/h^4)$ with $\Omega_K=\sqrt{m_1/r^3}$ the Keplerian orbital frequency~\cite{2025PhRvD.111h4006D}. Additionally, the semi-latus rectum $p$ is related to $a_m$ and $e$ as $a_m = p/(1-e^2)$. 

Since we are interested in the evolution of the orbital parameters over the orbit-averaged radiation-reaction timescale, we can approximate the instantaneous radial position of the secondary as $r \approx a_m$. Note that in Eq.~\eqref{eq:semimajorevolve}, the semi-major axis evolution changes sign when the secondary transitions from the supersonic ($e \gg h$) to the subsonic ($e \lesssim h$) regime. This happens because the dominant effect transitions from local dynamical friction in the supersonic regime (which leads to a net outward migration for non-zero eccentricities) to planetary migration in the subsonic regime (which always causes inward migration)~\cite{2025PhRvD.111h4006D}.

In Ref.~\cite{2025PhRvD.111h4006D}, a secondary of mass $m_2 = 50 M_\odot$ was injected over a 4-year long signal, for which the accretion effect was shown to be measurable in some of the example cases. However, the computational expense of the loosely-coherent inference pipeline in this paper restricts us to the analysis of significantly shorter signals. We therefore increase the secondary mass to amplify the environmental perturbation, allowing significant dephasing to build up over a much shorter observation duration. Furthermore, the time spent by the signal in the LISA sensitivity band scales inversely with the mass ratio because higher mass-ratio systems radiate GWs and hence evolve more rapidly. As a result, we find that an observation time of $T=0.5$ years suffices for our analysis. 

\begin{table}[tbp]
    \centering
    \caption{\justifying Summary of the injected \texttt{KerrAccretion} IMRI parameters for Example \rom{3} (Sec.~\ref{sec:results_example3}). The total signal duration is $T=0.5$ years. The evolving parameters are reported at the initial time $t=0$.}
    \label{tab:example3_params}
    \renewcommand{\arraystretch}{1.2}
    \begin{tabular}{@{}lll@{}}
        \toprule
        \textbf{Parameter (units)} & \textbf{Symbol} & \textbf{Value} \\
        \midrule
        \multicolumn{3}{@{}l}{\textbf{Intrinsic static parameters}} \\
        \midrule
        Primary mass ($M_\odot$) & $m_1$ & $10^6$  \\
        Secondary mass ($M_\odot$) & $m_2$ & $2\times10^3$ \\
        Primary spin (--) & $a$ & $0.0$ \\
        Surface density ($g/cm^2$) & $\Sigma_0$ & $5.25\times 10^{5}$ \\
        Aspect ratio (--) & $h_0$ & 0.025 \\
        Disk profile (--) & $\Sigma_p$ & -3/2 \\
        \midrule
        \multicolumn{3}{@{}l}{\textbf{Intrinsic evolving parameters (at $\boldsymbol{t=0.0}$)}} \\
        \midrule
        Semi-latus rectum ($m_1$)& $p_0$ & $24.442$ \\
        Eccentricity (--) & $e_0$ & $0.04$ \\
        Cosine of inclination (--) & $x_{I0}$ & $1.0$ \\
        Initial azimuthal phase (rad) & $\Phi_{\varphi,0}$ & $1.0$ \\
        Initial polar phase (rad) & $\Phi_{\theta,0}$ & $0.0$ \\
        Initial radial phase (rad) & $\Phi_{r,0}$ & $1.0$ \\
        \midrule
        \multicolumn{3}{@{}l}{\textbf{Extrinsic static parameters}} \\
        \midrule
        Luminosity distance (Gpc) & $d_L$ & $16.344$ \\
        Source polar angle (rad) & $\theta_S$ & $\pi/4$ \\
        Source azimuthal angle (rad) & $\phi_S$ & $\pi/3$ \\
        Spin polar angle (rad) & $\theta_K$ & $\pi/6$ \\
        Spin azimuthal angle (rad) & $\phi_K$ & $\pi/8$ \\
        \bottomrule
    \end{tabular}
\end{table}

\begin{figure}[tbp]
    \centering
    \includegraphics[width=0.98\linewidth]{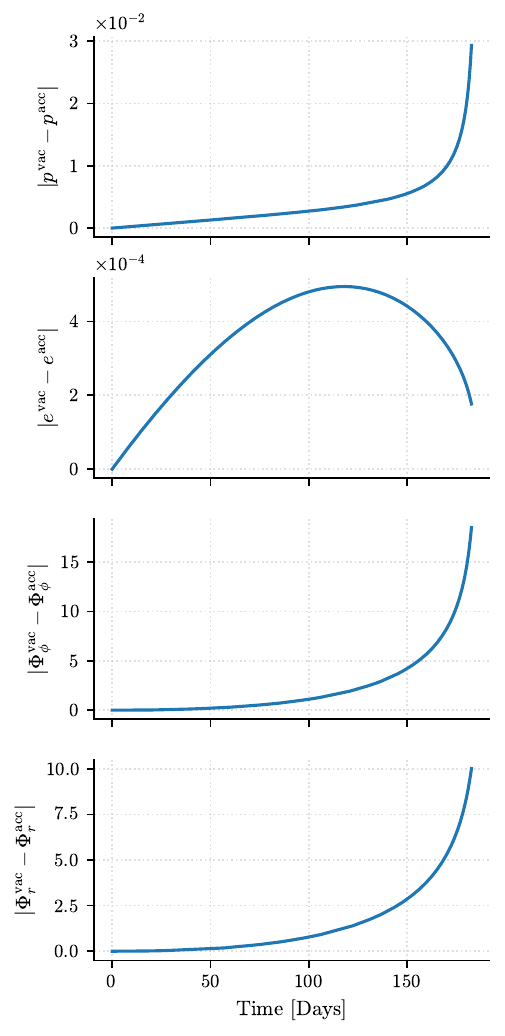}
    \caption{\justifying Absolute difference $|\varphi_j^{\rm vac} - \varphi_j^{\rm acc}|$ between the $j$th evolving parameter's trajectory in the environment-rich IMRI injection and the signal's vacuum-GR restriction. The accretion model is adapted from Ref.~\cite{2025PhRvD.111h4006D}. From top to bottom, we plot the differences in $p$, $e$, $\Phi_\phi$, and $\Phi_r$, respectively, along the full $T=0.5$ year inspiral. Note the unique evolution in the orbital eccentricity case (second panel) due to the supersonic to subsonic transition of the secondary within the observation window (see corresponding text and Ref.~\cite{2025PhRvD.111h4006D} more details).}
    \label{fig:trajcompare_example3}
\end{figure}

To model the environment-rich IMRI, we define a custom waveform class called \texttt{FastKerrEccentricEquatorialAccretionFlux} (hereafter \texttt{KerrAccretion}) built upon the fully-relativistic adiabatic Kerr inspiral model native to \textsc{FEW}~\cite{2025PhRvD.112j4023C}. Such custom waveform classes can be directly fed into the \textsc{SPLIT} pipeline. The injected parameters of the IMRI source are summarized in Tab.~\ref{tab:example3_params}. Correspondingly, in Fig.~\ref{fig:trajcompare_example3}, we plot the absolute difference $|\varphi_j^{\rm vac} - \varphi_j^{\rm acc}|$ comparing the trajectory of the $j$th evolving parameter in the injected environment-rich IMRI signal and its corresponding vacuum-GR restriction (obtained by setting $\Sigma_0 = h_0 = 0$). We find that both the radial and azimuthal phases (bottom two panels in Fig.~\ref{fig:trajcompare_example3}) incur a maximum dephasings $\gtrsim 10$ radians, hinting towards a significant deviation from pure vacuum-GR dynamics. Additionally, we note the presence of a maximum in the absolute difference of the eccentricity trajectories at $\sim 110$ days (second panel from top). This is because the secondary undergoes a supersonic to subsonic transition during the observation such that the eccentricity in the accretion trajectory is initially damped more quickly and then less rapidly than in the vacuum-GR case~\cite{2025PhRvD.111h4006D}. 

We employ the same \texttt{KerrAccretion} waveform model as the analysis template, but with the accretion parameters $\Sigma_0$, $h_0$, $\Sigma_p$ fixed to 0 such that the template space is restricted to the vacuum-GR limit. We infer only two static parameters in this example: $\vec\vartheta = (m_1, m_2)$ with the primary spin $a=0$ fixed in the analysis, and the same set of four evolving parameters $\vec\varphi_0=(p_0,e_0,\Phi_{\phi_0},\Phi_{r_0})$ as in the previous examples. At first, we again segment the data into 5 blocks such that from Eq.~\eqref{eq:inferencedim}, $(D^{\rm FC}_{\rm infer},D^{\rm LC}_{\rm infer})=(6,22)$. Later, in Sec.~\ref{sec:results_example3_Nblocks3}, we reduce $N_{\rm blocks}$ to 3 to qualitatively demonstrate the impact of the number of blocks on systematic biases. The uniform priors on the inferred parameters are summarized in Tab.~\ref{tab:example3_priors}. To accommodate the trajectory's departure from vacuum-GR evolution, we introduce a more conservative $\Sigma_{\rm prior} = {\rm diag}(10^{-2}, 10^{-3}, 1.0, 1.0)$ for the prior transition probability~\eqref{eq:priortransition} in this example. 

Parallelized over 4 NVIDIA H100 GPUs with 2 subprocesses per GPU, each likelihood evaluation took $\approx 44$ ms on average. Chains in the fully-coherent analysis, with 15 walkers across 3 temperatures, converged after $\approx 9\times10^3$ iterations, or equivalently a total walltime of $\approx 5$ hours. The loosely-coherent analysis chains (45 walkers across 5 temperatures) also reached convergence after $\approx 3\times 10^4$ iterations or $\approx 3.4$ days of walltime (see Sec.~\ref{sec:discussion_future} for a discussion on future cost mitigation strategies in the loosely-coherent analysis).

\begin{figure}[tbp]
    \centering
    \includegraphics[width=0.98\linewidth]{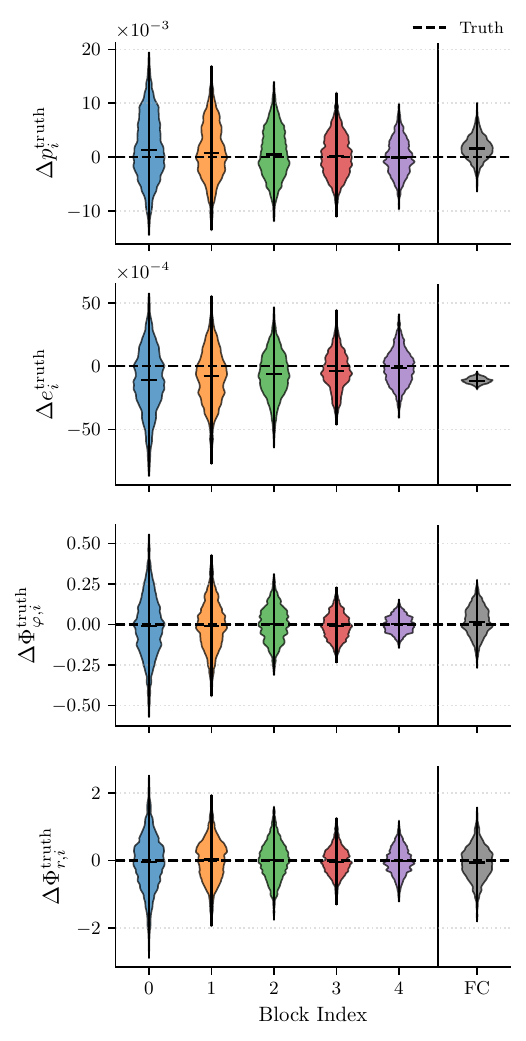}
    \caption{\justifying 1D marginal posterior residuals ($\Delta{\varphi}^{\rm truth}_{ij} = \varphi_{ij}^{\rm truth} - \hat{\varphi}_{ij}$) constructed from the 99.7 percentile evolving parameter samples for the environment-rich IMRI injection (Example \rom{3}, Sec.~\ref{sec:results_example3}). From top to bottom, the panels show the residuals for the semi-latus rectum ($\Delta p_i$), eccentricity ($\Delta e_i$), azimuthal phase ($\Delta \Phi_{\varphi, i}$), and radial phase ($\Delta \Phi_{r, i}$), respectively. In each panel, the first five violin plots display the distributions inferred within blocks $i=0, \dots, 4$ of the loosely-coherent analysis. The rightmost violin plot (FC) corresponds to the residual from the standard fully-coherent analysis. The true injected parameter values lie exactly at zero (horizontal dashed lines).}
    \label{fig:1d_sc_vs_fc_example3}
\end{figure}

\begin{table}[tbp]
    \centering
    \caption{\justifying Summary of the uniform prior bounds for the fully-coherent (FC) and loosely-coherent (LC) inference configurations for Example~\rom{3} (Sec.~\ref{sec:results_example3}). The bounds on the evolving parameters in subsequent blocks $i>0$ of the loosely-coherent analysis are the same as in the $i=0$ block reported here.}
    \label{tab:example3_priors}
    \renewcommand{\arraystretch}{1.2}
    \begin{tabular}{@{}lll@{}}
        \toprule
        \textbf{Parameter} & \textbf{FC} & \textbf{LC} \\
        \midrule
        \multicolumn{3}{@{}l}{\textbf{Static parameters}} \\
        \midrule
        $m_1$ & $10^6\times[0.999, 1.001]$ & $10^6\times[0.999, 1.001]$ \\
        $m_2$ & $[1998, 2002]$ & $[1998, 2002]$ \\
        \midrule
        \multicolumn{3}{@{}l}{\textbf{Evolving parameters}} \\
        \midrule
        $p_0$ & $[24.197, 24.686]$ & $[16.946, 24.686]$ \\
        $e_0$ & $[0.035,0.045]$ & $[0.01,0.05]$ \\
        $\Phi_{\varphi,0}$ & $[0, 2\pi]$ & $[0, 2\pi]$ \\
        $\Phi_{r,0}$ & $[0, 2\pi]$ & $[0, 2\pi]$ \\
        \bottomrule
    \end{tabular}
\end{table}

The 1D marginal residuals $\Delta\varphi_{ij}^{\rm truth}$ of the recovered evolving parameters are visualized in Fig.~\ref{fig:1d_sc_vs_fc_example3}. In both, the fully- and loosely-coherent analyses, the azimuthal and radial phases do not incur significant biases. On the other hand, non-zero systematic biases are incurred in $p$ and $e$ in both frameworks, more substantially in the fully-coherent case as we will quantify shortly. In the loosely-coherent case, these biases are larger in earlier time blocks $(i\leq2)$. This time-dependence reflects the nature of the injected accretion modification~\eqref{eq:eccevolve},~\eqref{eq:semimajorevolve}, which is an effective negative post-Newtonian order modification~\cite{2025PhRvD.111h4006D}, inducing a larger perturbation onto the secondary in the early inspiral phase. The corresponding 6D joint posteriors at the initial time $t=0$ are presented in Fig.~\ref{fig:full_sc_vs_fc_example3} of Appendix~\ref{app:plots}.

\begin{table*}[tbp]
    \centering
    \caption{\justifying Statistical biases in the posterior recovery of the environment-rich IMRI injection (Example~\rom{3}, Sec.~\ref{sec:results_example3}) in the fully-coherent (FC) and loosely-coherent (LC) analyses. The 6D Mahalanobis distance ($D_{\rm Maha}$) and 1D marginal $z$-scores quantify the deviation of the inferred parameters from the true injected values.}
    \label{tab:bias_metrics_example3}
    \renewcommand{\arraystretch}{1.2}
    \begin{tabular}{@{}lcccccc@{}}
        \toprule
        \textbf{Parameters} & \textbf{LC ($i=0$)} & \textbf{LC ($i=1$)} & \textbf{LC ($i=2$)} & \textbf{LC ($i=3$)} & \textbf{LC ($i=4$)} & \textbf{FC} \\
        \midrule
        \multicolumn{7}{@{}l}{\textbf{Mahalanobis distance}} \\
        \midrule
        (Full-dim) & 0.646 & 0.693 & 0.784 & 0.858 & 0.639 & 1207.835 \\
        \midrule
        \multicolumn{7}{@{}l}{\textbf{1D $z$-scores}} \\
        \midrule
        $m_1$ & 0.017 & - & - & - & - & 0.154 \\
        $m_2$ & 0.031 & - & - & - & - & 0.075 \\
        $p_i$ & 0.324 & 0.246 & 0.176 & 0.085 & 0.027 & 0.689 \\
        $e_i$ & 0.469 & 0.401 & 0.355 & 0.222 & 0.065 & 4.802 \\
        $\Phi_{\varphi,i}$ & 0.041 & 0.036 & 0.004 & 0.120 & 0.029 & 0.213 \\
        $\Phi_{r,i}$ & 0.071 & 0.015 & 0.019 & 0.032 & 0.039 & 0.116 \\
        \bottomrule
    \end{tabular}
\end{table*}

To quantify the systematic biases incurred in the posterior recovery, we calculate the 1D $z$-scores for each modeled parameter $\vec\theta_i=(\vec\vartheta,\vec\varphi_i)$ as $z_{ij}=|\theta^{\rm truth}_{ij}-\mu_{ij}|/\sigma_{ij}$ where $\mu_{ij}$ is the mean of the $j$th parameter's samples in the $i$th block and $\sigma_{ij}$ is the corresponding standard deviation. Additionally, to gauge the full 6-dimensional bias in each block, we calculate the Mahalanobis distance~\cite{mahalanobis} $D_{\rm Maha,i} = \sqrt{(\vec\theta^{\rm truth}_{i}-\vec{\mu}_i)^T\Sigma_i^{-1}(\vec\theta^{\rm truth}_{i}-\vec{\mu}_i)}$ where $\Sigma_i^{-1}$ is the sample covariance matrix in the $i$th block. For the fully-coherent case, we calculate the $z$-scores and Mahalanobis distances identically, except we drop the block index $i$. 

The $z$-scores and Mahalanobis distances are reported in Tab.~\ref{tab:bias_metrics_example3}, and reveal a sharp contrast between the two approaches. The loosely-coherent analysis consistently incurs significantly smaller systematic biases than the fully-coherent case across all inferred parameters. Most notably, while the fully-coherent posterior lies far away from the true values with $D_{\rm Maha} \approx 1200$, all blocks in the loosely-coherent analysis remain fully consistent with the injection, yielding $D_{\rm Maha, i} \lesssim 0.9$. From the 1D marginal $z$-scores, we find that the largest individual biases in both analyses are incurred in the orbital eccentricity $e$. This is physically consistent with the unique accretion-induced evolution of $e$ discussed earlier, making its trajectory harder to fit with a pure vacuum-GR template, especially over the full observation window. Consequently, this effect manifests as a $\approx 4.8\sigma$ bias in $e$ in the fully-coherent case. The loosely-coherent pipeline, inferring the trajectory over the much shorter $T_{\rm block}=0.1$ year window per block, is able to largely absorb the resulting deviations and yields only a maximum bias of $\approx 0.5\sigma$.

\subsubsection{Reanalyzing Example~\rom{3} with $N_{\rm blocks} = 3$ to assess the dependence on the number of blocks}~\label{sec:results_example3_Nblocks3}

To assess how the results depend on the number of blocks, here we repeat the analysis of Example~\rom{3} (Sec.~\ref{sec:results_example3}) with $N_{\rm blocks} = 3$, keeping the injection, the analysis template, the priors, and all hyperparameters unchanged. From Eq.~\eqref{eq:inferencedim}, the inference dimensionality reduces from $D_{\rm infer} = 22$ to $14$. We use 30 MCMC walkers across 5 parallel-tempering steps, and the chains satisfy the same convergence criteria. We achieve convergence after $\approx 10^3$ iterations (or $\approx 2$ hours), more than an order-of-magnitude faster than the $N_{\rm blocks} = 5$ case, and consistently with the dimensionality scaling of the convergence rates discussed in Sec.~\ref{sec:results_setup}.

Across the three blocks, we obtain Mahalanobis distances of $D_{{\rm Maha}, i} = 2.518$, $4.438$, and $20.371$, and a maximum $z$-score of $0.888$, again incurred in the orbital eccentricity of the earliest block. The recovered metrics are thus higher, i.e. worse, than the $N_{\rm blocks} = 5$ case, but still significantly better than the FC case. This is expected since the phase coherence now must be maintained over a longer $T_{\rm block} = 0.5/3 \approx 0.17$ years baseline, allowing the unmodeled accretion perturbation to accumulate further within each block. The trend is consistent with Eq.~\eqref{eq:zscorescaling}: the maximum $z$-score falls by a factor $\approx 1.9$ between $N_{\rm blocks} = 3$ and $5$, which is larger than the factor $\sqrt{5/3} \approx 1.3$ expected from posterior broadening alone~\eqref{eq:lossofsnrposterior}, indicating that the bias $\Delta\vec\theta_i$ in Eq.~\eqref{eq:zscorescaling} is also suppressed. Conversely, the block-wise posteriors are also narrower in the $N_{\rm blocks} = 3$ case, since each of the longer time-length blocks accumulates a larger fraction of the optimal SNR~\eqref{eq:lossofsnr}. Overall, we find that increasing $N_{\rm blocks}$ improves consistency with the truth at the cost of statistical precision and higher computational costs, as anticipated in Sec.~\ref{sec:snrloss}. A study determining the optimal number of blocks that balances the costs and the method's advantages is beyond the scope of this work, but should be performed in the future. The full 6D joint posteriors for this analysis at the initial time $t=0$ are presented in Fig.~\ref{fig:full_sc_vs_fc_example3_Nblocks3} of Appendix~\ref{app:plots}.

\section{Discussion}\label{sec:discussion}

\subsection{Summary and outlook}

LISA is a next-generation space-based detector which will unlock the GW universe in the milli-Hz band. Among other sources, LISA will observe dozens of EMRIs and IMRIs which will stay in-band for months to years. Standard fully-coherent analyses of these sources impose stringent waveform accuracy requirements and can lead to severe systematic biases in the presence of data outliers or unmodeled physical effects~\cite{2007PhRvD..76j4018C,2008PhRvD..78l4020L,2024PhRvD.109l4048B,2025PhRvD.111h2010K,2025PhRvD.112j4023C,2025LRR....28....9L,2014PhRvD..89j4059B,2023PhRvX..13b1035S,Kejriwal:2023djc,2026PhRvD.113b3036S,2025CQGra..42f5018C,2025arXiv251216322B,2025PhRvD.111l4053B,2026arXiv260317072S}.

In this paper, we proposed a loosely-coherent inference framework to mitigate such challenges. Our method is implemented in a GPU-accelerated, highly modular Python package called \textsc{SPLIT}~\cite{kejriwal_2026_20290209}. It employs a heavy-tailed Student-t likelihood (Sec.~\ref{sec:studenttlikelihood}) to intrinsically down-weight the impact of data outliers and a Markovian Student-t prior (Sec.~\ref{sec:studenttprior}) to allow for sufficient flexibility against unmodeled secular perturbations. We quantified the fractional loss of optimal SNR~\eqref{eq:lossofsnr}, derived a relation between the posterior variance and the size of $N_{\rm blocks}$~\eqref{eq:lossofsnrposterior} (Sec.~\ref{sec:snrloss}), and quantified the scaling of systematic biases through the 1D $z$-scores with $N_{\rm blocks}~\eqref{eq:zscorescaling}$. We also proposed a sequential Gibbs sampling algorithm to efficiently traverse the large dimensionality of the joint posterior space (Sec.~\ref{sec:results_setup}). 

We performed the loosely-coherent analysis on three distinct examples, comparing against fully-coherent baselines in each case. First, as a consistency check, we injected a vacuum-GR EMRI signal with a zero-noise realization (Sec.~\ref{sec:results_example1}), and recovered the true parameters without bias. In the second example, we injected a vacuum-GR EMRI, this time with a zero-mean Gaussian noise process (Sec.~\ref{sec:results_example2}). The true parameters fell consistently within 1$\sigma$ confidence intervals of the posteriors and were found to agree with the fully-coherent results. This example validated the block-independence assumption employed in the construction of the likelihood~\eqref{eq:localstationary}. In the third example, we injected an IMRI signal with an accretion disk secularly perturbing the evolution trajectory of the secondary (Sec.~\ref{sec:results_example3}). We found that across all inferred parameters, the loosely-coherent posteriors with $N_{\rm blocks} = 5$ were significantly more consistent with the true values compared to the fully-coherent case. In particular, the fully-coherent analysis incurred a maximum bias $\approx 4.8\sigma$ in eccentricity and a full-dimensional Mahalanobis distance of $\approx 1200$. In comparison, the maximum bias in the loosely-coherent case was $\approx 0.5\sigma$, also in eccentricity, and a Mahalanobis distance of $\lesssim 0.9$ across all blocks. Finally, in Sec.~\ref{sec:results_example3_Nblocks3}, we reanalyzed Example~\rom{3} with $N_{\rm blocks} = 3$, achieving chain convergence an order-of-magnitude more quickly than the $N_{\rm blocks} = 5$ case but simultaneously incurring larger systematic biases. This sub-example also verified the scaling of posterior widths and $z$-scores with $N_{\rm blocks}$ as formulated in Sec.~\ref{sec:snrloss}. Overall, we found that the results are consistent between the two approaches under ideal conditions while the loosely-coherent framework was considerably more robust when an unmodeled astrophysical effect was injected. The choice of analysis hyperparameters such as $N_{\rm blocks}$, $\nu_{\rm like}$, $\nu_{\rm prior}$, $\Sigma_{\rm prior}$, etc. are expected to impact convergence rates and the degree of incurred systematic biases, and should be studied in much more detail in the future.

Ultimately, fully-coherent analysis remains the gold standard for GW parameter estimation. Indeed, provided sufficient control over systematic biases caused by data outliers, waveform inaccuracies, and unmodeled physical effects, it is this fully-coherent precision that will enable constraints on astrophysical environments, fundamental physics, and cosmology promised by EMRI/IMRI observations in the LISA band. However, even for the transient signals observed by current ground-based detector networks like LIGO-Virgo-KAGRA~\cite{2025arXiv250818082T}, systematic errors remain a significant concern~\cite{2024PhRvD.109d3037D,2024arXiv240502197G,2025PhRvX..15c1036D}. Such challenges will be drastically exacerbated for the much longer duration EMRI/IMRI signals in the LISA band as we discussed in this paper. In this context, the loosely-coherent pipeline may best serve as a bridge between an unconstrained prior space and the fully-coherent posteriors: \textsc{SPLIT} first accurately and robustly pinpoints the narrow, loosely-coherent posterior region around the source parameters and subsequently, fully-coherent follow-ups, utilizing more descriptive waveforms and refined physical assumptions, further constrain the posterior to the $\sim 10^{-4}-10^{-5}$ relative precision necessary for maximizing the scientific potential of EMRIs/IMRIs in the LISA band~\cite{2024arXiv240207571C}. 

\subsection{Future directions}~\label{sec:discussion_future}

Despite all of its advantages, the loosely-coherent framework's primary bottleneck remains the substantial computational cost associated with inferring the large $D_{\rm infer}$-dimensional posterior. Here, we emphasize that this cost is specific to our formulation; indeed computational efficiency has been one of the main motivations for developing semi-coherent methods in the CW search context. There are two main differences between these formalisms. First, our application is to the \textit{inference} stage, unlike the CW context where segmented methods are typically utilized for search. There, relaxing the phase coherence assumption broadens the peak of the detection statistic, so that a given parameter-space region can be covered by a sparser grid of templates~\cite{1998PhRvD..57.2101B,2000PhRvD..61h2001B,2005PhRvD..72d2004C}. For EMRI/IMRI \textit{inference} within the LISA data analysis pipeline, template banks may be computationally restrictive~\cite{2004CQGra..21S1595G}, and the model parameters will be sampled from the continuous posterior space instead. Second, as discussed in Sec.~\ref{sec:introduction}, semi- and loosely-coherent methods typically maximize or marginalize over the segment-level (local) parameters so that the sampled space remains independent of the number of segments, whereas here these evolving parameters are jointly inferred with the global static parameters leading to the linear scaling shown in Eq.~\eqref{eq:inferencedim}. Although the custom sequential Gibbs sampling moves introduced in this work (Sec.~\ref{sec:results_setup}) largely mitigate the low acceptance ratios that plague standard global-proposal algorithms, in all tested examples, the chain convergence rate is still significantly slower than in the fully-coherent case. This large difference can at least in part be attributed to the convergence criteria specified in Sec.~\ref{sec:results_setup} which were constructed with the fully-coherent analysis in mind following standard references~\cite{2013PASP..125..306F,1992StaSc...7..457G}, but which may be too strict for a loosely-coherent analysis. We also initialized the MCMC walkers within a tight $D_{\rm infer}$-sphere with the same volume in the two approaches, which may have disadvantaged the loosely-coherent case due to its larger posterior widths.

Nevertheless, slow convergence is a generic limitation of hierarchical frameworks of such kind, in which a large number of state variables (the set of evolving parameters in each block) are jointly inferred with the global parameters with which they strongly correlate (the static parameters). See, e.g., Refs.~\cite{2000physics...9028N,2007arXiv0708.3797P,2013arXiv1312.0906B}. Since the sampling space's dimensionality scales linearly with the number of data blocks $N_{\rm blocks}$~\eqref{eq:inferencedim}, traversing the full inference space becomes significantly more expensive than the fully-coherent case. To mitigate these costs, future iterations of the \textsc{SPLIT} pipeline can benefit, for example, from the integration of simulation-based inference (SBI) methods~\cite{2020PNAS..11730055C,prince2023understanding}, particularly conditional normalizing flows (CNFs)~\cite{2015arXiv150505770J,2019arXiv191200042W}. One promising approach that builds on the current implementation is to construct a CNF-accelerated Gibbs sampler, in which the sequential MCMC updates of the local evolving parameters $\vec\varphi_i$ in the $i$th block are replaced by direct draws from a trained CNF's posterior estimate for that block, $q(\vec\varphi_i|\vec{d}_i,\vec\vartheta,\{\vec\varphi_{j\neq i}\})$, conditioned on the local data $\vec{d}_i$, the current sample value of the static parameters $\vec\vartheta$, and the current state of the evolving parameters in all $j\neq i$ blocks $\{\vec\varphi_{j\neq i}\}$. By circumventing the costly likelihood evaluation during the sequential updates and achieving an effective sample acceptance rate near unity, this technique may significantly improve convergence rates. Additional gains may also come from accelerated evaluations of the inner product itself for long-duration signals, such as the short Fourier transform framework for compact binary inspirals proposed in Ref.~\cite{2025PhRvD.111j4044T}. The segmented inner product approach extended to EMRIs/IMRIs and directly applied to the per-block likelihood evaluation may reduce the per-likelihood cost reported here.

A primary advantage of the Student-t likelihood constructed in Sec.~\ref{sec:studenttlikelihood} is its heavy-tailed nature, which offers robustness against significant data outliers within any single block. A notable example of such outliers are instrumental artifacts, specifically non-Gaussian noise transients known as glitches~\cite{2022PhRvD.106f2001A}. A recent study~\cite{2025arXiv251216322B} empirically demonstrated that standard fully-coherent EMRI inference is largely robust to weakly-mitigated glitches, introducing biases $\lesssim 1\sigma$, while unmitigated glitches can introduce larger inconsistencies. Compared to similar analyses for MBH binaries~\cite{2023PhRvD.108l3029S}, the $T=2.0$-year long EMRI signals analyzed in Ref.~\cite{2025arXiv251216322B} fared well against glitches. This may be attributed to the much longer duration of EMRI signals compared to glitch durations. However, for IMRIs and especially in the loosely-coherent pipeline where the observation window is segmented into shorter blocks bringing the signal timescales much closer to that of instrumental glitches, strong systematic biases from noise transients may potentially be reintroduced in the local block-wise posteriors. Thus, future studies should validate whether the current loosely-coherent framework, or extensions thereof, can help mitigate such biases and maintain consistency with the injected values. A rigorous study quantifying the impact of glitches on the loosely-coherent framework would necessitate projecting the solar system barycenter (SSB) frame waveform onto time-delay interferometry (TDI) observables~\cite{2021LRR....24....1T,2025PhRvD.112f3041M}, roughly doubling the computational cost per likelihood evaluation compared to the LWA noise model employed in this work. Nevertheless, the \textsc{SPLIT} pipeline is already fully equipped to apply these projections with a simple toggle interfacing with the \textsc{LISAAnalysisTools}~\cite{michael_katz_2025_17138723} and \textsc{FastLISAResponse}~\cite{2022PhRvD.106j3001K,michael_katz_2025_17162632} packages, making detailed studies possible in the near future.

In all but one example, we uniformly segmented the data into $N_{\rm blocks} = 5$ equal-length blocks. This value was chosen empirically to balance the cost of inferring a parameter space of dimensionality $D_{\rm infer} \propto N_{\rm blocks}$~\eqref{eq:inferencedim} against the robustness advantages of the loosely-coherent framework. Future works should study the broader systematic impact of varying $N_{\rm blocks}$ on posterior recovery and convergence rates. They should also combine these insights with the results derived in this paper (Sec.~\ref{sec:snrloss} and Appendix~\ref{app:blocindependence}) to provide practical heuristics over the range of viable $N_{\rm blocks}$. Another important avenue for exploration is the equal-length constraint. Formally, the loosely-coherent framework allows each block to span an arbitrary length. Future renditions of the \textsc{SPLIT} pipeline could gain advantage from this freedom, for example, by assigning longer durations to earlier blocks to compensate for the lower SNR accumulation in the early inspiral phase, or by defining highly localized blocks to isolate identified noise glitches within the LISA data stream. 

A longer-term initiative could be the integration of the loosely-coherent framework within the broader LISA global fit pipeline. In realistic LISA data, multiple source types (indexed by $k=1,\ldots,K$), each with an \textit{a priori} unknown count $N_k$, will overlap in both time and frequency. The goal of the global fit is to simultaneously estimate the count $N_k$ and infer the $D_k\times N_k$-dimensional space spanned by each source category, where $D_k$ is the parameter space dimensionality of the $k$th source type. Several implementations have emerged to undertake this computationally formidable task~\cite{2023PhRvD.107f3004L,2025PhRvD.111j3014D,2025PhRvD.111b4060K}, most of which employ Bayesian inference techniques such as reversible-jump MCMC (RJMCMC)~\cite{Green:1995mxx}. While RJMCMC allows seamless ``jumps'' between analysis spaces with different $N_k$ and $D_k$, such techniques can be prohibitively expensive when generalized over the full range of LISA-band sources. Furthermore, analyzing the full observation window coherently maintains, if not exacerbates, the impact of unmodeled instrumental and astrophysical effects on posterior recovery. By segmenting the data into shorter-duration blocks, each of which can be analyzed much more robustly in the presence of noise and modeling artifacts, the loosely-coherent framework thus offers a natural solution to this problem. 

Finally, beyond the milli-Hz band of LISA, the fundamental principles underlying the loosely-coherent pipeline are broadly applicable to the next generation of ground-based GW observatories. Third-generation (3G) detectors, such as the Einstein Telescope~\cite{2020JCAP...03..050M,2026JCAP...03..081A} and Cosmic Explorer~\cite{2021arXiv210909882E}, will be sensitive in the broad frequency range of $\sim 1-1000$ Hz, and will enable observations of the early inspiral phase of stellar-mass compact binaries months or even years before their coalescence. These signals will be exceptionally rich scientifically, offering unprecedented insights, e.g., into formation channels of the compact binary population~\cite{2026JCAP...03..081A}. However, the long duration of these observations will also introduce some practical challenges---such as the accumulation of phase error over the entire signal duration---that directly mirror the difficulties of EMRI/IMRI inference discussed in this paper~\cite{2025PhRvD.112j2004B}. Similar systematics are pervasive in the analysis of other 3G band sources, such as sub-solar mass IMRIs, due to their rich harmonic content~\cite{2022PhRvL.128k1104B,2025arXiv251113324G}. Consequently, for the same reasons highlighted throughout this work, a loosely-coherent approach for source inference in the 3G detector band should be explored as a natural and robust way forward.

\section*{Acknowledgements}
I am grateful to Alvin J. K. Chua, Jonathan Gair, and Ilya Mandel for useful comments during the preparation of this manuscript. I also acknowledge the support of the NUS Research Scholarship and the computational resources provided by the NUS IT Research and Computing Group.


\bibliography{bibliography.bib}

@ARTICLE{2011PhRvD..84l2004R,
       author = {{R{\"o}ver}, Christian},
        title = "{Student-t based filter for robust signal detection}",
      journal = {\prd},
         year = 2011,
        month = dec,
       volume = {84},
       number = {12},
          eid = {122004},
        pages = {122004},
          doi = {10.1103/PhysRevD.84.122004},
archivePrefix = {arXiv},
       eprint = {1109.0442},
 primaryClass = {physics.data-an},
       adsurl = {https://ui.adsabs.harvard.edu/abs/2011PhRvD..84l2004R}
}

@ARTICLE{2023PhRvD.108j3005S,
       author = {{Sasli}, Argyro and {Karnesis}, Nikolaos and {Stergioulas}, Nikolaos},
        title = "{Heavy-tailed likelihoods for robustness against data outliers: Applications to the analysis of gravitational wave data}",
      journal = {\prd},
         year = 2023,
        month = nov,
       volume = {108},
       number = {10},
          eid = {103005},
        pages = {103005},
          doi = {10.1103/PhysRevD.108.103005},
archivePrefix = {arXiv},
       eprint = {2305.04709},
 primaryClass = {gr-qc},
       adsurl = {https://ui.adsabs.harvard.edu/abs/2023PhRvD.108j3005S}
}

@ARTICLE{2011CQGra..28a5010R,
       author = {{R{\"o}ver}, Christian and {Meyer}, Renate and {Christensen}, Nelson},
        title = "{Modelling coloured residual noise in gravitational-wave signal processing}",
      journal = {Classical and Quantum Gravity},
         year = 2011,
        month = jan,
       volume = {28},
       number = {1},
          eid = {015010},
        pages = {015010},
          doi = {10.1088/0264-9381/28/1/015010},
archivePrefix = {arXiv},
       eprint = {0804.3853},
 primaryClass = {stat.ME},
       adsurl = {https://ui.adsabs.harvard.edu/abs/2011CQGra..28a5010R}
}

@ARTICLE{2022PhRvL.128k1104B,
       author = {{Barsanti}, Susanna and {De Luca}, Valerio and {Maselli}, Andrea and {Pani}, Paolo},
        title = "{Detecting Subsolar-Mass Primordial Black Holes in Extreme Mass-Ratio Inspirals with LISA and Einstein Telescope}",
      journal = {\prl},
         year = 2022,
        month = mar,
       volume = {128},
       number = {11},
          eid = {111104},
        pages = {111104},
          doi = {10.1103/PhysRevLett.128.111104},
archivePrefix = {arXiv},
       eprint = {2109.02170},
 primaryClass = {gr-qc},
       adsurl = {https://ui.adsabs.harvard.edu/abs/2022PhRvL.128k1104B}
}

@ARTICLE{2025arXiv251113324G,
       author = {{Giri}, Devesh and {Gadre}, Bhooshan},
        title = "{Sub-Solar Mass Intermediate Mass Ratio Inspirals: Waveform Systematics and Detection Prospects with Gravitational Waves}",
      journal = {arXiv e-prints},
         year = 2025,
        month = nov,
          eid = {arXiv:2511.13324},
        pages = {arXiv:2511.13324},
          doi = {10.48550/arXiv.2511.13324},
archivePrefix = {arXiv},
       eprint = {2511.13324},
 primaryClass = {gr-qc},
       adsurl = {https://ui.adsabs.harvard.edu/abs/2025arXiv251113324G}
}

@ARTICLE{1992PhRvD..46.5236F,
       author = {{Finn}, Lee S.},
        title = "{Detection, measurement, and gravitational radiation}",
      journal = {\prd},
         year = 1992,
        month = dec,
       volume = {46},
       number = {12},
        pages = {5236-5249},
          doi = {10.1103/PhysRevD.46.5236},
archivePrefix = {arXiv},
       eprint = {gr-qc/9209010},
 primaryClass = {gr-qc},
       adsurl = {https://ui.adsabs.harvard.edu/abs/1992PhRvD..46.5236F}
}

@ARTICLE{1994PhRvD..49.2658C,
       author = {{Cutler}, Curt and {Flanagan}, {\'E}anna E.},
        title = "{Gravitational waves from merging compact binaries: How accurately can one extract the binary's parameters from the inspiral waveform\textbackslash?}",
      journal = {\prd},
         year = 1994,
        month = mar,
       volume = {49},
       number = {6},
        pages = {2658-2697},
          doi = {10.1103/PhysRevD.49.2658},
archivePrefix = {arXiv},
       eprint = {gr-qc/9402014},
 primaryClass = {gr-qc},
       adsurl = {https://ui.adsabs.harvard.edu/abs/1994PhRvD..49.2658C}
}

@book{CaseBerg:01,
  author = {Casella, George and Berger, Roger},
  howpublished = {{Textbook Binding}},
  isbn = {0534243126},
  month = {June},
  opturl = {http://www.amazon.fr/exec/obidos/ASIN/0534243126/citeulike04-21},
  publisher = {{Duxbury Resource Center}},
  title = {Statistical Inference},
  year = 2001
}

@ARTICLE{2019CQGra..36j5011R,
       author = {{Robson}, Travis and {Cornish}, Neil J. and {Liu}, Chang},
        title = "{The construction and use of LISA sensitivity curves}",
      journal = {Classical and Quantum Gravity},
         year = 2019,
        month = may,
       volume = {36},
       number = {10},
          eid = {105011},
        pages = {105011},
          doi = {10.1088/1361-6382/ab1101},
archivePrefix = {arXiv},
       eprint = {1803.01944},
 primaryClass = {astro-ph.HE},
       adsurl = {https://ui.adsabs.harvard.edu/abs/2019CQGra..36j5011R}
}

@ARTICLE{2016PhRvD..93l3009S,
       author = {{Suvorova}, S. and {Sun}, L. and {Melatos}, A. and {Moran}, W. and {Evans}, R.~J.},
        title = "{Hidden Markov model tracking of continuous gravitational waves from a neutron star with wandering spin}",
      journal = {\prd},
         year = 2016,
        month = jun,
       volume = {93},
       number = {12},
          eid = {123009},
        pages = {123009},
          doi = {10.1103/PhysRevD.93.123009},
archivePrefix = {arXiv},
       eprint = {1606.02412},
 primaryClass = {astro-ph.IM},
       adsurl = {https://ui.adsabs.harvard.edu/abs/2016PhRvD..93l3009S}
}

@ARTICLE{2021PhRvL.126e1102C,
       author = {{Chua}, Alvin J.~K. and {Katz}, Michael L. and {Warburton}, Niels and {Hughes}, Scott A.},
        title = "{Rapid Generation of Fully Relativistic Extreme-Mass-Ratio-Inspiral Waveform Templates for LISA Data Analysis}",
      journal = {\prl},
         year = 2021,
        month = feb,
       volume = {126},
       number = {5},
          eid = {051102},
        pages = {051102},
          doi = {10.1103/PhysRevLett.126.051102},
archivePrefix = {arXiv},
       eprint = {2008.06071},
 primaryClass = {gr-qc},
       adsurl = {https://ui.adsabs.harvard.edu/abs/2021PhRvL.126e1102C}
}

@ARTICLE{2021PhRvD.104f4047K,
       author = {{Katz}, Michael L. and {Chua}, Alvin J.~K. and {Speri}, Lorenzo and {Warburton}, Niels and {Hughes}, Scott A.},
        title = "{Fast extreme-mass-ratio-inspiral waveforms: New tools for millihertz gravitational-wave data analysis}",
      journal = {\prd},
         year = 2021,
        month = sep,
       volume = {104},
       number = {6},
          eid = {064047},
        pages = {064047},
          doi = {10.1103/PhysRevD.104.064047},
archivePrefix = {arXiv},
       eprint = {2104.04582},
 primaryClass = {gr-qc},
       adsurl = {https://ui.adsabs.harvard.edu/abs/2021PhRvD.104f4047K}
}

@ARTICLE{2023arXiv230712585S,
       author = {{Speri}, Lorenzo and {Katz}, Michael L. and {Chua}, Alvin J.~K. and {Hughes}, Scott A. and {Warburton}, Niels and {Thompson}, Jonathan E. and {Chapman-Bird}, Christian E.~A. and {Gair}, Jonathan R.},
        title = "{Fast and Fourier: Extreme Mass Ratio Inspiral Waveforms in the Frequency Domain}",
      journal = {arXiv e-prints},
         year = 2023,
        month = jul,
          eid = {arXiv:2307.12585},
        pages = {arXiv:2307.12585},
          doi = {10.48550/arXiv.2307.12585},
archivePrefix = {arXiv},
       eprint = {2307.12585},
 primaryClass = {gr-qc},
       adsurl = {https://ui.adsabs.harvard.edu/abs/2023arXiv230712585S}
}

@ARTICLE{2025PhRvD.112j4023C,
       author = {{Chapman-Bird}, Christian E.~A. and {Speri}, Lorenzo and {Nasipak}, Zachary and {Burke}, Ollie and {Katz}, Michael L. and {Santini}, Alessandro and {Kejriwal}, Shubham and {Lynch}, Philip and {Mathews}, Josh and {Khalvati}, Hassan and {Thompson}, Jonathan E. and {Isoyama}, Soichiro and {Hughes}, Scott A. and {Warburton}, Niels and {Chua}, Alvin J.~K. and {Pigou}, Maxime},
        title = "{Efficient waveforms for asymmetric-mass eccentric equatorial inspirals into rapidly spinning black holes}",
      journal = {\prd},
         year = 2025,
        month = nov,
       volume = {112},
       number = {10},
          eid = {104023},
        pages = {104023},
          doi = {10.1103/scbp-75pf},
archivePrefix = {arXiv},
       eprint = {2506.09470},
 primaryClass = {gr-qc},
       adsurl = {https://ui.adsabs.harvard.edu/abs/2025PhRvD.112j4023C}
}

@ARTICLE{2023MNRAS.526.4814K,
       author = {{Karnesis}, Nikolaos and {Katz}, Michael L. and {Korsakova}, Natalia and {Gair}, Jonathan R. and {Stergioulas}, Nikolaos},
        title = "{Eryn: a multipurpose sampler for Bayesian inference}",
      journal = {\mnras},
         year = 2023,
        month = dec,
       volume = {526},
       number = {4},
        pages = {4814-4830},
          doi = {10.1093/mnras/stad2939},
archivePrefix = {arXiv},
       eprint = {2303.02164},
 primaryClass = {astro-ph.IM},
       adsurl = {https://ui.adsabs.harvard.edu/abs/2023MNRAS.526.4814K}
}

@software{michael_katz_2025_17162828,
  author       = {Michael Katz and
                  Nikolaos Karnesis and
                  Antonio Coín and
                  Casey McGrath and
                  Natalia Korsakova and
                  Rodrigo Tenorio},
  title        = {mikekatz04/Eryn: v1.2.1},
  month        = sep,
  year         = 2025,
  publisher    = {Zenodo},
  version      = {v1.2.1},
  doi          = {10.5281/zenodo.17162828},
  url          = {https://doi.org/10.5281/zenodo.17162828},
  swhid        = {swh:1:dir:ad490bdf164431347ac3db9e309603318619f0af
                   ;origin=https://doi.org/10.5281/zenodo.7688361;vis
                   it=swh:1:snp:050b8385c7cceb9f3a6eab50bcd91f66b53a7
                   ae7;anchor=swh:1:rel:f94ccf81487a51a05f44df1bf1100
                   f1b3b2a29e1;path=mikekatz04-Eryn-2525830
                  },
}

@ARTICLE{2013PASP..125..306F,
       author = {{Foreman-Mackey}, Daniel and {Hogg}, David W. and {Lang}, Dustin and {Goodman}, Jonathan},
        title = "{emcee: The MCMC Hammer}",
      journal = {\pasp},
         year = 2013,
        month = mar,
       volume = {125},
       number = {925},
        pages = {306},
          doi = {10.1086/670067},
archivePrefix = {arXiv},
       eprint = {1202.3665},
 primaryClass = {astro-ph.IM},
       adsurl = {https://ui.adsabs.harvard.edu/abs/2013PASP..125..306F}
}

@ARTICLE{2020JCAP...03..050M,
       author = {{Maggiore}, Michele and {Van Den Broeck}, Chris and {Bartolo}, Nicola and {Belgacem}, Enis and {Bertacca}, Daniele and {Bizouard}, Marie Anne and {Branchesi}, Marica and {Clesse}, Sebastien and {Foffa}, Stefano and {Garc{\'\i}a-Bellido}, Juan and {Grimm}, Stefan and {Harms}, Jan and {Hinderer}, Tanja and {Matarrese}, Sabino and {Palomba}, Cristiano and {Peloso}, Marco and {Ricciardone}, Angelo and {Sakellariadou}, Mairi},
        title = "{Science case for the Einstein telescope}",
      journal = {\jcap},
         year = 2020,
        month = mar,
       volume = {2020},
       number = {3},
          eid = {050},
        pages = {050},
          doi = {10.1088/1475-7516/2020/03/050},
archivePrefix = {arXiv},
       eprint = {1912.02622},
 primaryClass = {astro-ph.CO},
       adsurl = {https://ui.adsabs.harvard.edu/abs/2020JCAP...03..050M}
}

@ARTICLE{2021arXiv210909882E,
       author = {{Evans}, Matthew and {Adhikari}, Rana X and {Afle}, Chaitanya and {Ballmer}, Stefan W. and {Biscoveanu}, Sylvia and {Borhanian}, Ssohrab and {Brown}, Duncan A. and {Chen}, Yanbei and {Eisenstein}, Robert and {Gruson}, Alexandra and {Gupta}, Anuradha and {Hall}, Evan D. and {Huxford}, Rachael and {Kamai}, Brittany and {Kashyap}, Rahul and {Kissel}, Jeff S. and {Kuns}, Kevin and {Landry}, Philippe and {Lenon}, Amber and {Lovelace}, Geoffrey and {McCuller}, Lee and {Ng}, Ken K.~Y. and {Nitz}, Alexander H. and {Read}, Jocelyn and {Sathyaprakash}, B.~S. and {Shoemaker}, David H. and {Slagmolen}, Bram J.~J. and {Smith}, Joshua R. and {Srivastava}, Varun and {Sun}, Ling and {Vitale}, Salvatore and {Weiss}, Rainer},
        title = "{A Horizon Study for Cosmic Explorer: Science, Observatories, and Community}",
      journal = {arXiv e-prints},
         year = 2021,
        month = sep,
          eid = {arXiv:2109.09882},
        pages = {arXiv:2109.09882},
          doi = {10.48550/arXiv.2109.09882},
archivePrefix = {arXiv},
       eprint = {2109.09882},
 primaryClass = {astro-ph.IM},
       adsurl = {https://ui.adsabs.harvard.edu/abs/2021arXiv210909882E}
}

@ARTICLE{2010PhRvD..81j4014G,
       author = {{Gair}, Jonathan R. and {Tang}, Christopher and {Volonteri}, Marta},
        title = "{LISA extreme-mass-ratio inspiral events as probes of the black hole mass function}",
      journal = {\prd},
         year = 2010,
        month = may,
       volume = {81},
       number = {10},
          eid = {104014},
        pages = {104014},
          doi = {10.1103/PhysRevD.81.104014},
archivePrefix = {arXiv},
       eprint = {1004.1921},
 primaryClass = {astro-ph.GA},
       adsurl = {https://ui.adsabs.harvard.edu/abs/2010PhRvD..81j4014G}
}

@ARTICLE{2013LRR....16....7G,
       author = {{Gair}, Jonathan R. and {Vallisneri}, Michele and {Larson}, Shane L. and {Baker}, John G.},
        title = "{Testing General Relativity with Low-Frequency, Space-Based Gravitational-Wave Detectors}",
      journal = {Living Reviews in Relativity},
         year = 2013,
        month = dec,
       volume = {16},
       number = {1},
          eid = {7},
        pages = {7},
          doi = {10.12942/lrr-2013-7},
archivePrefix = {arXiv},
       eprint = {1212.5575},
 primaryClass = {gr-qc},
       adsurl = {https://ui.adsabs.harvard.edu/abs/2013LRR....16....7G}
}

@ARTICLE{2017PhRvD..96h4039C,
       author = {{Chamberlain}, Katie and {Yunes}, Nicol{\'a}s},
        title = "{Theoretical physics implications of gravitational wave observation with future detectors}",
      journal = {\prd},
         year = 2017,
        month = oct,
       volume = {96},
       number = {8},
          eid = {084039},
        pages = {084039},
          doi = {10.1103/PhysRevD.96.084039},
archivePrefix = {arXiv},
       eprint = {1704.08268},
 primaryClass = {gr-qc},
       adsurl = {https://ui.adsabs.harvard.edu/abs/2017PhRvD..96h4039C}
}

@ARTICLE{2011PhRvD..84b4032K,
       author = {{Kocsis}, Bence and {Yunes}, Nicol{\'a}s and {Loeb}, Abraham},
        title = "{Observable signatures of extreme mass-ratio inspiral black hole binaries embedded in thin accretion disks}",
      journal = {\prd},
         year = 2011,
        month = jul,
       volume = {84},
       number = {2},
          eid = {024032},
        pages = {024032},
          doi = {10.1103/PhysRevD.84.024032},
archivePrefix = {arXiv},
       eprint = {1104.2322},
 primaryClass = {astro-ph.GA},
       adsurl = {https://ui.adsabs.harvard.edu/abs/2011PhRvD..84b4032K}
}

@ARTICLE{2015PhRvD..91j2003L,
       author = {{Leaci}, Paola and {Prix}, Reinhard},
        title = "{Directed searches for continuous gravitational waves from binary systems: Parameter-space metrics and optimal Scorpius X-1 sensitivity}",
      journal = {\prd},
         year = 2015,
        month = may,
       volume = {91},
       number = {10},
          eid = {102003},
        pages = {102003},
          doi = {10.1103/PhysRevD.91.102003},
archivePrefix = {arXiv},
       eprint = {1502.00914},
 primaryClass = {gr-qc},
       adsurl = {https://ui.adsabs.harvard.edu/abs/2015PhRvD..91j2003L}
}

@ARTICLE{2011PhRvD..84b3007P,
       author = {{Prix}, R. and {Giampanis}, S. and {Messenger}, C.},
        title = "{Search method for long-duration gravitational-wave transients from neutron stars}",
      journal = {\prd},
         year = 2011,
        month = jul,
       volume = {84},
       number = {2},
          eid = {023007},
        pages = {023007},
          doi = {10.1103/PhysRevD.84.023007},
archivePrefix = {arXiv},
       eprint = {1104.1704},
 primaryClass = {gr-qc},
       adsurl = {https://ui.adsabs.harvard.edu/abs/2011PhRvD..84b3007P}
}

@ARTICLE{2017PhRvD..96f3004A,
       author = {{Ashton}, G. and {Prix}, R. and {Jones}, D.~I.},
        title = "{Statistical characterization of pulsar glitches and their potential impact on searches for continuous gravitational waves}",
      journal = {\prd},
         year = 2017,
        month = sep,
       volume = {96},
       number = {6},
          eid = {063004},
        pages = {063004},
          doi = {10.1103/PhysRevD.96.063004},
archivePrefix = {arXiv},
       eprint = {1704.00742},
 primaryClass = {gr-qc},
       adsurl = {https://ui.adsabs.harvard.edu/abs/2017PhRvD..96f3004A}
}

@ARTICLE{2016PhRvD..93h4024K,
       author = {{Keitel}, David},
        title = "{Robust semicoherent searches for continuous gravitational waves with noise and signal models including hours to days long transients}",
      journal = {\prd},
         year = 2016,
        month = apr,
       volume = {93},
       number = {8},
          eid = {084024},
        pages = {084024},
          doi = {10.1103/PhysRevD.93.084024},
archivePrefix = {arXiv},
       eprint = {1509.02398},
 primaryClass = {gr-qc},
       adsurl = {https://ui.adsabs.harvard.edu/abs/2016PhRvD..93h4024K}
}

@ARTICLE{2026PhRvD.114b4004S,
       author = {{Sasli}, Argyro and {Karamanis}, Minas and {Karnesis}, Nikolaos and {Coughlin}, Michael W. and {Mandic}, Vuk and {Seljak}, Uro{\v{s}} and {Stergioulas}, Nikolaos},
        title = "{Beyond Gaussian assumptions: A new robust statistical framework for gravitational-wave data analysis}",
      journal = {\prd},
         year = 2026,
        month = jul,
       volume = {114},
       number = {2},
          eid = {024004},
        pages = {024004},
          doi = {10.1103/n5vd-kzp1},
archivePrefix = {arXiv},
       eprint = {2602.22074},
 primaryClass = {gr-qc},
       adsurl = {https://ui.adsabs.harvard.edu/abs/2026PhRvD.114b4004S}
}

@ARTICLE{2025PhRvD.111j4044T,
       author = {{Tenorio}, Rodrigo and {Gerosa}, Davide},
        title = "{Scalable data-analysis framework for long-duration gravitational waves from compact binaries using short Fourier transforms}",
      journal = {\prd},
         year = 2025,
        month = may,
       volume = {111},
       number = {10},
          eid = {104044},
        pages = {104044},
          doi = {10.1103/PhysRevD.111.104044},
archivePrefix = {arXiv},
       eprint = {2502.11823},
 primaryClass = {gr-qc},
       adsurl = {https://ui.adsabs.harvard.edu/abs/2025PhRvD.111j4044T}
}

@ARTICLE{2026arXiv260115198S,
       author = {{Singh}, Shashwat and {Chapman-Bird}, Christian E.~A. and {Berry}, Christopher P.~L. and {Veitch}, John},
        title = "{Revealing massive black hole astrophysics: The potential of hierarchical inference with extreme mass-ratio inspiral observations}",
      journal = {arXiv e-prints},
         year = 2026,
        month = jan,
          eid = {arXiv:2601.15198},
        pages = {arXiv:2601.15198},
          doi = {10.48550/arXiv.2601.15198},
archivePrefix = {arXiv},
       eprint = {2601.15198},
 primaryClass = {astro-ph.HE},
       adsurl = {https://ui.adsabs.harvard.edu/abs/2026arXiv260115198S}
}

@ARTICLE{2026PhRvD.113b3036S,
       author = {{Speri}, Lorenzo and {Barsanti}, Susanna and {Maselli}, Andrea and {Sotiriou}, Thomas P. and {Warburton}, Niels and {van de Meent}, Maarten and {Chua}, Alvin J.~K. and {Burke}, Ollie and {Gair}, Jonathan},
        title = "{Probing fundamental physics with extreme mass ratio inspirals: Full Bayesian inference for scalar charge}",
      journal = {\prd},
         year = 2026,
        month = jan,
       volume = {113},
       number = {2},
          eid = {023036},
        pages = {023036},
          doi = {10.1103/cnhz-6zlk},
archivePrefix = {arXiv},
       eprint = {2406.07607},
 primaryClass = {gr-qc},
       adsurl = {https://ui.adsabs.harvard.edu/abs/2026PhRvD.113b3036S}
}

@ARTICLE{2020PhRvD.102l4054B,
       author = {{Burke}, Ollie and {Gair}, Jonathan R. and {Sim{\'o}n}, Joan and {Edwards}, Matthew C.},
        title = "{Constraining the spin parameter of near-extremal black holes using LISA}",
      journal = {\prd},
         year = 2020,
        month = dec,
       volume = {102},
       number = {12},
          eid = {124054},
        pages = {124054},
          doi = {10.1103/PhysRevD.102.124054},
archivePrefix = {arXiv},
       eprint = {2010.05932},
 primaryClass = {gr-qc},
       adsurl = {https://ui.adsabs.harvard.edu/abs/2020PhRvD.102l4054B}
}

@ARTICLE{2009CQGra..26i4034G,
       author = {{Gair}, Jonathan R.},
        title = "{Probing black holes at low redshift using LISA EMRI observations}",
      journal = {Classical and Quantum Gravity},
         year = 2009,
        month = may,
       volume = {26},
       number = {9},
          eid = {094034},
        pages = {094034},
          doi = {10.1088/0264-9381/26/9/094034},
archivePrefix = {arXiv},
       eprint = {0811.0188},
 primaryClass = {gr-qc},
       adsurl = {https://ui.adsabs.harvard.edu/abs/2009CQGra..26i4034G}
}

@ARTICLE{2023PhRvX..13b1035S,
       author = {{Speri}, Lorenzo and {Antonelli}, Andrea and {Sberna}, Laura and {Babak}, Stanislav and {Barausse}, Enrico and {Gair}, Jonathan R. and {Katz}, Michael L.},
        title = "{Probing Accretion Physics with Gravitational Waves}",
      journal = {Physical Review X},
         year = 2023,
        month = apr,
       volume = {13},
       number = {2},
          eid = {021035},
        pages = {021035},
          doi = {10.1103/PhysRevX.13.021035},
archivePrefix = {arXiv},
       eprint = {2207.10086},
 primaryClass = {gr-qc},
       adsurl = {https://ui.adsabs.harvard.edu/abs/2023PhRvX..13b1035S}
}

@ARTICLE{2021arXiv210602053L,
       author = {{Laghi}, Danny},
        title = "{Gravitational wave cosmology with EMRIs}",
      journal = {arXiv e-prints},
         year = 2021,
        month = jun,
          eid = {arXiv:2106.02053},
        pages = {arXiv:2106.02053},
          doi = {10.48550/arXiv.2106.02053},
archivePrefix = {arXiv},
       eprint = {2106.02053},
 primaryClass = {astro-ph.CO},
       adsurl = {https://ui.adsabs.harvard.edu/abs/2021arXiv210602053L}
}

@ARTICLE{2023LRR....26....5A,
       author = {{Auclair}, Pierre and {Bacon}, David and {Baker}, Tessa and {Barreiro}, Tiago and {Bartolo}, Nicola and {Belgacem}, Enis and {Bellomo}, Nicola and {Ben-Dayan}, Ido and {Bertacca}, Daniele and {Besancon}, Marc and {Blanco-Pillado}, Jose J. and {Blas}, Diego and {Boileau}, Guillaume and {Calcagni}, Gianluca and {Caldwell}, Robert and {Caprini}, Chiara and {Carbone}, Carmelita and {Chang}, Chia-Feng and {Chen}, Hsin-Yu and {Christensen}, Nelson and {Clesse}, Sebastien and {Comelli}, Denis and {Congedo}, Giuseppe and {Contaldi}, Carlo and {Crisostomi}, Marco and {Croon}, Djuna and {Cui}, Yanou and {Cusin}, Giulia and {Cutting}, Daniel and {Dalang}, Charles and {De Luca}, Valerio and {Pozzo}, Walter Del and {Desjacques}, Vincent and {Dimastrogiovanni}, Emanuela and {Dorsch}, Glauber C. and {Ezquiaga}, Jose Maria and {Fasiello}, Matteo and {Figueroa}, Daniel G. and {Flauger}, Raphael and {Franciolini}, Gabriele and {Frusciante}, Noemi and {Fumagalli}, Jacopo and {Garc{\'\i}a-Bellido}, Juan and {Gould}, Oliver and {Holz}, Daniel and {Iacconi}, Laura and {Jain}, Rajeev Kumar and {Jenkins}, Alexander C. and {Jinno}, Ryusuke and {Joana}, Cristian and {Karnesis}, Nikolaos and {Konstandin}, Thomas and {Koyama}, Kazuya and {Kozaczuk}, Jonathan and {Kuroyanagi}, Sachiko and {Laghi}, Danny and {Lewicki}, Marek and {Lombriser}, Lucas and {Madge}, Eric and {Maggiore}, Michele and {Malhotra}, Ameek and {Mancarella}, Michele and {Mandic}, Vuk and {Mangiagli}, Alberto and {Matarrese}, Sabino and {Mazumdar}, Anupam and {Mukherjee}, Suvodip and {Musco}, Ilia and {Nardini}, Germano and {No}, Jose Miguel and {Papanikolaou}, Theodoros and {Peloso}, Marco and {Pieroni}, Mauro and {Pilo}, Luigi and {Raccanelli}, Alvise and {Renaux-Petel}, S{\'e}bastien and {Renzini}, Arianna I. and {Ricciardone}, Angelo and {Riotto}, Antonio and {Romano}, Joseph D. and {Rollo}, Rocco and {Pol}, Alberto Roper and {Morales}, Ester Ruiz and {Sakellariadou}, Mairi and {Saltas}, Ippocratis D. and {Scalisi}, Marco and {Schmitz}, Kai and {Schwaller}, Pedro and {Sergijenko}, Olga and {Servant}, Geraldine and {Simakachorn}, Peera and {Sorbo}, Lorenzo and {Sousa}, Lara and {Speri}, Lorenzo and {Steer}, Dani{\`e}le A. and {Tamanini}, Nicola and {Tasinato}, Gianmassimo and {Torrado}, Jes{\'u}s and {Unal}, Caner and {Vennin}, Vincent and {Vernieri}, Daniele and {Vernizzi}, Filippo and {Volonteri}, Marta and {Wachter}, Jeremy M. and {Wands}, David and {Witkowski}, Lukas T. and {Zumalac{\'a}rregui}, Miguel and {Annis}, James and {Ares}, F{\"e}anor Reuben and {Avelino}, Pedro P. and {Avgoustidis}, Anastasios and {Barausse}, Enrico and {Bonilla}, Alexander and {Bonvin}, Camille and {Bosso}, Pasquale and {Calabrese}, Matteo and {{\c{C}}al{\i}{\textcommabelow s}kan}, Mesut and {Cembranos}, Jose A.~R. and {Chala}, Mikael and {Chernoff}, David and {Clough}, Katy and {Criswell}, Alexander and {Das}, Saurya and {Silva}, Antonio da and {Dayal}, Pratika and {Domcke}, Valerie and {Durrer}, Ruth and {Easther}, Richard and {Escoffier}, Stephanie and {Ferrans}, Sandrine and {Fryer}, Chris and {Gair}, Jonathan and {Gordon}, Chris and {Hendry}, Martin and {Hindmarsh}, Mark and {Hooper}, Deanna C. and {Kajfasz}, Eric and {Kopp}, Joachim and {Koushiappas}, Savvas M. and {Kumar}, Utkarsh and {Kunz}, Martin and {Lagos}, Macarena and {Lilley}, Marc and {Lizarraga}, Joanes and {Lobo}, Francisco S.~N. and {Maleknejad}, Azadeh and {Martins}, C.~J.~A.~P. and {Meerburg}, P. Daniel and {Meyer}, Renate and {Mimoso}, Jos{\'e} Pedro and {Nesseris}, Savvas and {Nunes}, Nelson and {Oikonomou}, Vasilis and {Orlando}, Giorgio and {{\"O}zsoy}, Ogan and {Pacucci}, Fabio and {Palmese}, Antonella and {Petiteau}, Antoine and {Pinol}, Lucas and {Zwart}, Simon Portegies and {Pratten}, Geraint and {Prokopec}, Tomislav and {Quenby}, John and {Rastgoo}, Saeed and {Roest}, Diederik and {Rummukainen}, Kari and {Schimd}, Carlo and {Secroun}, Aur{\'e}lia and {Sesana}, Alberto and {Sopuerta}, Carlos F. and {Tereno}, Ismael and {Tolley}, Andrew and {Urrestilla}, Jon and {Vagenas}, Elias C. and {van de Vis}, Jorinde and {van de Weygaert}, Rien and {Wardell}, Barry and {Weir}, David J. and {White}, Graham and {{\'S}wie{\.z}ewska}, Bogumi{\l}a and {Zhdanov}, Valery I. and {The LISA Cosmology Working Group}},
        title = "{Cosmology with the Laser Interferometer Space Antenna}",
      journal = {Living Reviews in Relativity},
         year = 2023,
        month = dec,
       volume = {26},
       number = {1},
          eid = {5},
        pages = {5},
          doi = {10.1007/s41114-023-00045-2},
archivePrefix = {arXiv},
       eprint = {2204.05434},
 primaryClass = {astro-ph.CO},
       adsurl = {https://ui.adsabs.harvard.edu/abs/2023LRR....26....5A}
}

@ARTICLE{2007PhRvD..76j4018C,
       author = {{Cutler}, Curt and {Vallisneri}, Michele},
        title = "{LISA detections of massive black hole inspirals: Parameter extraction errors due to inaccurate template waveforms}",
      journal = {\prd},
         year = 2007,
        month = nov,
       volume = {76},
       number = {10},
          eid = {104018},
        pages = {104018},
          doi = {10.1103/PhysRevD.76.104018},
archivePrefix = {arXiv},
       eprint = {0707.2982},
 primaryClass = {gr-qc},
       adsurl = {https://ui.adsabs.harvard.edu/abs/2007PhRvD..76j4018C}
}

@ARTICLE{2008PhRvD..78l4020L,
       author = {{Lindblom}, Lee and {Owen}, Benjamin J. and {Brown}, Duncan A.},
        title = "{Model waveform accuracy standards for gravitational wave data analysis}",
      journal = {\prd},
         year = 2008,
        month = dec,
       volume = {78},
       number = {12},
          eid = {124020},
        pages = {124020},
          doi = {10.1103/PhysRevD.78.124020},
archivePrefix = {arXiv},
       eprint = {0809.3844},
 primaryClass = {gr-qc},
       adsurl = {https://ui.adsabs.harvard.edu/abs/2008PhRvD..78l4020L}
}

@ARTICLE{2024PhRvD.109l4048B,
       author = {{Burke}, Ollie and {Piovano}, Gabriel Andres and {Warburton}, Niels and {Lynch}, Philip and {Speri}, Lorenzo and {Kavanagh}, Chris and {Wardell}, Barry and {Pound}, Adam and {Durkan}, Leanne and {Miller}, Jeremy},
        title = "{Assessing the importance of first postadiabatic terms for small-mass-ratio binaries}",
      journal = {\prd},
         year = 2024,
        month = jun,
       volume = {109},
       number = {12},
          eid = {124048},
        pages = {124048},
          doi = {10.1103/PhysRevD.109.124048},
archivePrefix = {arXiv},
       eprint = {2310.08927},
 primaryClass = {gr-qc},
       adsurl = {https://ui.adsabs.harvard.edu/abs/2024PhRvD.109l4048B}
}

@article{Kejriwal:2023djc,
    author = "Kejriwal, Shubham and Speri, Lorenzo and Chua, Alvin J. K.",
    title = "{Impact of correlations on the modeling and inference of beyond vacuum{\textendash}general relativistic effects in extreme-mass-ratio inspirals}",
    eprint = "2312.13028",
    archivePrefix = "arXiv",
    primaryClass = "gr-qc",
    doi = "10.1103/PhysRevD.110.084060",
    journal = "Phys. Rev. D",
    volume = "110",
    number = "8",
    pages = "084060",
    year = "2024"
}

@ARTICLE{2025PhRvD.111h2010K,
       author = {{Khalvati}, Hassan and {Santini}, Alessandro and {Duque}, Francisco and {Speri}, Lorenzo and {Gair}, Jonathan and {Yang}, Huan and {Brito}, Richard},
        title = "{Impact of relativistic waveforms in LISA's science objectives with extreme-mass-ratio inspirals}",
      journal = {\prd},
         year = 2025,
        month = apr,
       volume = {111},
       number = {8},
          eid = {082010},
        pages = {082010},
          doi = {10.1103/PhysRevD.111.082010},
archivePrefix = {arXiv},
       eprint = {2410.17310},
 primaryClass = {gr-qc},
       adsurl = {https://ui.adsabs.harvard.edu/abs/2025PhRvD.111h2010K}
}

@ARTICLE{2025LRR....28....9L,
       author = {{LISA Consortium Waveform Working Group} and {Afshordi}, Niayesh and {Ak{\c{c}}ay}, Sarp and {Seoane}, Pau Amaro and {Antonelli}, Andrea and {Aurrekoetxea}, Josu C. and {Barack}, Leor and {Barausse}, Enrico and {Benkel}, Robert and {Bernard}, Laura and {Bernuzzi}, Sebastiano and {Berti}, Emanuele and {Bonetti}, Matteo and {Bonga}, B{\'e}atrice and {Bozzola}, Gabriele and {Brito}, Richard and {Buonanno}, Alessandra and {C{\'a}rdenas-Avenda{\~n}o}, Alejandro and {Casals}, Marc and {Chernoff}, David F. and {Chua}, Alvin J.~K. and {Clough}, Katy and {Colleoni}, Marta and {Comp{\`e}re}, Geoffrey and {Dhesi}, Mekhi and {Druart}, Adrien and {Durkan}, Leanne and {Faye}, Guillaume and {Ferguson}, Deborah and {Field}, Scott E. and {Gabella}, William E. and {Garc{\'\i}a-Bellido}, Juan and {Gracia-Linares}, Miguel and {Gerosa}, Davide and {Green}, Stephen R. and {Haney}, Maria and {Hannam}, Mark and {Heffernan}, Anna and {Hinderer}, Tanja and {Helfer}, Thomas and {Hughes}, Scott A. and {Husa}, Sascha and {Isoyama}, Soichiro and {Katz}, Michael L. and {Kavanagh}, Chris and {Khanna}, Gaurav and {Kidder}, Larry E. and {Korol}, Valeriya and {K{\"u}chler}, Lorenzo and {Laguna}, Pablo and {Larrouturou}, Fran{\c{c}}ois and {Tiec}, Alexandre Le and {Leather}, Benjamin and {Lim}, Eugene A. and {Lim}, Hyun and {Littenberg}, Tyson B. and {Long}, Oliver and {Lousto}, Carlos O. and {Lovelace}, Geoffrey and {Lukes-Gerakopoulos}, Georgios and {Lynch}, Philip and {Macedo}, Rodrigo P. and {Markakis}, Charalampos and {Maggio}, Elisa and {Mandel}, Ilya and {Maselli}, Andrea and {Mathews}, Josh and {Mourier}, Pierre and {Neilsen}, David and {Nagar}, Alessandro and {Nichols}, David A. and {Nov{\'a}k}, Jan and {Okounkova}, Maria and {O'Shaughnessy}, Richard and {Oshita}, Naritaka and {O'Toole}, Conor and {Pan}, Zhen and {Pani}, Paolo and {Pappas}, George and {Paschalidis}, Vasileios and {Pfeiffer}, Harald P. and {Pompili}, Lorenzo and {Pound}, Adam and {Pratten}, Geraint and {R{\"u}ter}, Hannes R. and {Ruiz}, Milton and {Sam}, Zeyd and {Sberna}, Laura and {Shapiro}, Stuart L. and {Shoemaker}, Deirdre M. and {Sopuerta}, Carlos F. and {Spiers}, Andrew and {Sundar}, Hari and {Tamanini}, Nicola and {Thompson}, Jonathan E. and {Toubiana}, Alexandre and {Tsokaros}, Antonios and {Upton}, Samuel D. and {van de Meent}, Maarten and {Vernieri}, Daniele and {Wachter}, Jeremy M. and {Warburton}, Niels and {Wardell}, Barry and {Witek}, Helvi and {Witzany}, Vojt{\v{e}}ch and {Yang}, Huan and {Zilh{\~a}o}, Miguel and {Albertini}, Angelica and {Arun}, K.~G. and {Bezares}, Miguel and {Bonilla}, Alexander and {Chapman-Bird}, Christian and {Cownden}, Bradley and {Cunningham}, Kevin and {Devitt}, Chris and {Dolan}, Sam and {Duque}, Francisco and {Dyson}, Conor and {Fryer}, Chris L. and {Gair}, Jonathan R. and {Giacomazzo}, Bruno and {Gupta}, Priti and {Han}, Wen-Biao and {Haas}, Roland and {Hirschmann}, Eric W. and {Huerta}, E.~A. and {Jetzer}, Philippe and {Kelly}, Bernard and {Khalil}, Mohammed and {Lewis}, Jack and {Lloyd-Ronning}, Nicole and {Marsat}, Sylvain and {Nardini}, Germano and {Neef}, Jakob and {Ottewill}, Adrian and {Pantelidou}, Christiana and {Piovano}, Gabriel Andres and {Redondo-Yuste}, Jaime and {Sagunski}, Laura and {Stein}, Leo C. and {Skoup{\'y}}, Viktor and {Sperhake}, Ulrich and {Speri}, Lorenzo and {Spieksma}, Thomas F.~M. and {Stevens}, Chris and {Trestini}, David and {Va{\~n}{\'o}-Vi{\~n}uales}, Alex},
        title = "{Waveform modelling for the Laser Interferometer Space Antenna}",
      journal = {Living Reviews in Relativity},
         year = 2025,
        month = oct,
       volume = {28},
       number = {1},
          eid = {9},
        pages = {9},
          doi = {10.1007/s41114-025-00056-1},
archivePrefix = {arXiv},
       eprint = {2311.01300},
 primaryClass = {gr-qc},
       adsurl = {https://ui.adsabs.harvard.edu/abs/2025LRR....28....9L}
}

@ARTICLE{2025CQGra..42f5018C,
       author = {{Castelli}, Eleonora and {Baghi}, Quentin and {Baker}, John G. and {Slutsky}, Jacob and {Bobin}, J{\'e}r{\^o}me and {Karnesis}, Nikolaos and {Petiteau}, Antoine and {Sauter}, Orion and {Wass}, Peter and {Weber}, William J.},
        title = "{Extracting gravitational wave signals from LISA data in the presence of artifacts}",
      journal = {Classical and Quantum Gravity},
         year = 2025,
        month = mar,
       volume = {42},
       number = {6},
          eid = {065018},
        pages = {065018},
          doi = {10.1088/1361-6382/adb931},
archivePrefix = {arXiv},
       eprint = {2411.13402},
 primaryClass = {gr-qc},
       adsurl = {https://ui.adsabs.harvard.edu/abs/2025CQGra..42f5018C}
}

@ARTICLE{2023PhRvD.108l3029S,
       author = {{Spadaro}, Alice and {Buscicchio}, Riccardo and {Vetrugno}, Daniele and {Klein}, Antoine and {Gerosa}, Davide and {Vitale}, Stefano and {Dolesi}, Rita and {Weber}, William Joseph and {Colpi}, Monica},
        title = "{Glitch systematics on the observation of massive black-hole binaries with LISA}",
      journal = {\prd},
         year = 2023,
        month = dec,
       volume = {108},
       number = {12},
          eid = {123029},
        pages = {123029},
          doi = {10.1103/PhysRevD.108.123029},
archivePrefix = {arXiv},
       eprint = {2306.03923},
 primaryClass = {gr-qc},
       adsurl = {https://ui.adsabs.harvard.edu/abs/2023PhRvD.108l3029S}
}

@ARTICLE{2025PhRvD.111l4053B,
       author = {{Burke}, Ollie and {Marsat}, Sylvain and {Gair}, Jonathan R. and {Katz}, Michael L.},
        title = "{Addressing data gaps and assessing noise mismodeling in LISA}",
      journal = {\prd},
         year = 2025,
        month = jun,
       volume = {111},
       number = {12},
          eid = {124053},
        pages = {124053},
          doi = {10.1103/5jr8-k2ss},
archivePrefix = {arXiv},
       eprint = {2502.17426},
 primaryClass = {gr-qc},
       adsurl = {https://ui.adsabs.harvard.edu/abs/2025PhRvD.111l4053B}
}

@ARTICLE{2025ApJ...983...99A,
       author = {{Abac}, A.~G. and {Abbott}, R. and {Abouelfettouh}, I. and {Acernese}, F. and {Ackley}, K. and {Adhicary}, S. and {Adhikari}, N. and {Adhikari}, R.~X. and {Adkins}, V.~K. and {Agarwal}, D. and {Agathos}, M. and {Aghaei Abchouyeh}, M. and {Aguiar}, O.~D. and {Aguilar}, I. and {Aiello}, L. and {Ain}, A. and {Ajith}, P. and {Akutsu}, T. and {Albanesi}, S. and {Alfaidi}, R.~A. and {Al-Jodah}, A. and {All{\'e}n{\'e}}, C. and {Allocca}, A. and {Al-Shammari}, S. and {Altin}, P.~A. and {Alvarez-Lopez}, S. and {Amato}, A. and {Amez-Droz}, L. and {Amorosi}, A. and {Amra}, C. and {Ananyeva}, A. and {Anderson}, S.~B. and {Anderson}, W.~G. and {Andia}, M. and {Ando}, M. and {Andrade}, T. and {Andres}, N. and {Andr{\'e}s-Carcasona}, M. and {Andri{\'c}}, T. and {Anglin}, J. and {Ansoldi}, S. and {Antelis}, J.~M. and {Antier}, S. and {Aoumi}, M. and {Appavuravther}, E.~Z. and {Appert}, S. and {Apple}, S.~K. and {Arai}, K. and {Araya}, A. and {Araya}, M.~C. and {Areeda}, J.~S. and {Argianas}, L. and {Aritomi}, N. and {Armato}, F. and {Arnaud}, N. and {Arogeti}, M. and {Aronson}, S.~M. and {Ashton}, G. and {Aso}, Y. and {Assiduo}, M. and {Assis de Souza Melo}, S. and {Aston}, S.~M. and {Astone}, P. and {Attadio}, F. and {Aubin}, F. and {AultONeal}, K. and {Avallone}, G. and {Babak}, S. and {Badaracco}, F. and {Badger}, C. and {Bae}, S. and {Bagnasco}, S. and {Bagui}, E. and {Baier}, J.~G. and {Baiotti}, L. and {Bajpai}, R. and {Baka}, T. and {Ball}, M. and {Ballardin}, G. and {Ballmer}, S.~W. and {Banagiri}, S. and {Banerjee}, B. and {Bankar}, D. and {Baral}, P. and {Barayoga}, J.~C. and {Barish}, B.~C. and {Barker}, D. and {Barneo}, P. and {Barone}, F. and {Barr}, B. and {Barsotti}, L. and {Barsuglia}, M. and {Barta}, D. and {Bartoletti}, A.~M. and {Barton}, M.~A. and {Bartos}, I. and {Basak}, S. and {Basalaev}, A. and {Bassiri}, R. and {Basti}, A. and {Bates}, D.~E. and {Bawaj}, M. and {Baxi}, P. and {Bayley}, J.~C. and {Baylor}, A.~C. and {Baynard}, II, P.~A. and {Bazzan}, M. and {Bedakihale}, V.~M. and {Beirnaert}, F. and {Bejger}, M. and {Belardinelli}, D. and {Bell}, A.~S. and {Benedetto}, V. and {Benoit}, W. and {Bentley}, J.~D. and {Ben Yaala}, M. and {Bera}, S. and {Berbel}, M. and {Bergamin}, F. and {Berger}, B.~K. and {Bernuzzi}, S. and {Beroiz}, M. and {Bersanetti}, D. and {Bertolini}, A. and {Betzwieser}, J. and {Beveridge}, D. and {Bevins}, N. and {Bhandare}, R. and {Bhardwaj}, U. and {Bhatt}, R. and {Bhattacharjee}, D. and {Bhaumik}, S. and {Bhowmick}, S. and {Bianchi}, A. and {Bilenko}, I.~A. and {Billingsley}, G. and {Binetti}, A. and {Bini}, S. and {Birnholtz}, O. and {Biscoveanu}, S. and {Bisht}, A. and {Bitossi}, M. and {Bizouard}, M.-A. and {Blackburn}, J.~K. and {Blagg}, L.~A. and {Blair}, C.~D. and {Blair}, D.~G. and {Bobba}, F. and {Bode}, N. and {Boileau}, G. and {Boldrini}, M. and {Bolingbroke}, G.~N. and {Bolliand}, A. and {Bonavena}, L.~D. and {Bondarescu}, R. and {Bondu}, F. and {Bonilla}, E. and {Bonilla}, M.~S. and {Bonino}, A. and {Bonnand}, R. and {Booker}, P. and {Borchers}, A. and {Boschi}, V. and {Bose}, S. and {Bossilkov}, V. and {Boudart}, V. and {Boudon}, A. and {Bozzi}, A. and {Bradaschia}, C. and {Brady}, P.~R. and {Braglia}, M. and {Branch}, A. and {Branchesi}, M. and {Brandt}, J. and {Braun}, I. and {Breschi}, M. and {Briant}, T. and {Brillet}, A. and {Brinkmann}, M. and {Brockill}, P. and {Brockmueller}, E. and {Brooks}, A.~F. and {Brown}, B.~C. and {Brown}, D.~D. and {Brozzetti}, M.~L. and {Brunett}, S. and {Bruno}, G. and {Bruntz}, R. and {Bryant}, J. and {Bucci}, F. and {Buchanan}, J. and {Bulashenko}, O. and {Bulik}, T. and {Bulten}, H.~J. and {Buonanno}, A. and {Burtnyk}, K. and {Buscicchio}, R. and {Buskulic}, D. and {Buy}, C. and {Byer}, R.~L.},
        title = "{Search for Continuous Gravitational Waves from Known Pulsars in the First Part of the Fourth LIGO-Virgo-KAGRA Observing Run}",
      journal = {\apj},
         year = 2025,
        month = apr,
       volume = {983},
       number = {2},
          eid = {99},
        pages = {99},
          doi = {10.3847/1538-4357/adb3a0},
archivePrefix = {arXiv},
       eprint = {2501.01495},
 primaryClass = {astro-ph.HE},
       adsurl = {https://ui.adsabs.harvard.edu/abs/2025ApJ...983...99A}
}

@ARTICLE{1979PhRvD..20..351Z,
       author = {{Zimmermann}, M. and {Szedenits}, Jr., E.},
        title = "{Gravitational waves from rotating and precessing rigid bodies: Simple models and applications to pulsars}",
      journal = {\prd},
         year = 1979,
        month = jul,
       volume = {20},
       number = {2},
        pages = {351-355},
          doi = {10.1103/PhysRevD.20.351},
       adsurl = {https://ui.adsabs.harvard.edu/abs/1979PhRvD..20..351Z}
}

@ARTICLE{2014ApJ...785..119A,
       author = {{Aasi}, J. and {Abadie}, J. and {Abbott}, B.~P. and {Abbott}, R. and {Abbott}, T. and {Abernathy}, M.~R. and {Accadia}, T. and {Acernese}, F. and {Adams}, C. and {Adams}, T. and {Adhikari}, R.~X. and {Affeldt}, C. and {Agathos}, M. and {Aggarwal}, N. and {Aguiar}, O.~D. and {Ajith}, P. and {Allen}, B. and {Allocca}, A. and {Amador Ceron}, E. and {Amariutei}, D. and {Anderson}, R.~A. and {Anderson}, S.~B. and {Anderson}, W.~G. and {Arai}, K. and {Araya}, M.~C. and {Arceneaux}, C. and {Areeda}, J. and {Ast}, S. and {Aston}, S.~M. and {Astone}, P. and {Aufmuth}, P. and {Aulbert}, C. and {Austin}, L. and {Aylott}, B.~E. and {Babak}, S. and {Baker}, P.~T. and {Ballardin}, G. and {Ballmer}, S.~W. and {Barayoga}, J.~C. and {Barker}, D. and {Barnum}, S.~H. and {Barone}, F. and {Barr}, B. and {Barsotti}, L. and {Barsuglia}, M. and {Barton}, M.~A. and {Bartos}, I. and {Bassiri}, R. and {Basti}, A. and {Batch}, J. and {Bauchrowitz}, J. and {Bauer}, Th. S. and {Bebronne}, M. and {Behnke}, B. and {Bejger}, M. and {Beker}, M.~G. and {Bell}, A.~S. and {Bell}, C. and {Belopolski}, I. and {Bergmann}, G. and {Berliner}, J.~M. and {Bersanetti}, D. and {Bertolini}, A. and {Bessis}, D. and {Betzwieser}, J. and {Beyersdorf}, P.~T. and {Bhadbhade}, T. and {Bilenko}, I.~A. and {Billingsley}, G. and {Birch}, J. and {Bitossi}, M. and {Bizouard}, M.~A. and {Black}, E. and {Blackburn}, J.~K. and {Blackburn}, L. and {Blair}, D. and {Blom}, M. and {Bock}, O. and {Bodiya}, T.~P. and {Boer}, M. and {Bogan}, C. and {Bond}, C. and {Bondu}, F. and {Bonelli}, L. and {Bonnand}, R. and {Bork}, R. and {Born}, M. and {Boschi}, V. and {Bose}, S. and {Bosi}, L. and {Bowers}, J. and {Bradaschia}, C. and {Brady}, P.~R. and {Braginsky}, V.~B. and {Branchesi}, M. and {Brannen}, C.~A. and {Brau}, J.~E. and {Breyer}, J. and {Briant}, T. and {Bridges}, D.~O. and {Brillet}, A. and {Brinkmann}, M. and {Brisson}, V. and {Britzger}, M. and {Brooks}, A.~F. and {Brown}, D.~A. and {Brown}, D.~D. and {Br{\"u}ckner}, F. and {Bulik}, T. and {Bulten}, H.~J. and {Buonanno}, A. and {Buskulic}, D. and {Buy}, C. and {Byer}, R.~L. and {Cadonati}, L. and {Cagnoli}, G. and {Calder{\'o}n Bustillo}, J. and {Calloni}, E. and {Camp}, J.~B. and {Campsie}, P. and {Cannon}, K.~C. and {Canuel}, B. and {Cao}, J. and {Capano}, C.~D. and {Carbognani}, F. and {Carbone}, L. and {Caride}, S. and {Castiglia}, A. and {Caudill}, S. and {Cavagli{\`a}}, M. and {Cavalier}, F. and {Cavalieri}, R. and {Cella}, G. and {Cepeda}, C. and {Cesarini}, E. and {Chakraborty}, R. and {Chalermsongsak}, T. and {Chao}, S. and {Charlton}, P. and {Chassande-Mottin}, E. and {Chen}, X. and {Chen}, Y. and {Chincarini}, A. and {Chiummo}, A. and {Cho}, H.~S. and {Chow}, J. and {Christensen}, N. and {Chu}, Q. and {Chua}, S.~S.~Y. and {Chung}, S. and {Ciani}, G. and {Clara}, F. and {Clark}, D.~E. and {Clark}, J.~A. and {Cleva}, F. and {Coccia}, E. and {Cohadon}, P.-F. and {Colla}, A. and {Colombini}, M. and {Constancio}, Jr., M. and {Conte}, A. and {Conte}, R. and {Cook}, D. and {Corbitt}, T.~R. and {Cordier}, M. and {Cornish}, N. and {Corsi}, A. and {Costa}, C.~A. and {Coughlin}, M.~W. and {Coulon}, J.-P. and {Countryman}, S. and {Couvares}, P. and {Coward}, D.~M. and {Cowart}, M. and {Coyne}, D.~C. and {Craig}, K. and {Creighton}, J.~D.~E. and {Creighton}, T.~D. and {Crowder}, S.~G. and {Cumming}, A. and {Cunningham}, L. and {Cuoco}, E. and {Dahl}, K. and {Dal Canton}, T. and {Damjanic}, M. and {Danilishin}, S.~L. and {D'Antonio}, S. and {Danzmann}, K. and {Dattilo}, V. and {Daudert}, B. and {Daveloza}, H. and {Davier}, M. and {Davies}, G.~S. and {Daw}, E.~J. and {Day}, R. and {Dayanga}, T. and {De Rosa}, R. and {Debreczeni}, G. and {Degallaix}, J. and {Del Pozzo}, W.},
        title = "{Gravitational Waves from Known Pulsars: Results from the Initial Detector Era}",
      journal = {\apj},
         year = 2014,
        month = apr,
       volume = {785},
       number = {2},
          eid = {119},
        pages = {119},
          doi = {10.1088/0004-637X/785/2/119},
archivePrefix = {arXiv},
       eprint = {1309.4027},
 primaryClass = {astro-ph.HE},
       adsurl = {https://ui.adsabs.harvard.edu/abs/2014ApJ...785..119A}
}

@ARTICLE{2017ApJ...839...12A,
       author = {{Abbott}, B.~P. and {Abbott}, R. and {Abbott}, T.~D. and {Abernathy}, M.~R. and {Acernese}, F. and {Ackley}, K. and {Adams}, C. and {Adams}, T. and {Addesso}, P. and {Adhikari}, R.~X. and {Adya}, V.~B. and {Affeldt}, C. and {Agathos}, M. and {Agatsuma}, K. and {Aggarwal}, N. and {Aguiar}, O.~D. and {Aiello}, L. and {Ain}, A. and {Ajith}, P. and {Allen}, B. and {Allocca}, A. and {Altin}, P.~A. and {Ananyeva}, A. and {Anderson}, S.~B. and {Anderson}, W.~G. and {Appert}, S. and {Arai}, K. and {Araya}, M.~C. and {Areeda}, J.~S. and {Arnaud}, N. and {Arun}, K.~G. and {Ascenzi}, S. and {Ashton}, G. and {Ast}, M. and {Aston}, S.~M. and {Astone}, P. and {Aufmuth}, P. and {Aulbert}, C. and {Avila-Alvarez}, A. and {Babak}, S. and {Bacon}, P. and {Bader}, M.~K.~M. and {Baker}, P.~T. and {Baldaccini}, F. and {Ballardin}, G. and {Ballmer}, S.~W. and {Barayoga}, J.~C. and {Barclay}, S.~E. and {Barish}, B.~C. and {Barker}, D. and {Barone}, F. and {Barr}, B. and {Barsotti}, L. and {Barsuglia}, M. and {Barta}, D. and {Bartlett}, J. and {Bartos}, I. and {Bassiri}, R. and {Basti}, A. and {Batch}, J.~C. and {Baune}, C. and {Bavigadda}, V. and {Bazzan}, M. and {Beer}, C. and {Bejger}, M. and {Belahcene}, I. and {Belgin}, M. and {Bell}, A.~S. and {Berger}, B.~K. and {Bergmann}, G. and {Berry}, C.~P.~L. and {Bersanetti}, D. and {Bertolini}, A. and {Betzwieser}, J. and {Bhagwat}, S. and {Bhandare}, R. and {Bilenko}, I.~A. and {Billingsley}, G. and {Billman}, C.~R. and {Birch}, J. and {Birney}, R. and {Birnholtz}, O. and {Biscans}, S. and {Bisht}, A. and {Bitossi}, M. and {Biwer}, C. and {Bizouard}, M.~A. and {Blackburn}, J.~K. and {Blackman}, J. and {Blair}, C.~D. and {Blair}, D.~G. and {Blair}, R.~M. and {Bloemen}, S. and {Bock}, O. and {Boer}, M. and {Bogaert}, G. and {Bohe}, A. and {Bondu}, F. and {Bonnand}, R. and {Boom}, B.~A. and {Bork}, R. and {Boschi}, V. and {Bose}, S. and {Bouffanais}, Y. and {Bozzi}, A. and {Bradaschia}, C. and {Brady}, P.~R. and {Braginsky}, V.~B. and {Branchesi}, M. and {Brau}, J.~E. and {Briant}, T. and {Brillet}, A. and {Brinkmann}, M. and {Brisson}, V. and {Brockill}, P. and {Broida}, J.~E. and {Brooks}, A.~F. and {Brown}, D.~A. and {Brown}, D.~D. and {Brown}, N.~M. and {Brunett}, S. and {Buchanan}, C.~C. and {Buikema}, A. and {Bulik}, T. and {Bulten}, H.~J. and {Buonanno}, A. and {Buskulic}, D. and {Buy}, C. and {Byer}, R.~L. and {Cabero}, M. and {Cadonati}, L. and {Cagnoli}, G. and {Cahillane}, C. and {Calder{\'o}n Bustillo}, J. and {Callister}, T.~A. and {Calloni}, E. and {Camp}, J.~B. and {Canepa}, M. and {Cannon}, K.~C. and {Cao}, H. and {Cao}, J. and {Capano}, C.~D. and {Capocasa}, E. and {Carbognani}, F. and {Caride}, S. and {Casanueva Diaz}, J. and {Casentini}, C. and {Caudill}, S. and {Cavagli{\`a}}, M. and {Cavalier}, F. and {Cavalieri}, R. and {Cella}, G. and {Cepeda}, C.~B. and {Cerboni Baiardi}, L. and {Cerretani}, G. and {Cesarini}, E. and {Chamberlin}, S.~J. and {Chan}, M. and {Chao}, S. and {Charlton}, P. and {Chassande-Mottin}, E. and {Cheeseboro}, B.~D. and {Chen}, H.~Y. and {Chen}, Y. and {Cheng}, H.-P. and {Chincarini}, A. and {Chiummo}, A. and {Chmiel}, T. and {Cho}, H.~S. and {Cho}, M. and {Chow}, J.~H. and {Christensen}, N. and {Chu}, Q. and {Chua}, A.~J.~K. and {Chua}, S. and {Chung}, S. and {Ciani}, G. and {Clara}, F. and {Clark}, J.~A. and {Cleva}, F. and {Cocchieri}, C. and {Coccia}, E. and {Cohadon}, P.-F. and {Colla}, A. and {Collette}, C.~G. and {Cominsky}, L. and {Constancio}, Jr., M. and {Conti}, L. and {Cooper}, S.~J. and {Corbitt}, T.~R. and {Cornish}, N. and {Corsi}, A. and {Cortese}, S. and {Costa}, C.~A. and {Coughlin}, M.~W. and {Coughlin}, S.~B. and {Coulon}, J.-P. and {Countryman}, S.~T. and {Couvares}, P. and {Covas}, P.~B.},
        title = "{First Search for Gravitational Waves from Known Pulsars with Advanced LIGO}",
      journal = {\apj},
         year = 2017,
        month = apr,
       volume = {839},
       number = {1},
          eid = {12},
        pages = {12},
          doi = {10.3847/1538-4357/aa677f},
archivePrefix = {arXiv},
       eprint = {1701.07709},
 primaryClass = {astro-ph.HE},
       adsurl = {https://ui.adsabs.harvard.edu/abs/2017ApJ...839...12A}
}

@ARTICLE{2023LRR....26....3R,
       author = {{Riles}, Keith},
        title = "{Searches for continuous-wave gravitational radiation}",
      journal = {Living Reviews in Relativity},
         year = 2023,
        month = dec,
       volume = {26},
       number = {1},
          eid = {3},
        pages = {3},
          doi = {10.1007/s41114-023-00044-3},
archivePrefix = {arXiv},
       eprint = {2206.06447},
 primaryClass = {astro-ph.HE},
       adsurl = {https://ui.adsabs.harvard.edu/abs/2023LRR....26....3R}
}

@ARTICLE{2023APh...15302880W,
       author = {{Wette}, Karl},
        title = "{Searches for continuous gravitational waves from neutron stars: A twenty-year retrospective}",
      journal = {Astroparticle Physics},
         year = 2023,
        month = nov,
       volume = {153},
          eid = {102880},
        pages = {102880},
          doi = {10.1016/j.astropartphys.2023.102880},
archivePrefix = {arXiv},
       eprint = {2305.07106},
 primaryClass = {gr-qc},
       adsurl = {https://ui.adsabs.harvard.edu/abs/2023APh...15302880W}
}

@ARTICLE{2025PhRvD.111h3016C,
       author = {{Carlin}, Julian B. and {Melatos}, Andrew},
        title = "{How much spin wandering can continuous gravitational wave search algorithms handle?}",
      journal = {\prd},
         year = 2025,
        month = apr,
       volume = {111},
       number = {8},
          eid = {083016},
        pages = {083016},
          doi = {10.1103/PhysRevD.111.083016},
archivePrefix = {arXiv},
       eprint = {2504.08163},
 primaryClass = {gr-qc},
       adsurl = {https://ui.adsabs.harvard.edu/abs/2025PhRvD.111h3016C}
}

@ARTICLE{1998PhRvD..57.2101B,
       author = {{Brady}, Patrick R. and {Creighton}, Teviet and {Cutler}, Curt and {Schutz}, Bernard F.},
        title = "{Searching for periodic sources with LIGO}",
      journal = {\prd},
         year = 1998,
        month = feb,
       volume = {57},
       number = {4},
        pages = {2101-2116},
          doi = {10.1103/PhysRevD.57.2101},
archivePrefix = {arXiv},
       eprint = {gr-qc/9702050},
 primaryClass = {gr-qc},
       adsurl = {https://ui.adsabs.harvard.edu/abs/1998PhRvD..57.2101B}
}

@ARTICLE{2025PhRvD.112j2004B,
       author = {{Baker}, A. Makai and {Lasky}, Paul D. and {Thrane}, Eric and {Golomb}, Jacob},
        title = "{Significant challenges for astrophysical inference with next-generation gravitational-wave observatories}",
      journal = {\prd},
         year = 2025,
        month = nov,
       volume = {112},
       number = {10},
          eid = {102004},
        pages = {102004},
          doi = {10.1103/n6t6-5wn3},
archivePrefix = {arXiv},
       eprint = {2503.04073},
 primaryClass = {gr-qc},
       adsurl = {https://ui.adsabs.harvard.edu/abs/2025PhRvD.112j2004B}
}

@ARTICLE{2008PhRvD..77d2001V,
       author = {{Vallisneri}, Michele},
        title = "{Use and abuse of the Fisher information matrix in the assessment of gravitational-wave parameter-estimation prospects}",
      journal = {\prd},
         year = 2008,
        month = feb,
       volume = {77},
       number = {4},
          eid = {042001},
        pages = {042001},
          doi = {10.1103/PhysRevD.77.042001},
archivePrefix = {arXiv},
       eprint = {gr-qc/0703086},
 primaryClass = {gr-qc},
       adsurl = {https://ui.adsabs.harvard.edu/abs/2008PhRvD..77d2001V}
}

@software{kejriwal_2026_20290209,
  author       = {Kejriwal, Shubham},
  title        = {SPLIT - Segmented Posteriors for Long-Inspiral
                   Templates
                  },
  month        = may,
  year         = 2026,
  publisher    = {Zenodo},
  version      = {v0.0.1alpha},
  doi          = {10.5281/zenodo.20290209},
  url          = {https://doi.org/10.5281/zenodo.20290209},
}

@article{wiener1930generalized,
  title={Generalized harmonic analysis},
  author={Wiener, Norbert},
  journal={Acta mathematica},
  volume={55},
  number={1},
  pages={117--258},
  year={1930},
  publisher={Springer}
}

@article{Khintchine1934,
author = {Khintchine, A.},
journal = {Mathematische Annalen},
pages = {604-615},
title = {Korrelationstheorie der stationären stochastischen Prozesse},
url = {http://eudml.org/doc/159698},
volume = {109},
year = {1934},
}

@ARTICLE{2024PhRvD.109d3037D,
       author = {{Divyajyoti}, Sumit, Kumar and {Tibrewal}, Snehal and {Romero-Shaw}, Isobel M. and {Mishra}, Chandra Kant},
        title = "{Blind spots and biases: The dangers of ignoring eccentricity in gravitational-wave signals from binary black holes}",
      journal = {\prd},
         year = 2024,
        month = feb,
       volume = {109},
       number = {4},
          eid = {043037},
        pages = {043037},
          doi = {10.1103/PhysRevD.109.043037},
archivePrefix = {arXiv},
       eprint = {2309.16638},
 primaryClass = {gr-qc},
       adsurl = {https://ui.adsabs.harvard.edu/abs/2024PhRvD.109d3037D}
}

@ARTICLE{2024arXiv240502197G,
       author = {{Gupta}, Anuradha and {Arun}, K.~G. and {Barausse}, Enrico and {Bernard}, Laura and {Berti}, Emanuele and {Bhat}, Sajad A. and {Buonanno}, Alessandra and {Cardoso}, Vitor and {Cheung}, Shun Yin and {Clarke}, Teagan A. and {Datta}, Sayantani and {Dhani}, Arnab and {Mar{\'\i}a Ezquiaga}, Jose and {Gupta}, Ish and {Guttman}, Nir and {Hinderer}, Tanja and {Hu}, Qian and {Janquart}, Justin and {Johnson-McDaniel}, Nathan K. and {Kashyap}, Rahul and {Krishnendu}, N.~V. and {Lasky}, Paul D. and {Lundgren}, Andrew and {Maggio}, Elisa and {Mahapatra}, Parthapratim and {Maselli}, Andrea and {Narayan}, Purnima and {Nielsen}, Alex B. and {Nuttall}, Laura K. and {Pani}, Paolo and {Passenger}, Lachlan and {Payne}, Ethan and {Pompili}, Lorenzo and {Reali}, Luca and {Saini}, Pankaj and {Samajdar}, Anuradha and {Tiwari}, Shubhanshu and {Tong}, Hui and {Van Den Broeck}, Chris and {Yagi}, Kent and {Yang}, Huan and {Yunes}, Nicol{\'a}s and {Sathyaprakash}, B.~S.},
        title = "{Possible Causes of False General Relativity Violations in Gravitational Wave Observations}",
      journal = {arXiv e-prints},
         year = 2024,
        month = may,
          eid = {arXiv:2405.02197},
        pages = {arXiv:2405.02197},
          doi = {10.48550/arXiv.2405.02197},
archivePrefix = {arXiv},
       eprint = {2405.02197},
 primaryClass = {gr-qc},
       adsurl = {https://ui.adsabs.harvard.edu/abs/2024arXiv240502197G}
}

@ARTICLE{2025PhRvX..15c1036D,
       author = {{Dhani}, Arnab and {V{\"o}lkel}, Sebastian H. and {Buonanno}, Alessandra and {Estelles}, Hector and {Gair}, Jonathan and {Pfeiffer}, Harald P. and {Pompili}, Lorenzo and {Toubiana}, Alexandre},
        title = "{Systematic Biases in Estimating the Properties of Black Holes Due to Inaccurate Gravitational-Wave Models}",
      journal = {Physical Review X},
         year = 2025,
        month = jul,
       volume = {15},
       number = {3},
          eid = {031036},
        pages = {031036},
          doi = {10.1103/5pks-qz6b},
archivePrefix = {arXiv},
       eprint = {2404.05811},
 primaryClass = {gr-qc},
       adsurl = {https://ui.adsabs.harvard.edu/abs/2025PhRvX..15c1036D}
}

@ARTICLE{2025arXiv250818082T,
       author = {{The LIGO Scientific Collaboration} and {the Virgo Collaboration} and {the KAGRA Collaboration} and {Abac}, A.~G. and {Abouelfettouh}, I. and {Acernese}, F. and {Ackley}, K. and {Adamcewicz}, C. and {Adhicary}, S. and {Adhikari}, D. and {Adhikari}, N. and {Adhikari}, R.~X. and {Adkins}, V.~K. and {Afroz}, S. and {Agapito}, A. and {Agarwal}, D. and {Agathos}, M. and {Aggarwal}, N. and {Aggarwal}, S. and {Aguiar}, O.~D. and {Ahrend}, I.-L. and {Aiello}, L. and {Ain}, A. and {Ajith}, P. and {Akutsu}, T. and {Albanesi}, S. and {Ali}, W. and {Al-Kershi}, S. and {All{\'e}n{\'e}}, C. and {Allocca}, A. and {Al-Shammari}, S. and {Altin}, P.~A. and {Alvarez-Lopez}, S. and {Amar}, W. and {Amarasinghe}, O. and {Amato}, A. and {Amicucci}, F. and {Amra}, C. and {Ananyeva}, A. and {Anderson}, S.~B. and {Anderson}, W.~G. and {Andia}, M. and {Ando}, M. and {Andr{\'e}s-Carcasona}, M. and {Andri{\'c}}, T. and {Anglin}, J. and {Ansoldi}, S. and {Antelis}, J.~M. and {Antier}, S. and {Aoumi}, M. and {Appavuravther}, E.~Z. and {Appert}, S. and {Apple}, S.~K. and {Arai}, K. and {Araya}, A. and {Araya}, M.~C. and {Arca Sedda}, M. and {Areeda}, J.~S. and {Aritomi}, N. and {Armato}, F. and {Armstrong}, S. and {Arnaud}, N. and {Arogeti}, M. and {Aronson}, S.~M. and {Arun}, K.~G. and {Ashton}, G. and {Aso}, Y. and {Asprea}, L. and {Assiduo}, M. and {Assis de Souza Melo}, S. and {Aston}, S.~M. and {Astone}, P. and {Attadio}, F. and {Aubin}, F. and {AultONeal}, K. and {Avallone}, G. and {Avila}, E.~A. and {Babak}, S. and {Badger}, C. and {Bae}, S. and {Bagnasco}, S. and {Baiotti}, L. and {Bajpai}, R. and {Baka}, T. and {Baker}, A.~M. and {Baker}, K.~A. and {Baker}, T. and {Baldi}, G. and {Baldicchi}, N. and {Ball}, M. and {Ballardin}, G. and {Ballmer}, S.~W. and {Banagiri}, S. and {Banerjee}, B. and {Bankar}, D. and {Baptiste}, T.~M. and {Baral}, P. and {Baratti}, M. and {Barayoga}, J.~C. and {Barish}, B.~C. and {Barker}, D. and {Barman}, N. and {Barneo}, P. and {Barone}, F. and {Barr}, B. and {Barsotti}, L. and {Barsuglia}, M. and {Barta}, D. and {Bartoletti}, A.~M. and {Barton}, M.~A. and {Bartos}, I. and {Basalaev}, A. and {Bassiri}, R. and {Basti}, A. and {Bawaj}, M. and {Baxi}, P. and {Bayley}, J.~C. and {Baylor}, A.~C. and {Baynard}, II, P.~A. and {Bazzan}, M. and {Bedakihale}, V.~M. and {Beirnaert}, F. and {Bejger}, M. and {Belardinelli}, D. and {Bell}, A.~S. and {Bellie}, D.~S. and {Bellizzi}, L. and {Benoit}, W. and {Bentara}, I. and {Bentley}, J.~D. and {Ben Yaala}, M. and {Bera}, S. and {Bergamin}, F. and {Berger}, B.~K. and {Bernuzzi}, S. and {Beroiz}, M. and {Berry}, C.~P.~L. and {Bersanetti}, D. and {Bertheas}, T. and {Bertolini}, A. and {Betzwieser}, J. and {Beveridge}, D. and {Bevilacqua}, G. and {Bevins}, N. and {Bhandare}, R. and {Bhatt}, R. and {Bhattacharjee}, D. and {Bhattacharyya}, S. and {Bhaumik}, S. and {Biancalana}, V. and {Bianchi}, A. and {Bilenko}, I.~A. and {Billingsley}, G. and {Binetti}, A. and {Bini}, S. and {Binu}, C. and {Biot}, S. and {Birnholtz}, O. and {Biscoveanu}, S. and {Bisht}, A. and {Bitossi}, M. and {Bizouard}, M.-A. and {Blaber}, S. and {Blackburn}, J.~K. and {Blagg}, L.~A. and {Blair}, C.~D. and {Blair}, D.~G. and {Bode}, N. and {Boettner}, N. and {Boileau}, G. and {Boldrini}, M. and {Bolingbroke}, G.~N. and {Bolliand}, A. and {Bonavena}, L.~D. and {Bondarescu}, R. and {Bondu}, F. and {Bonilla}, E. and {Bonilla}, M.~S. and {Bonino}, A. and {Bonnand}, R. and {Borchers}, A. and {Borhanian}, S. and {Boschi}, V. and {Bose}, S. and {Bossilkov}, V. and {Bothra}, Y. and {Boudon}, A. and {Bourg}, L. and {Boyle}, M. and {Bozzi}, A. and {Bradaschia}, C. and {Brady}, P.~R. and {Branch}, A. and {Branchesi}, M. and {Braun}, I. and {Briant}, T. and {Brillet}, A. and {Brinkmann}, M. and {Brockill}, P. and {Brockmueller}, E.},
        title = "{GWTC-4.0: Updating the Gravitational-Wave Transient Catalog with Observations from the First Part of the Fourth LIGO-Virgo-KAGRA Observing Run}",
      journal = {arXiv e-prints},
         year = 2025,
        month = aug,
          eid = {arXiv:2508.18082},
        pages = {arXiv:2508.18082},
          doi = {10.48550/arXiv.2508.18082},
archivePrefix = {arXiv},
       eprint = {2508.18082},
 primaryClass = {gr-qc},
       adsurl = {https://ui.adsabs.harvard.edu/abs/2025arXiv250818082T}
}

@ARTICLE{2020SciPy-NMeth,
  author  = {Virtanen, Pauli and Gommers, Ralf and Oliphant, Travis E. and
            Haberland, Matt and Reddy, Tyler and Cournapeau, David and
            Burovski, Evgeni and Peterson, Pearu and Weckesser, Warren and
            Bright, Jonathan and {van der Walt}, St{\'e}fan J. and
            Brett, Matthew and Wilson, Joshua and Millman, K. Jarrod and
            Mayorov, Nikolay and Nelson, Andrew R. J. and Jones, Eric and
            Kern, Robert and Larson, Eric and Carey, C J and
            Polat, {\.I}lhan and Feng, Yu and Moore, Eric W. and
            {VanderPlas}, Jake and Laxalde, Denis and Perktold, Josef and
            Cimrman, Robert and Henriksen, Ian and Quintero, E. A. and
            Harris, Charles R. and Archibald, Anne M. and
            Ribeiro, Ant{\^o}nio H. and Pedregosa, Fabian and
            {van Mulbregt}, Paul and {SciPy 1.0 Contributors}},
  title   = {{{SciPy} 1.0: Fundamental Algorithms for Scientific
            Computing in Python}},
  journal = {Nature Methods},
  year    = {2020},
  volume  = {17},
  pages   = {261--272},
  adsurl  = {https://rdcu.be/b08Wh},
  doi     = {10.1038/s41592-019-0686-2},
}

@ARTICLE{2021PhRvD.104d2003M,
       author = {{Melatos}, A. and {Clearwater}, P. and {Suvorova}, S. and {Sun}, L. and {Moran}, W. and {Evans}, R.~J.},
        title = "{Hidden Markov model tracking of continuous gravitational waves from a binary neutron star with wandering spin. III. Rotational phase tracking}",
      journal = {\prd},
         year = 2021,
        month = aug,
       volume = {104},
       number = {4},
          eid = {042003},
        pages = {042003},
          doi = {10.1103/PhysRevD.104.042003},
archivePrefix = {arXiv},
       eprint = {2107.12822},
 primaryClass = {gr-qc},
       adsurl = {https://ui.adsabs.harvard.edu/abs/2021PhRvD.104d2003M}
}

@ARTICLE{2026arXiv260314168T,
       author = {{The LIGO Scientific Collaboration} and {the Virgo Collaboration} and {the KAGRA Collaboration} and {Abac}, A.~G. and {Abouelfettouh}, I. and {Acernese}, F. and {Ackley}, K. and {Adam}, A. and {Adamcewicz}, C. and {Adhicary}, S. and {Adhikari}, D. and {Adhikari}, N. and {Adhikari}, R.~X. and {Adkins}, V.~K. and {Afroz}, S. and {Agapito}, A. and {Agarwal}, D. and {Agathos}, M. and {Aggarwal}, N. and {Aggarwal}, S. and {Aguiar}, O.~D. and {Ahrend}, I.-L. and {Aiello}, L. and {Ain}, A. and {Ajith}, P. and {Akutsu}, T. and {Albanesi}, S. and {Albers}, L. and {Ali}, W. and {Al-Kershi}, S. and {All{\'e}n{\'e}}, C. and {Allocca}, A. and {Al-Shammari}, S. and {Altin}, P.~A. and {Alvarez-Lopez}, S. and {Amarasinghe}, O. and {Amato}, A. and {Amicucci}, F. and {Amra}, C. and {Anand}, C. and {Ananyeva}, A. and {Anderson}, S.~B. and {Anderson}, W.~G. and {Andia}, M. and {Ando}, M. and {Andr{\'e}s-Carcasona}, M. and {Andrey}, J.~L. and {Andri{\'c}}, T. and {Anglin}, J. and {Anna}, J. and {Ansoldi}, S. and {Antelis}, J.~M. and {Antier}, S. and {Aoumi}, M. and {Appavuravther}, E.~Z. and {Appert}, S. and {Apple}, S.~K. and {Arai}, K. and {Araya}, A. and {Araya}, M.~C. and {Arca Sedda}, M. and {Arciprete}, F. and {Areeda}, J.~S. and {Aritomi}, N. and {Armato}, F. and {Armstrong}, S. and {Arnaud}, N. and {Arogeti}, M. and {Aronson}, S.~M. and {Ashton}, G. and {Aso}, Y. and {Asprea}, L. and {Assiduo}, M. and {Assis de Souza Melo}, S. and {Aston}, S.~M. and {Astone}, P. and {Attadio}, F. and {Aubin}, F. and {AultONeal}, K. and {Avallone}, G. and {Avila}, E.~A. and {Babak}, S. and {Badger}, C. and {Bae}, S. and {Bagnasco}, S. and {Baiotti}, L. and {Bajpai}, R. and {Baka}, T. and {Baker}, K.~A. and {Baker}, T. and {Balbi}, G. and {Baldi}, G. and {Baldicchi}, N. and {Ball}, M. and {Ballardin}, G. and {Ballmer}, S.~W. and {Banagiri}, S. and {Banerjee}, B. and {Bankar}, D. and {Baptiste}, T.~M. and {Baral}, P. and {Baratti}, M. and {Barayoga}, J.~C. and {Baric}, K. and {Barish}, B.~C. and {Barker}, D. and {Barman}, N. and {Barneo}, P. and {Barone}, F. and {Barr}, B. and {Barrios}, M. and {Barsotti}, L. and {Barsuglia}, M. and {Barta}, D. and {Barton}, M.~A. and {Bartos}, I. and {Basalaev}, A. and {Bassiri}, R. and {Basti}, A. and {Bawaj}, M. and {Baxi}, P. and {Bayley}, J.~C. and {Baylor}, A.~C. and {Baynard}, II, P.~A. and {Bazzan}, M. and {Bedakihale}, V.~M. and {Beirnaert}, F. and {Bejger}, M. and {Belardinelli}, D. and {Bell}, A.~S. and {Bellani}, C. and {Bellizzi}, L. and {Beltran-Martinez}, D. and {Benoit}, W. and {Bentara}, I. and {Ben Yaala}, M. and {Bera}, S. and {Bergamin}, F. and {Berger}, B.~K. and {Bernuzzi}, S. and {Beroiz}, M. and {Berry}, I. and {Bersanetti}, D. and {Bertheas}, T. and {Bertolini}, A. and {Betzwieser}, J. and {Beveridge}, D. and {Bevilacqua}, G. and {Bevins}, N. and {Bhandare}, R. and {Bhatt}, R. and {Bhattacharjee}, A. and {Bhattacharjee}, D. and {Bhattacharyya}, S. and {Bhaumik}, S. and {Biancalana}, V. and {Bianchi}, A. and {Bianchi}, F. and {Bilenko}, I.~A. and {Billingsley}, G. and {Binetti}, A. and {Bini}, S. and {Binu}, C. and {Biot}, S. and {Birnholtz}, O. and {Biscoveanu}, S. and {Bisht}, A. and {Bitossi}, M. and {Bizouard}, M.-A. and {Blaber}, S. and {Blackburn}, J.~K. and {Blagg}, L.~A. and {Blair}, C.~D. and {Blair}, D.~G. and {Bode}, N. and {Boettner}, N. and {Bogdan}, P. and {Boileau}, G. and {Boldrini}, M. and {Bolingbroke}, G.~N. and {Bolliand}, A. and {Bonavena}, L.~D. and {Bondarescu}, R. and {Bondu}, F. and {Bonhomme}, V.~A. and {Bonilla}, E. and {Bonilla}, M.~S. and {Bonino}, A. and {Bonnand}, R. and {Borchers}, A. and {Borghi}, N. and {Boschi}, V. and {Bose}, S. and {Bossilkov}, V. and {Bothra}, Y. and {Boudon}, A. and {Boyle}, M. and {Bozzi}, A. and {Bradaschia}, C. and {Brady}, M.~J.},
        title = "{All-sky Searches for Continuous Gravitational Waves from Isolated Neutron Stars in the Data from the First Part of the Fourth LIGO-Virgo-KAGRA Observing Run}",
      journal = {arXiv e-prints},
         year = 2026,
        month = mar,
          eid = {arXiv:2603.14168},
        pages = {arXiv:2603.14168},
          doi = {10.48550/arXiv.2603.14168},
archivePrefix = {arXiv},
       eprint = {2603.14168},
 primaryClass = {gr-qc},
       adsurl = {https://ui.adsabs.harvard.edu/abs/2026arXiv260314168T}
}

@ARTICLE{2025PhRvD.111h4006D,
       author = {{Duque}, Francisco and {Kejriwal}, Shubham and {Sberna}, Laura and {Speri}, Lorenzo and {Gair}, Jonathan},
        title = "{Constraining accretion physics with gravitational waves from eccentric extreme-mass-ratio inspirals}",
      journal = {\prd},
         year = 2025,
        month = apr,
       volume = {111},
       number = {8},
          eid = {084006},
        pages = {084006},
          doi = {10.1103/PhysRevD.111.084006},
archivePrefix = {arXiv},
       eprint = {2411.03436},
 primaryClass = {gr-qc},
       adsurl = {https://ui.adsabs.harvard.edu/abs/2025PhRvD.111h4006D}
}

@ARTICLE{2022PhRvL.128w1101I,
       author = {{Isoyama}, Soichiro and {Fujita}, Ryuichi and {Chua}, Alvin J.~K. and {Nakano}, Hiroyuki and {Pound}, Adam and {Sago}, Norichika},
        title = "{Adiabatic Waveforms from Extreme-Mass-Ratio Inspirals: An Analytical Approach}",
      journal = {\prl},
         year = 2022,
        month = jun,
       volume = {128},
       number = {23},
          eid = {231101},
        pages = {231101},
          doi = {10.1103/PhysRevLett.128.231101},
archivePrefix = {arXiv},
       eprint = {2111.05288},
 primaryClass = {gr-qc},
       adsurl = {https://ui.adsabs.harvard.edu/abs/2022PhRvL.128w1101I}
}

@ARTICLE{2010CAMCS...5...65G,
       author = {{Goodman}, Jonathan and {Weare}, Jonathan},
        title = "{Ensemble samplers with affine invariance}",
      journal = {Communications in Applied Mathematics and Computational Science},
         year = 2010,
        month = jan,
       volume = {5},
       number = {1},
        pages = {65-80},
          doi = {10.2140/camcos.2010.5.65},
       adsurl = {https://ui.adsabs.harvard.edu/abs/2010CAMCS...5...65G}
}

@ARTICLE{2022PhRvD.106j4051C,
       author = {{Chua}, Alvin J.~K.},
        title = "{One-stop function for gravitational-wave detection, identification, and inference}",
      journal = {\prd},
         year = 2022,
        month = nov,
       volume = {106},
       number = {10},
          eid = {104051},
        pages = {104051},
          doi = {10.1103/PhysRevD.106.104051},
archivePrefix = {arXiv},
       eprint = {2205.08702},
 primaryClass = {gr-qc},
       adsurl = {https://ui.adsabs.harvard.edu/abs/2022PhRvD.106j4051C}
}

@ARTICLE{2025PhRvD.111b2005K,
       author = {{Karnesis}, N. and {Sasli}, A. and {Buscicchio}, R. and {Stergioulas}, N.},
        title = "{Characterization of non-Gaussian stochastic signals with heavier-tailed likelihoods}",
      journal = {\prd},
         year = 2025,
        month = jan,
       volume = {111},
       number = {2},
          eid = {022005},
        pages = {022005},
          doi = {10.1103/PhysRevD.111.022005},
archivePrefix = {arXiv},
       eprint = {2410.14354},
 primaryClass = {gr-qc},
       adsurl = {https://ui.adsabs.harvard.edu/abs/2025PhRvD.111b2005K}
}

@ARTICLE{2024EPJC...84.1023M,
       author = {{Martini}, Alessandro and {Schmidt}, Stefano and {Ashton}, Gregory and {Del Pozzo}, Walter},
        title = "{Maximum entropy spectral analysis: an application to gravitational waves data analysis}",
      journal = {European Physical Journal C},
         year = 2024,
        month = oct,
       volume = {84},
       number = {10},
          eid = {1023},
        pages = {1023},
          doi = {10.1140/epjc/s10052-024-13400-6},
archivePrefix = {arXiv},
       eprint = {2106.09499},
 primaryClass = {stat.ME},
       adsurl = {https://ui.adsabs.harvard.edu/abs/2024EPJC...84.1023M}
}

@book{bretthorst2013bayesian,
  title={Bayesian Spectrum Analysis and Parameter Estimation},
  author={Bretthorst, G.L.},
  isbn={9781468493993},
  series={Lecture Notes in Statistics},
  url={https://books.google.com.bn/books?id=imX1BwAAQBAJ},
  year={2013},
  publisher={Springer New York}
}

@article{10.1111/j.2517-6161.1957.tb00242.x,
    author = {Whittle, P.},
    title = {Curve and Periodogram Smoothing},
    journal = {Journal of the Royal Statistical Society: Series B (Methodological)},
    volume = {19},
    number = {1},
    pages = {38-47},
    year = {1957},
    month = {01},
    issn = {0035-9246},
    doi = {10.1111/j.2517-6161.1957.tb00242.x},
    url = {https://doi.org/10.1111/j.2517-6161.1957.tb00242.x},
    eprint = {https://academic.oup.com/jrsssb/article-pdf/19/1/38/49094277/jrsssb\_19\_1\_38.pdf},
}

@ARTICLE{2025PhRvD.111j4002T,
       author = {{Tenorio}, Rodrigo and {M{\'e}rou}, Joan-Ren{\'e} and {Sintes}, Alicia M.},
        title = "{One-stop strategy to search for long-duration gravitational-wave signals}",
      journal = {\prd},
         year = 2025,
        month = may,
       volume = {111},
       number = {10},
          eid = {104002},
        pages = {104002},
          doi = {10.1103/PhysRevD.111.104002},
archivePrefix = {arXiv},
       eprint = {2411.18370},
 primaryClass = {gr-qc},
       adsurl = {https://ui.adsabs.harvard.edu/abs/2025PhRvD.111j4002T}
}

@ARTICLE{2024PhRvD.110j3026B,
       author = {{Bandopadhyay}, Diganta and {Moore}, Christopher J.},
        title = "{GPU-accelerated semicoherent hierarchical search for stellar-mass binary inspiral signals in LISA}",
      journal = {\prd},
         year = 2024,
        month = nov,
       volume = {110},
       number = {10},
          eid = {103026},
        pages = {103026},
          doi = {10.1103/PhysRevD.110.103026},
archivePrefix = {arXiv},
       eprint = {2408.13170},
 primaryClass = {gr-qc},
       adsurl = {https://ui.adsabs.harvard.edu/abs/2024PhRvD.110j3026B}
}

@ARTICLE{2026PhRvD.113b4061S,
       author = {{Speri}, Lorenzo and {Tenorio}, Rodrigo and {Chapman-Bird}, Christian and {Gerosa}, Davide},
        title = "{Single-harmonic search for extreme mass-ratio inspirals}",
      journal = {\prd},
         year = 2026,
        month = jan,
       volume = {113},
       number = {2},
          eid = {024061},
        pages = {024061},
          doi = {10.1103/dh3j-ksfl},
archivePrefix = {arXiv},
       eprint = {2510.20891},
 primaryClass = {gr-qc},
       adsurl = {https://ui.adsabs.harvard.edu/abs/2026PhRvD.113b4061S}
}

@ARTICLE{2023PhRvD.108h4014B,
       author = {{Bandopadhyay}, Diganta and {Moore}, Christopher J.},
        title = "{LISA stellar-mass black hole searches with semicoherent and particle-swarm methods}",
      journal = {\prd},
         year = 2023,
        month = oct,
       volume = {108},
       number = {8},
          eid = {084014},
        pages = {084014},
          doi = {10.1103/PhysRevD.108.084014},
archivePrefix = {arXiv},
       eprint = {2305.18048},
 primaryClass = {gr-qc},
       adsurl = {https://ui.adsabs.harvard.edu/abs/2023PhRvD.108h4014B}
}

@ARTICLE{2010CQGra..27t5017D,
       author = {{Dergachev}, V.},
        title = "{On blind searches for noise dominated signals: a loosely coherent approach}",
      journal = {Classical and Quantum Gravity},
         year = 2010,
        month = oct,
       volume = {27},
       number = {20},
          eid = {205017},
        pages = {205017},
          doi = {10.1088/0264-9381/27/20/205017},
archivePrefix = {arXiv},
       eprint = {1003.2178},
 primaryClass = {gr-qc},
       adsurl = {https://ui.adsabs.harvard.edu/abs/2010CQGra..27t5017D}
}

@ARTICLE{1992StaSc...7..457G,
       author = {{Gelman}, Andrew and {Rubin}, Donald B.},
        title = "{Inference from Iterative Simulation Using Multiple Sequences}",
      journal = {Statistical Science},
         year = 1992,
        month = jan,
       volume = {7},
        pages = {457-472},
          doi = {10.1214/ss/1177011136},
       adsurl = {https://ui.adsabs.harvard.edu/abs/1992StaSc...7..457G}
}

@ARTICLE{2017PhRvD..95j3012B,
       author = {{Babak}, Stanislav and {Gair}, Jonathan and {Sesana}, Alberto and {Barausse}, Enrico and {Sopuerta}, Carlos F. and {Berry}, Christopher P.~L. and {Berti}, Emanuele and {Amaro-Seoane}, Pau and {Petiteau}, Antoine and {Klein}, Antoine},
        title = "{Science with the space-based interferometer LISA. V. Extreme mass-ratio inspirals}",
      journal = {\prd},
         year = 2017,
        month = may,
       volume = {95},
       number = {10},
          eid = {103012},
        pages = {103012},
          doi = {10.1103/PhysRevD.95.103012},
archivePrefix = {arXiv},
       eprint = {1703.09722},
 primaryClass = {gr-qc},
       adsurl = {https://ui.adsabs.harvard.edu/abs/2017PhRvD..95j3012B}
}

@BOOK{2007nras.book.....P,
       author = {{Press}, William H. and {Teukolsky}, Saul A. and {Vetterling}, William T. and {Flannery}, Brian P.},
        title = "{Numerical Recipes: The Art of Scientific Computing}",
         year = 2007,
       adsurl = {https://ui.adsabs.harvard.edu/abs/2007nras.book.....P}
}

@ARTICLE{2024arXiv240207571C,
       author = {{Colpi}, Monica and {Danzmann}, Karsten and {Hewitson}, Martin and {Holley-Bockelmann}, Kelly and {Jetzer}, Philippe and {Nelemans}, Gijs and {Petiteau}, Antoine and {Shoemaker}, David and {Sopuerta}, Carlos and {Stebbins}, Robin and {Tanvir}, Nial and {Ward}, Henry and {Weber}, William Joseph and {Thorpe}, Ira and {Daurskikh}, Anna and {Deep}, Atul and {Fern{\'a}ndez N{\'u}{\~n}ez}, Ignacio and {Garc{\'\i}a Marirrodriga}, C{\'e}sar and {Gehler}, Martin and {Halain}, Jean-Philippe and {Jennrich}, Oliver and {Lammers}, Uwe and {Larra{\~n}aga}, Jonan and {Lieser}, Maike and {L{\"u}tzgendorf}, Nora and {Martens}, Waldemar and {Mondin}, Linda and {Piris Ni{\~n}o}, Ana and {Amaro-Seoane}, Pau and {Arca Sedda}, Manuel and {Auclair}, Pierre and {Babak}, Stanislav and {Baghi}, Quentin and {Baibhav}, Vishal and {Baker}, Tessa and {Bayle}, Jean-Baptiste and {Berry}, Christopher and {Berti}, Emanuele and {Boileau}, Guillaume and {Bonetti}, Matteo and {Brito}, Richard and {Buscicchio}, Riccardo and {Calcagni}, Gianluca and {Capelo}, Pedro R. and {Caprini}, Chiara and {Caputo}, Andrea and {Castelli}, Eleonora and {Chen}, Hsin-Yu and {Chen}, Xian and {Chua}, Alvin and {Davies}, Gareth and {Derdzinski}, Andrea and {Domcke}, Valerie Fiona and {Doneva}, Daniela and {Dvorkin}, Irna and {Mar{\'\i}a Ezquiaga}, Jose and {Gair}, Jonathan and {Haiman}, Zoltan and {Harry}, Ian and {Hartwig}, Olaf and {Hees}, Aurelien and {Heffernan}, Anna and {Husa}, Sascha and {Izquierdo-Villalba}, David and {Karnesis}, Nikolaos and {Klein}, Antoine and {Korol}, Valeriya and {Korsakova}, Natalia and {Kupfer}, Thomas and {Laghi}, Danny and {Lamberts}, Astrid and {Larson}, Shane and {Le Jeune}, Maude and {Lewicki}, Marek and {Littenberg}, Tyson and {Madge}, Eric and {Mangiagli}, Alberto and {Marsat}, Sylvain and {Vilchez}, Ivan Martin and {Maselli}, Andrea and {Mathews}, Josh and {van de Meent}, Maarten and {Muratore}, Martina and {Nardini}, Germano and {Pani}, Paolo and {Peloso}, Marco and {Pieroni}, Mauro and {Pound}, Adam and {Quelquejay-Leclere}, Hippolyte and {Ricciardone}, Angelo and {Rossi}, Elena Maria and {Sartirana}, Andrea and {Savalle}, Etienne and {Sberna}, Laura and {Sesana}, Alberto and {Shoemaker}, Deirdre and {Slutsky}, Jacob and {Sotiriou}, Thomas and {Speri}, Lorenzo and {Staab}, Martin and {Steer}, Dani{\`e}le and {Tamanini}, Nicola and {Tasinato}, Gianmassimo and {Torrado}, Jesus and {Torres-Orjuela}, Alejandro and {Toubiana}, Alexandre and {Vallisneri}, Michele and {Vecchio}, Alberto and {Volonteri}, Marta and {Yagi}, Kent and {Zwick}, Lorenz},
        title = "{LISA Definition Study Report}",
      journal = {arXiv e-prints},
         year = 2024,
        month = feb,
          eid = {arXiv:2402.07571},
        pages = {arXiv:2402.07571},
          doi = {10.48550/arXiv.2402.07571},
archivePrefix = {arXiv},
       eprint = {2402.07571},
 primaryClass = {astro-ph.CO},
       adsurl = {https://ui.adsabs.harvard.edu/abs/2024arXiv240207571C}
}

@ARTICLE{1973A&A....24..337S,
       author = {{Shakura}, N.~I. and {Sunyaev}, R.~A.},
        title = "{Black holes in binary systems. Observational appearance.}",
      journal = {\aap},
         year = 1973,
        month = jan,
       volume = {24},
        pages = {337-355},
       adsurl = {https://ui.adsabs.harvard.edu/abs/1973A&A....24..337S}
}

@ARTICLE{2013LRR....16....1A,
       author = {{Abramowicz}, Marek A. and {Fragile}, P. Chris},
        title = "{Foundations of Black Hole Accretion Disk Theory}",
      journal = {Living Reviews in Relativity},
         year = 2013,
        month = dec,
       volume = {16},
       number = {1},
          eid = {1},
        pages = {1},
          doi = {10.12942/lrr-2013-1},
archivePrefix = {arXiv},
       eprint = {1104.5499},
 primaryClass = {astro-ph.HE},
       adsurl = {https://ui.adsabs.harvard.edu/abs/2013LRR....16....1A}
}

@ARTICLE{2014PhRvD..89j4059B,
       author = {{Barausse}, Enrico and {Cardoso}, Vitor and {Pani}, Paolo},
        title = "{Can environmental effects spoil precision gravitational-wave astrophysics?}",
      journal = {\prd},
         year = 2014,
        month = may,
       volume = {89},
       number = {10},
          eid = {104059},
        pages = {104059},
          doi = {10.1103/PhysRevD.89.104059},
archivePrefix = {arXiv},
       eprint = {1404.7149},
 primaryClass = {gr-qc},
       adsurl = {https://ui.adsabs.harvard.edu/abs/2014PhRvD..89j4059B}
}

@Article{mahalanobis ,
author = {Mahalanobis, P C},
title={Reprint of: Mahalanobis, P.C. (1936) "On the Generalised Distance in Statistics."},
journal={Sankhya A},
year={2018},
month={Dec},
day={01},
volume={80},
number={1},
pages={1-7},
issn={0976-8378},
doi={10.1007/s13171-019-00164-5},
url={https://doi.org/10.1007/s13171-019-00164-5}
}

@ARTICLE{2019arXiv191200042W,
       author = {{Winkler}, Christina and {Worrall}, Daniel and {Hoogeboom}, Emiel and {Welling}, Max},
        title = "{Learning Likelihoods with Conditional Normalizing Flows}",
      journal = {arXiv e-prints},
         year = 2019,
        month = nov,
          eid = {arXiv:1912.00042},
        pages = {arXiv:1912.00042},
          doi = {10.48550/arXiv.1912.00042},
archivePrefix = {arXiv},
       eprint = {1912.00042},
 primaryClass = {cs.LG},
       adsurl = {https://ui.adsabs.harvard.edu/abs/2019arXiv191200042W}
}

@ARTICLE{2015arXiv150505770J,
       author = {{Jimenez Rezende}, Danilo and {Mohamed}, Shakir},
        title = "{Variational Inference with Normalizing Flows}",
      journal = {arXiv e-prints},
         year = 2015,
        month = may,
          eid = {arXiv:1505.05770},
        pages = {arXiv:1505.05770},
          doi = {10.48550/arXiv.1505.05770},
archivePrefix = {arXiv},
       eprint = {1505.05770},
 primaryClass = {stat.ML},
       adsurl = {https://ui.adsabs.harvard.edu/abs/2015arXiv150505770J}
}

@book{prince2023understanding,
  author = {Prince, Simon J.D.},
  description = {udlbook},
  publisher = {MIT Press},
  title = {Understanding Deep Learning},
  url = {http://udlbook.com},
  year = 2023
}

@ARTICLE{2025arXiv251216322B,
       author = {{Boumerdassi}, Amin and {Edwards}, Matthew C. and {Vajpeyi}, Avi and {Burke}, Ollie},
        title = "{First-time assessment of glitch-induced bias and uncertainty in inference of extreme mass ratio inspirals}",
      journal = {arXiv e-prints},
         year = 2025,
        month = dec,
          eid = {arXiv:2512.16322},
        pages = {arXiv:2512.16322},
          doi = {10.48550/arXiv.2512.16322},
archivePrefix = {arXiv},
       eprint = {2512.16322},
 primaryClass = {gr-qc},
       adsurl = {https://ui.adsabs.harvard.edu/abs/2025arXiv251216322B}
}

@ARTICLE{2025PhRvD.112f3041M,
       author = {{Muratore}, Martina and {Gair}, Jonathan and {Hartwig}, Olaf and {Katz}, Michael L. and {Toubiana}, Alexandre},
        title = "{Pipeline for searching and fitting instrumental glitches in LISA data}",
      journal = {\prd},
         year = 2025,
        month = sep,
       volume = {112},
       number = {6},
          eid = {063041},
        pages = {063041},
          doi = {10.1103/1sj2-219n},
archivePrefix = {arXiv},
       eprint = {2505.19870},
 primaryClass = {gr-qc},
       adsurl = {https://ui.adsabs.harvard.edu/abs/2025PhRvD.112f3041M}
}

@article{Green:1995mxx,
    author = "Green, Peter J.",
    title = "{Reversible jump Markov chain Monte Carlo computation and Bayesian model determination}",
    doi = "10.1093/biomet/82.4.711",
    journal = "Biometrika",
    volume = "82",
    number = "4",
    pages = "711--732",
    year = "1995"
}

@ARTICLE{2023PhRvD.107f3004L,
       author = {{Littenberg}, Tyson B. and {Cornish}, Neil J.},
        title = "{Prototype global analysis of LISA data with multiple source types}",
      journal = {\prd},
         year = 2023,
        month = mar,
       volume = {107},
       number = {6},
          eid = {063004},
        pages = {063004},
          doi = {10.1103/PhysRevD.107.063004},
archivePrefix = {arXiv},
       eprint = {2301.03673},
 primaryClass = {gr-qc},
       adsurl = {https://ui.adsabs.harvard.edu/abs/2023PhRvD.107f3004L}
}

@ARTICLE{2025PhRvD.111j3014D,
       author = {{Deng}, Senwen and {Babak}, Stanislav and {Le Jeune}, Maude and {Marsat}, Sylvain and {Plagnol}, {\'E}ric and {Sartirana}, Andrea},
        title = "{Modular global-fit pipeline for LISA data analysis}",
      journal = {\prd},
         year = 2025,
        month = may,
       volume = {111},
       number = {10},
          eid = {103014},
        pages = {103014},
          doi = {10.1103/PhysRevD.111.103014},
archivePrefix = {arXiv},
       eprint = {2501.10277},
 primaryClass = {gr-qc},
       adsurl = {https://ui.adsabs.harvard.edu/abs/2025PhRvD.111j3014D}
}

@ARTICLE{2026JCAP...03..081A,
       author = {{Abac}, Adrian and {Abramo}, Raul and {Albanesi}, Simone and {Albertini}, Angelica and {Agapito}, Alessandro and {Agathos}, Michalis and {Albertus}, Conrado and {Andersson}, Nils and {Andrade}, Tom{\'a}s and {Andreoni}, Igor and {Angeloni}, Federico and {Antonelli}, Marco and {Antoniadis}, John and {Antonini}, Fabio and {Arca Sedda}, Manuel and {Artale}, M. Celeste and {Ascenzi}, Stefano and {Auclair}, Pierre and {Bachetti}, Matteo and {Badger}, Charles and {Banerjee}, Biswajit and {Barba-Gonz{\'a}lez}, David and {Barta}, D{\'a}niel and {Bartolo}, Nicola and {Bauswein}, Andreas and {Begnoni}, Andrea and {Beirnaert}, Freija and {Bejger}, Micha{\l} and {Belgacem}, Enis and {Bellomo}, Nicola and {Bernard}, Laura and {Bernardini}, Maria Grazia and {Bernuzzi}, Sebastiano and {Berry}, Christopher P.~L. and {Berti}, Emanuele and {Bertone}, Gianfranco and {Bettoni}, Dario and {Bezares}, Miguel and {Bhagwat}, Swetha and {Bisero}, Sofia and {Bizouard}, Marie Anne and {Blanco-Pillado}, Jose J. and {Blasi}, Simone and {Bonino}, Alice and {Borghese}, Alice and {Borghi}, Nicola and {Borhanian}, Ssohrab and {Bortolas}, Elisa and {Botticella}, Maria Teresa and {Branchesi}, Marica and {Breschi}, Matteo and {Brito}, Richard and {Brocato}, Enzo and {Broekgaarden}, Floor S. and {Bulik}, Tomasz and {Buonanno}, Alessandra and {Burgio}, Fiorella and {Burrows}, Adam and {Calcagni}, Gianluca and {Canevarolo}, Sofia and {Cappellaro}, Enrico and {Capurri}, Giulia and {Carbone}, Carmelita and {Casadio}, Roberto and {Cayuso}, Ramiro and {Cerd{\'a}-Dur{\'a}n}, Pablo and {Char}, Prasanta and {Chaty}, Sylvain and {Chiarusi}, Tommaso and {Chruslinska}, Martyna and {Cireddu}, Francesco and {Cole}, Philippa and {Colombo}, Alberto and {Colpi}, Monica and {Comp{\`e}re}, Geoffrey and {Contaldi}, Carlo and {Corman}, Maxence and {Crescimbeni}, Francesco and {Cristallo}, Sergio and {Cuoco}, Elena and {Cusin}, Giulia and {Canton}, Tito Dal and {D{\'a}lya}, Gergely and {D'Avanzo}, Paolo and {Davari}, Nazanin and {De Luca}, Valerio and {De Renzis}, Viola and {Della Valle}, Massimo and {Del Pozzo}, Walter and {De Santi}, Federico and {De Santis}, Alessio Ludovico and {Dietrich}, Tim and {Dimastrogiovanni}, Ema and {Domenech}, Guillem and {Doneva}, Daniela and {Drago}, Marco and {Dupletsa}, Ulyana and {Duval}, Hannah and {Dvorkin}, Irina and {Elias-Rosa}, Nancy and {Fairhurst}, Stephen and {Fantina}, Anthea F. and {Fasiello}, Matteo and {Fays}, Maxime and {Fender}, Rob and {Fischer}, Tobias and {Foucart}, Fran{\c{c}}ois and {Fragos}, Tassos and {Foffa}, Stefano and {Franciolini}, Gabriele and {Fumagalli}, Jacopo and {Gair}, Jonathan and {Gamba}, Rossella and {Garcia-Bellido}, Juan and {Garc{\'\i}a-Quir{\'o}s}, Cecilio and {Gergely}, L{\'a}szl{\'o} {\'A}rp{\'a}d and {Ghirlanda}, Giancarlo and {Ghosh}, Archisman and {Giacomazzo}, Bruno and {Gittins}, Fabian and {Giudice}, Ines Francesca and {Goncharov}, Boris and {Gonzalez}, Alejandra and {Gori{\'e}ly}, St{\'e}phane and {Graziani}, Luca and {Greco}, Giuseppe and {Gualtieri}, Leonardo and {Guidi}, Gianluca Maria and {Gupta}, Ish and {Haney}, Maria and {Hannam}, Mark and {Harms}, Jan and {Harutyunyan}, Arus and {Haskell}, Brynmor and {Haungs}, Andreas and {Hazra}, Nandini and {Hemming}, Gary and {Heng}, Ik Siong and {Hinderer}, Tanja and {van der Horst}, Alexander and {Hu}, Qian and {Husa}, Sascha and {Iacovelli}, Francesco and {Illuminati}, Giulia and {Inguglia}, Gianluca and {Villalba}, David Izquierdo and {Janquart}, Justin and {Janssens}, Kamiel and {Jenkins}, Alexander C. and {Jones}, Ian and {Kacskovics}, Bal{\'a}zs and {Klessen}, Ralf S. and {Kokkotas}, Kostas and {Kuan}, Hao-Jui and {Kumar}, Sumit and {Kuroyanagi}, Sachiko and {Laghi}, Danny and {Lamberts}, Astrid and {Lambiase}, Gaetano and {Larrouturou}, Fran{\c{c}}ois and {Leaci}, Paola and {Lenzi}, Michele and {Levan}, Andrew and {Li}, T.~G.~F. and {Li}, Yufeng and {Liang}, Dicong and {Limongi}, Marco and {Liu}, Boyuan and {Llanes-Estrada}, Felipe J. and {Loffredo}, Eleonora and {Long}, Oliver and {Lope-Oter}, Eva and {Lukes-Gerakopoulos}, Georgios and {Maggio}, Elisa and {Maggiore}, Michele and {Mancarella}, Michele and {Mapelli}, Michela and {Marchant}, Pablo and {Margiotta}, Annarita and {Mariotti}, Alberto and {Marriott-Best}, Alisha and {Marsat}, Sylvain and {Mart{\'\i}nez-Pinedo}, Gabriel and {Maselli}, Andrea and {Mastrogiovanni}, Simone and {Matos}, Isabela and {Melandri}, Andrea and {Mendes}, Raissa F.~P. and {de Souza}, Josiel Mendon{\c{c}}a Soares and {Mentasti}, Giorgio and {Mezcua}, Mar and {M{\"o}sta}, Philipp and {Mondal}, Chiranjib and {Moresco}, Michele and {Mukherjee}, Tista and {Muttoni}, Niccol{\`o} and {Nagar}, Alessandro and {Narola}, Harsh and {Nava}, Lara and {Moreno}, Pablo Navarro},
        title = "{The Science of the Einstein Telescope}",
      journal = {\jcap},
         year = 2026,
        month = mar,
       volume = {2026},
       number = {3},
          eid = {081},
        pages = {081},
          doi = {10.1088/1475-7516/2026/03/081},
archivePrefix = {arXiv},
       eprint = {2503.12263},
 primaryClass = {gr-qc},
       adsurl = {https://ui.adsabs.harvard.edu/abs/2026JCAP...03..081A}
}

@software{michael_katz_2025_17138723,
  author       = {Michael Katz and
                  Christian Chapman-Bird and
                  Lorenzo Speri and
                  Nikolaos Karnesis and
                  Alex Correia},
  title        = {mikekatz04/LISAanalysistools: v1.1.0},
  month        = sep,
  year         = 2025,
  publisher    = {Zenodo},
  version      = {v1.1.0},
  doi          = {10.5281/zenodo.17138723},
  url          = {https://doi.org/10.5281/zenodo.17138723},
  swhid        = {swh:1:dir:04bfa131f293fe545d060e3697c151f8a9abf3b4
                   ;origin=https://doi.org/10.5281/zenodo.10930979;vi
                   sit=swh:1:snp:71dea015a2c4d2d4ee495000b5024f327779
                   bbf0;anchor=swh:1:rel:17c7c3b6d4bf7f52c971b9393ede
                   4911f8ce0345;path=mikekatz04-LISAanalysistools-d3d
                   62fb
                  },
}

@ARTICLE{2022PhRvD.106j3001K,
       author = {{Katz}, Michael L. and {Bayle}, Jean-Baptiste and {Chua}, Alvin J.~K. and {Vallisneri}, Michele},
        title = "{Assessing the data-analysis impact of LISA orbit approximations using a GPU-accelerated response model}",
      journal = {\prd},
         year = 2022,
        month = nov,
       volume = {106},
       number = {10},
          eid = {103001},
        pages = {103001},
          doi = {10.1103/PhysRevD.106.103001},
archivePrefix = {arXiv},
       eprint = {2204.06633},
 primaryClass = {gr-qc},
       adsurl = {https://ui.adsabs.harvard.edu/abs/2022PhRvD.106j3001K}
}

@software{michael_katz_2025_17162632,
  author       = {Michael Katz and
                  Jean-Baptiste Bayle},
  title        = {mikekatz04/lisa-on-gpu: v1.1.6},
  month        = sep,
  year         = 2025,
  publisher    = {Zenodo},
  version      = {v1.1.6},
  doi          = {10.5281/zenodo.17162632},
  url          = {https://doi.org/10.5281/zenodo.17162632},
  swhid        = {swh:1:dir:cec27a6217b491475631bd26781e90fa7df996ee
                   ;origin=https://doi.org/10.5281/zenodo.5867730;vis
                   it=swh:1:snp:889e6b8fa7a55fc2a08a55d22843d2c389401
                   d8e;anchor=swh:1:rel:cf6ac76b44adfe3e20999b4bd76a0
                   6fa420ad869;path=mikekatz04-lisa-on-gpu-b79965f
                  },
}

@ARTICLE{2025PhRvD.111b4060K,
       author = {{Katz}, Michael L. and {Karnesis}, Nikolaos and {Korsakova}, Natalia and {Gair}, Jonathan R. and {Stergioulas}, Nikolaos},
        title = "{Efficient GPU-accelerated multisource global fit pipeline for LISA data analysis}",
      journal = {\prd},
         year = 2025,
        month = jan,
       volume = {111},
       number = {2},
          eid = {024060},
        pages = {024060},
          doi = {10.1103/PhysRevD.111.024060},
archivePrefix = {arXiv},
       eprint = {2405.04690},
 primaryClass = {gr-qc},
       adsurl = {https://ui.adsabs.harvard.edu/abs/2025PhRvD.111b4060K}
}

@ARTICLE{2022PhRvD.106f2001A,
       author = {{Armano}, M. and {Audley}, H. and {Baird}, J. and {Binetruy}, P. and {Born}, M. and {Bortoluzzi}, D. and {Castelli}, E. and {Cavalleri}, A. and {Cesarini}, A. and {Chiavegato}, V. and {Cruise}, A.~M. and {Dal Bosco}, D. and {Danzmann}, K. and {De Deus Silva}, M. and {Diepholz}, I. and {Dixon}, G. and {Dolesi}, R. and {Ferraioli}, L. and {Ferroni}, V. and {Fitzsimons}, E.~D. and {Freschi}, M. and {Gesa}, L. and {Giardini}, D. and {Gibert}, F. and {Giusteri}, R. and {Grimani}, C. and {Grzymisch}, J. and {Harrison}, I. and {Hartig}, M.~S. and {Heinzel}, G. and {Hewitson}, M. and {Hollington}, D. and {Hoyland}, D. and {Hueller}, M. and {Inchausp{\'e}}, H. and {Jennrich}, O. and {Jetzer}, P. and {Johlander}, B. and {Karnesis}, N. and {Kaune}, B. and {Korsakova}, N. and {Killow}, C.~J. and {Lobo}, J.~A. and {L{\'o}pez-Zaragoza}, J.~P. and {Maarschalkerweerd}, R. and {Mance}, D. and {Mart{\'\i}n}, V. and {Martin-Polo}, L. and {Martin-Porqueras}, F. and {Martino}, J. and {McNamara}, P.~W. and {Mendes}, J. and {Mendes}, L. and {Meshksar}, N. and {Nofrarias}, M. and {Paczkowski}, S. and {Perreur-Lloyd}, M. and {Petiteau}, A. and {Plagnol}, E. and {Ramos-Castro}, J. and {Reiche}, J. and {Rivas}, F. and {Robertson}, D.~I. and {Russano}, G. and {Sala}, L. and {Sarra}, P. and {Slutsky}, J. and {Sopuerta}, C.~F. and {Sumner}, T. and {Texier}, D. and {Thorpe}, J.~I. and {Vetrugno}, D. and {Vitale}, S. and {Wanner}, G. and {Ward}, H. and {Wass}, P. and {Weber}, W.~J. and {Wissel}, L. and {Wittchen}, A. and {Zanoni}, C. and {Zweifel}, P. and {LISA Pathfinder Collaboration}},
        title = "{Transient acceleration events in LISA Pathfinder data: Properties and possible physical origin}",
      journal = {\prd},
         year = 2022,
        month = sep,
       volume = {106},
       number = {6},
          eid = {062001},
        pages = {062001},
          doi = {10.1103/PhysRevD.106.062001},
archivePrefix = {arXiv},
       eprint = {2205.11938},
 primaryClass = {astro-ph.IM},
       adsurl = {https://ui.adsabs.harvard.edu/abs/2022PhRvD.106f2001A}
}

@ARTICLE{2020PNAS..11730055C,
       author = {{Cranmer}, Kyle and {Brehmer}, Johann and {Louppe}, Gilles},
        title = "{The frontier of simulation-based inference}",
      journal = {Proceedings of the National Academy of Science},
         year = 2020,
        month = dec,
       volume = {117},
       number = {48},
        pages = {30055-30062},
          doi = {10.1073/pnas.1912789117},
archivePrefix = {arXiv},
       eprint = {1911.01429},
 primaryClass = {stat.ML},
       adsurl = {https://ui.adsabs.harvard.edu/abs/2020PNAS..11730055C}
}

@ARTICLE{2005PhRvD..72d2004C,
       author = {{Cutler}, Curt and {Gholami}, Iraj and {Krishnan}, Badri},
        title = "{Improved stack-slide searches for gravitational-wave pulsars}",
      journal = {\prd},
         year = 2005,
        month = aug,
       volume = {72},
       number = {4},
          eid = {042004},
        pages = {042004},
          doi = {10.1103/PhysRevD.72.042004},
archivePrefix = {arXiv},
       eprint = {gr-qc/0505082},
 primaryClass = {gr-qc},
       adsurl = {https://ui.adsabs.harvard.edu/abs/2005PhRvD..72d2004C}
}

@ARTICLE{2000PhRvD..61h2001B,
       author = {{Brady}, Patrick R. and {Creighton}, Teviet},
        title = "{Searching for periodic sources with LIGO. II. Hierarchical searches}",
      journal = {\prd},
         year = 2000,
        month = apr,
       volume = {61},
       number = {8},
          eid = {082001},
        pages = {082001},
          doi = {10.1103/PhysRevD.61.082001},
archivePrefix = {arXiv},
       eprint = {gr-qc/9812014},
 primaryClass = {gr-qc},
       adsurl = {https://ui.adsabs.harvard.edu/abs/2000PhRvD..61h2001B}
}

@ARTICLE{2018PhRvD..97j3020A,
       author = {{Ashton}, G. and {Prix}, R.},
        title = "{Hierarchical multistage MCMC follow-up of continuous gravitational wave candidates}",
      journal = {\prd},
         year = 2018,
        month = may,
       volume = {97},
       number = {10},
          eid = {103020},
        pages = {103020},
          doi = {10.1103/PhysRevD.97.103020},
archivePrefix = {arXiv},
       eprint = {1802.05450},
 primaryClass = {astro-ph.IM},
       adsurl = {https://ui.adsabs.harvard.edu/abs/2018PhRvD..97j3020A}
}

@ARTICLE{2021PhRvD.104h4012T,
       author = {{Tenorio}, Rodrigo and {Keitel}, David and {Sintes}, Alicia M.},
        title = "{Application of a hierarchical MCMC follow-up to Advanced LIGO continuous gravitational-wave candidates}",
      journal = {\prd},
         year = 2021,
        month = oct,
       volume = {104},
       number = {8},
          eid = {084012},
        pages = {084012},
          doi = {10.1103/PhysRevD.104.084012},
archivePrefix = {arXiv},
       eprint = {2105.13860},
 primaryClass = {gr-qc},
       adsurl = {https://ui.adsabs.harvard.edu/abs/2021PhRvD.104h4012T}
}

@ARTICLE{2024PhRvD.110l4049M,
       author = {{Mirasola}, Lorenzo and {Tenorio}, Rodrigo},
        title = "{Toward a computationally efficient follow-up pipeline for blind continuous gravitational-wave searches}",
      journal = {\prd},
         year = 2024,
        month = dec,
       volume = {110},
       number = {12},
          eid = {124049},
        pages = {124049},
          doi = {10.1103/PhysRevD.110.124049},
archivePrefix = {arXiv},
       eprint = {2405.18934},
 primaryClass = {gr-qc},
       adsurl = {https://ui.adsabs.harvard.edu/abs/2024PhRvD.110l4049M}
}

@ARTICLE{2014PhRvD..89f4023K,
       author = {{Keitel}, David and {Prix}, Reinhard and {Papa}, Maria Alessandra and {Leaci}, Paola and {Siddiqi}, Maham},
        title = "{Search for continuous gravitational waves: Improving robustness versus instrumental artifacts}",
      journal = {\prd},
         year = 2014,
        month = mar,
       volume = {89},
       number = {6},
          eid = {064023},
        pages = {064023},
          doi = {10.1103/PhysRevD.89.064023},
archivePrefix = {arXiv},
       eprint = {1311.5738},
 primaryClass = {gr-qc},
       adsurl = {https://ui.adsabs.harvard.edu/abs/2014PhRvD..89f4023K}
}

@ARTICLE{2004CQGra..21S1595G,
       author = {{Gair}, Jonathan R. and {Barack}, Leor and {Creighton}, Teviet and {Cutler}, Curt and {Larson}, Shane L. and {Phinney}, E. Sterl and {Vallisneri}, Michele},
        title = "{Event rate estimates for LISA extreme mass ratio capture sources}",
      journal = {Classical and Quantum Gravity},
         year = 2004,
        month = oct,
       volume = {21},
       number = {20},
        pages = {S1595-S1606},
          doi = {10.1088/0264-9381/21/20/003},
archivePrefix = {arXiv},
       eprint = {gr-qc/0405137},
 primaryClass = {gr-qc},
       adsurl = {https://ui.adsabs.harvard.edu/abs/2004CQGra..21S1595G}
}

@ARTICLE{2007arXiv0708.3797P,
       author = {{Papaspiliopoulos}, Omiros and {Roberts}, Gareth O. and {Sk{\"o}ld}, Martin},
        title = "{A General Framework for the Parametrization of Hierarchical Models}",
      journal = {arXiv e-prints},
         year = 2007,
        month = aug,
          eid = {arXiv:0708.3797},
        pages = {arXiv:0708.3797},
          doi = {10.48550/arXiv.0708.3797},
archivePrefix = {arXiv},
       eprint = {0708.3797},
 primaryClass = {stat.ME},
       adsurl = {https://ui.adsabs.harvard.edu/abs/2007arXiv0708.3797P}
}

@ARTICLE{2000physics...9028N,
       author = {{Neal}, Radford M.},
        title = "{Slice Sampling}",
      journal = {arXiv e-prints},
         year = 2000,
        month = sep,
          eid = {physics/0009028},
        pages = {physics/0009028},
          doi = {10.48550/arXiv.physics/0009028},
archivePrefix = {arXiv},
       eprint = {physics/0009028},
 primaryClass = {physics.data-an},
       adsurl = {https://ui.adsabs.harvard.edu/abs/2000physics...9028N}
}

@ARTICLE{2013arXiv1312.0906B,
       author = {{Betancourt}, M.~J. and {Girolami}, Mark},
        title = "{Hamiltonian Monte Carlo for Hierarchical Models}",
      journal = {arXiv e-prints},
         year = 2013,
        month = dec,
          eid = {arXiv:1312.0906},
        pages = {arXiv:1312.0906},
          doi = {10.48550/arXiv.1312.0906},
archivePrefix = {arXiv},
       eprint = {1312.0906},
 primaryClass = {stat.ME},
       adsurl = {https://ui.adsabs.harvard.edu/abs/2013arXiv1312.0906B}
}

@ARTICLE{2011PhRvL.107q1103Y,
       author = {{Yunes}, Nicol{\'a}s and {Kocsis}, Bence and {Loeb}, Abraham and {Haiman}, Zolt{\'a}n},
        title = "{Imprint of Accretion Disk-Induced Migration on Gravitational Waves from Extreme Mass Ratio Inspirals}",
      journal = {\prl},
         year = 2011,
        month = oct,
       volume = {107},
       number = {17},
          eid = {171103},
        pages = {171103},
          doi = {10.1103/PhysRevLett.107.171103},
archivePrefix = {arXiv},
       eprint = {1103.4609},
 primaryClass = {astro-ph.CO},
       adsurl = {https://ui.adsabs.harvard.edu/abs/2011PhRvL.107q1103Y}
}

@ARTICLE{2026arXiv260317072S,
       author = {{Speri}, Lorenzo and {Duque}, Francisco and {Barsanti}, Susanna and {Santini}, Alessandro and {Kejriwal}, Shubham and {Burke}, Ollie and {Chapman-Bird}, Christian E.~A.},
        title = "{Quantifying the Scientific Potential of Intermediate and Extreme Mass Ratio Inspirals with the Laser Interferometer Space Antenna}",
      journal = {arXiv e-prints},
         year = 2026,
        month = mar,
          eid = {arXiv:2603.17072},
        pages = {arXiv:2603.17072},
          doi = {10.48550/arXiv.2603.17072},
archivePrefix = {arXiv},
       eprint = {2603.17072},
 primaryClass = {astro-ph.IM},
       adsurl = {https://ui.adsabs.harvard.edu/abs/2026arXiv260317072S}
}

@ARTICLE{2021LRR....24....1T,
       author = {{Tinto}, Massimo and {Dhurandhar}, Sanjeev V.},
        title = "{Time-delay interferometry}",
      journal = {Living Reviews in Relativity},
         year = 2021,
        month = dec,
       volume = {24},
       number = {1},
          eid = {1},
        pages = {1},
          doi = {10.1007/s41114-020-00029-6},
archivePrefix = {arXiv},
       eprint = {gr-qc/0409034},
 primaryClass = {gr-qc},
       adsurl = {https://ui.adsabs.harvard.edu/abs/2021LRR....24....1T}
}

\appendix

\section{Validating the block-independence assumption}\label{app:blocindependence}

While the likelihood~\eqref{eq:localstationary} assumes that the noise is independent across blocks, realistic, highly coloured detector noise does not satisfy this condition exactly. However, this approximation holds well provided two conditions are met: (i) the duration of each block $T_{\rm block}$ is much larger than the noise decorrelation time $\tau_{\rm decorr}$, and (ii) the data at each block's boundaries is tapered such that at least $\tau_{\rm decorr}$ seconds of data is masked to zero to avoid leakage.

From the Wiener-Khinchin theorem~\cite{wiener1930generalized,Khintchine1934}, the one-sided noise power-spectral-density (PSD) $S_n(f)$ can be expressed as the Fourier transform of the autocorrelation function $R(\tau)$ (see, e.g.,~\cite[Chapter 13]{2007nras.book.....P}),
\begin{align}
    \int  {\rm d}\tau~R(\tau)\exp\left[-i2\pi f\tau\right] = \frac{1}{2}S_n(f).~\label{eq:wienerkhinchin}
\end{align}
We define the decorrelation time $\tau_{\rm decorr}$ as the time lag at which the envelope of $R(\tau)$ permanently drops below a threshold $\epsilon$, set here as $10^{-3}$.

\begin{figure}
    \centering
    \includegraphics[width=0.98\linewidth]{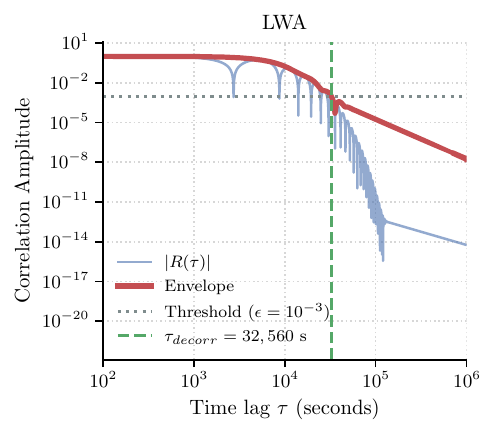}
    
    \vspace{0.1cm} 
    
    \includegraphics[width=0.98\linewidth]{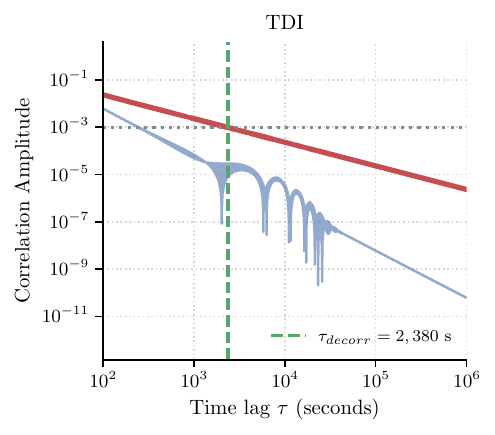}
    
    \caption{\justifying Evaluating the decorrelation time $\tau_{\rm decorr}$ from the autocorrelation function $R(\tau)$ and a given threshold $\epsilon=10^{-3}$ for (i) the long-wavelength approximation (LWA) PSD (top panel), and (ii) the A-channel PSD of 1st-generation time-delay interferometry (TDI) (bottom panel).}
    \label{fig:decorrelationtimes}
\end{figure}

\subsection{Long-Wavelength Approximation}~\label{app:sec_lwa}

In all examples explored in this paper, we model the detector noise using the Long-Wavelength Approximation (LWA)~\cite{2019CQGra..36j5011R} excluding the effect of galactic binary confusion noise. Similar to the treatment in our examples, we apply a high frequency band pass filter to the PSD with $(f_{\rm min}, f_{\rm max}) = (10^{-4}, 10^{-1})$ Hz before calculating the autocorrelation function $R(\tau)$ from Eq.~\eqref{eq:wienerkhinchin}. 

We plot the autocorrelation function for the LWA PSD in the top panel of Fig.~\ref{fig:decorrelationtimes}. As shown by the blue line, $R(\tau)$ is highly oscillatory and dips below the threshold multiple times before rising above it again. To ensure that we correctly identify the final time lag at which $R(\tau)$ permanently drops below $\epsilon$, we apply a Hilbert envelope using the \texttt{scipy.signal.envelope} function from Python's \textsc{SciPy} library~\cite{2020SciPy-NMeth} (red line in Fig.~\ref{fig:decorrelationtimes}). From the envelope, we deduce $\tau_{\rm decorr} \approx 32,560$ seconds or $10^{-3}$ years. We thus obtain an upper limit on $N_{\rm blocks} \leq T/\tau_{\rm decorr}$ which for a total signal duration of $T = 1.0~ (0.5)$ years gives $N_{\rm blocks} \leq 1000~ (500)$. In all our examples, we set $N_{\rm blocks} = 5$, satisfying the first condition. 

To satisfy the second condition, we apply a Tukey window to the data within each block using \texttt{scipy.signal.windows.tukey}. The Tukey window has a single shape parameter $\alpha \in [0,1]$, which sets the fraction of the total time window to be tapered to zero. For a block of time duration $T_{\rm block}$ seconds, the tapering applies symmetrically to the first and last $\alpha T_{\rm block}/2$ seconds. To suppress the correlations between the noise at the end of the $i$th block with the noise at the beginning of the $(i+1)$th block (up to time $\tau_{\rm decorr}$), we can apply a Tukey window that satisfies,
\begin{align}
    \frac{\alpha}{2}T_{\rm block} \geq \frac{\tau_{\rm decorr}}{2}.
\end{align}
Rearranging, we get a minimum threshold on $\alpha$,
\begin{align}
    \alpha \geq \frac{\tau_{\rm decorr}}{T_{\rm block}}.
\end{align}
For $\tau_{\rm decorr} \approx 10^{-3}$ years and $T = 1.0~ (0.5)$ years, we get $\alpha \gtrsim 0.005~ (0.01)$. In all examples, we set $\alpha = 0.02$ which is at least twice as large as the threshold. While this tapering implies a small $\approx 2\%$ loss of optimal SNR in each block, it is necessary to safely mitigate the risk of breaking the independent noise assumption, which can otherwise induce significant biases in the analysis.

\subsection{Time-delay interferometry}

While we have chosen the analytic LWA noise PSD from Ref.~\cite{2019CQGra..36j5011R} in this paper for its computational efficiency, future renditions of the loosely-coherent inference pipeline should adopt the PSD from time-delay interferometry (TDI)~\cite{2021LRR....24....1T} for more realistic results. In anticipation of such studies, here we repeat the $\tau_{\rm decorr}$ calculation from the previous section for 1st-generation TDI, in particular for the A-channel. We choose the same decorrelation time threshold of $\epsilon = 10^{-3}$. 

As shown in the bottom panel of Fig.~\ref{fig:decorrelationtimes}, for the A-channel in TDI 1st-generation we get $\tau_{\rm decorr} \approx 2,380$ seconds, more than an order-of-magnitude smaller than in the case of LWA. Correspondingly, for $T = 1.0~ (0.5)$ years, we get thresholds $N_{\rm blocks} \leq 13250~ (6600)$ and $\alpha \geq 0.0004~ (0.0008)$ to satisfy the noise independence condition. These bounds are even less stringent than the LWA case, ensuring that future loosely-coherent inference will not be limited by the independent noise assumption. \\

\section{Full-dimensional triangle plots at \texorpdfstring{$\boldsymbol{t=0}$}{t=0}} \label{app:plots}

\begin{figure*}[h]
    \centering
    \includegraphics[width=0.98\linewidth]{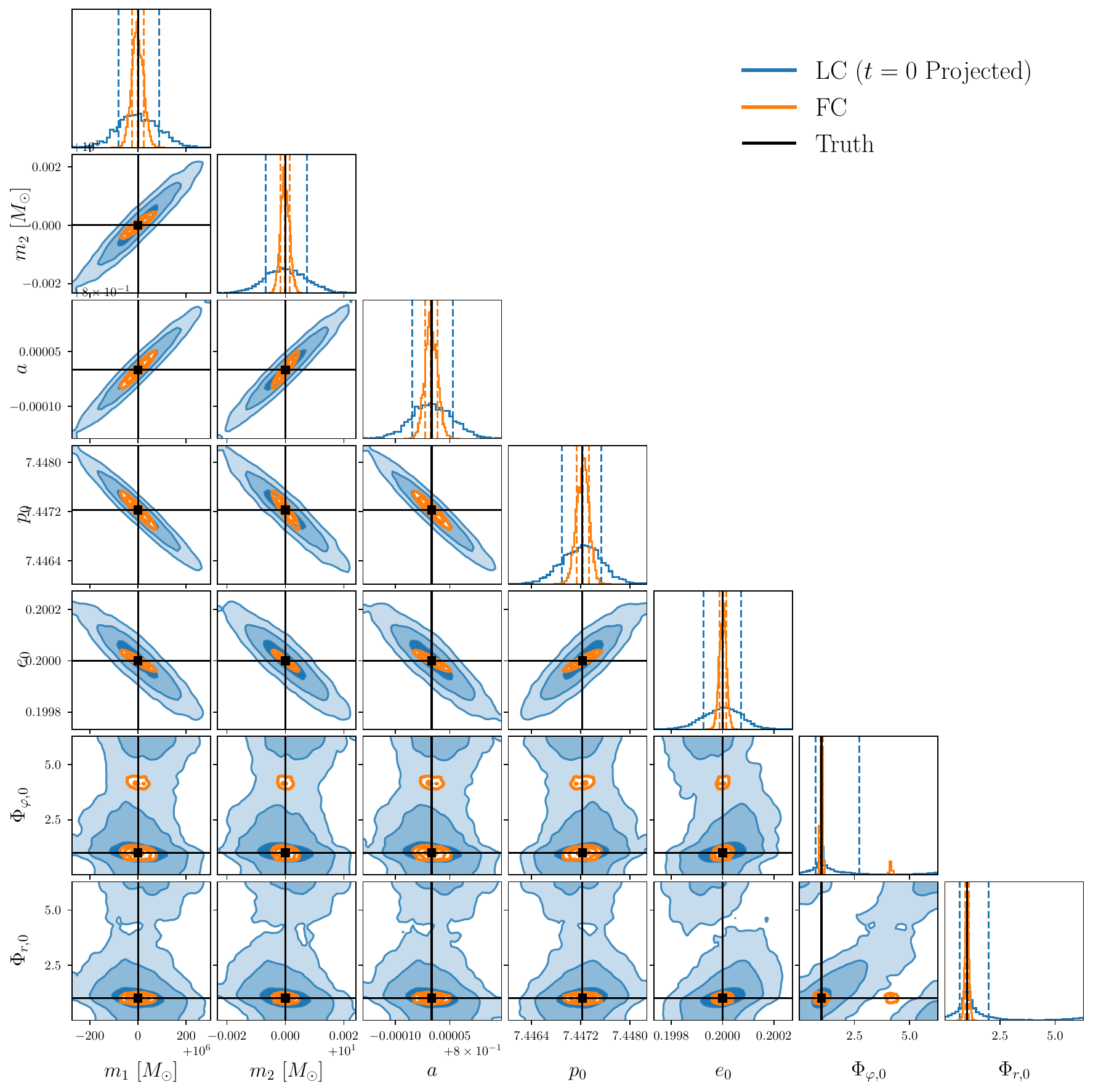}
    \caption{\justifying Full 7D triangle plot comparing the marginalized posterior distributions of the fully-coherent (FC, orange) and loosely-coherent (LC, blue) analyses for the zero-noise vacuum-GR injection (Example \rom{1}, Sec.~\ref{sec:results_example1}). The LC posterior samples for the evolving parameters have been projected back to the initial time $t=0$ to align with the FC parameter space (see Sec.~\ref{sec:t0_backprop}). The exact injected parameter values are indicated by the solid black lines.}
    \label{fig:full_sc_vs_fc_example1}
\end{figure*}

\paragraph{Example~\rom{1} (Sec.~\ref{sec:results_example1}).} We plot the full 7-dimensional triangle plot comparing the fully- and loosely-coherent posterior recoveries at the initial time $t=0$ in Fig.~\ref{fig:full_sc_vs_fc_example1}. For this direct comparison, we back-propagate the loosely-coherent samples from all $N_{\rm blocks} > 0$ to $t=0$ using the method described in Sec.~\ref{sec:t0_backprop}.

\begin{figure*}[h]
    \centering
    \includegraphics[width=0.98\linewidth]{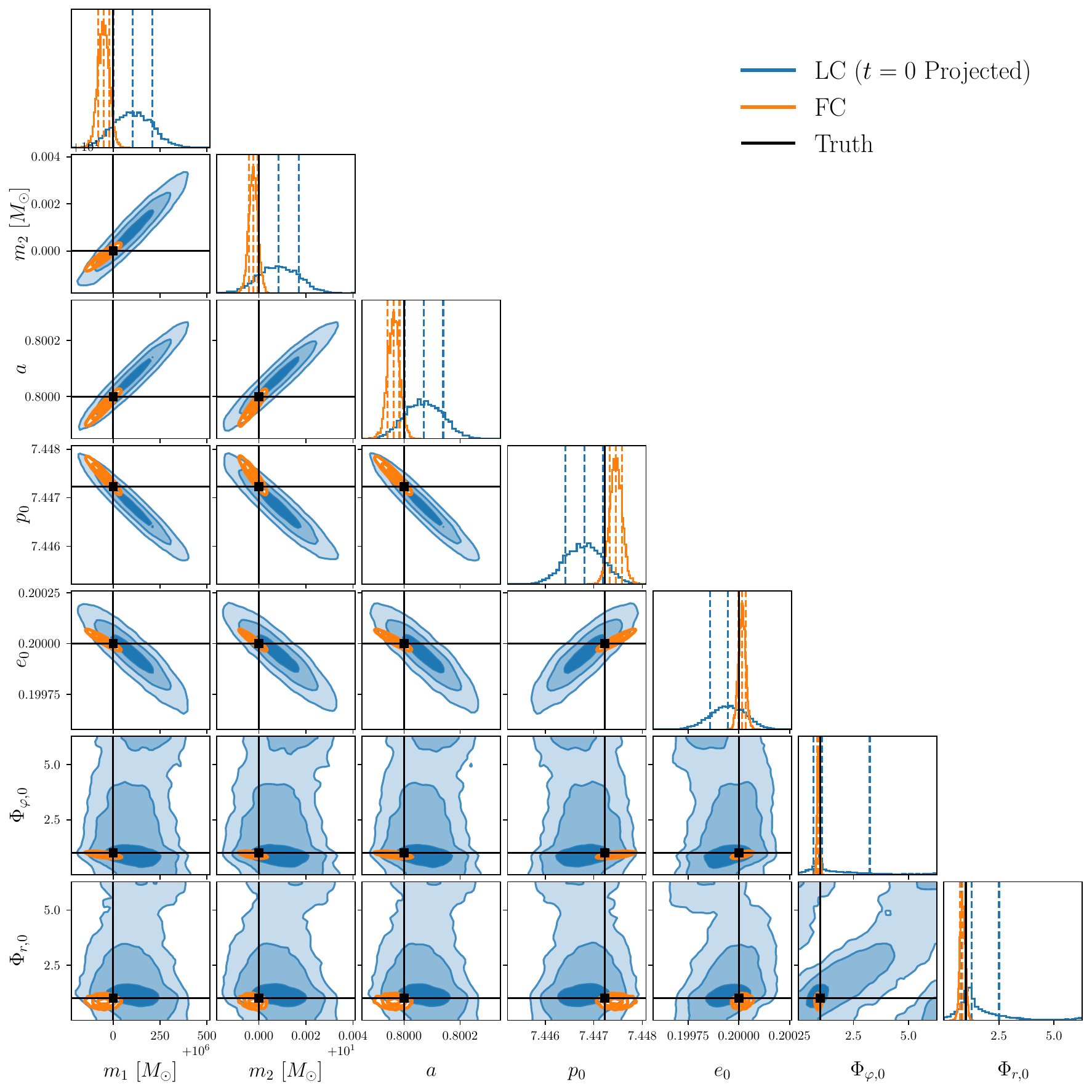}
    \caption{\justifying Full 7D triangle plot comparing the marginalized posterior distributions of the fully-coherent (FC, orange) and loosely-coherent (LC, blue) analyses for the Gaussian-noise vacuum-GR injection (Example \rom{2}, Sec.~\ref{sec:results_example2}). The LC posterior samples for the evolving parameters have been projected back to the initial time $t=0$ to align with the FC parameter space (see Sec.~\ref{sec:t0_backprop}). The exact injected parameter values are indicated by the solid black lines.}
    \label{fig:full_sc_vs_fc_example2}
\end{figure*}

\paragraph{Example~\rom{2} (Sec.~\ref{sec:results_example2}).} Correspondingly, the full 7-dimensional triangle plot at $t=0$ for Example~\rom{2} are presented in Fig.~\ref{fig:full_sc_vs_fc_example2}.

\begin{figure*}[h]
    \centering
    \includegraphics[width=0.98\linewidth]{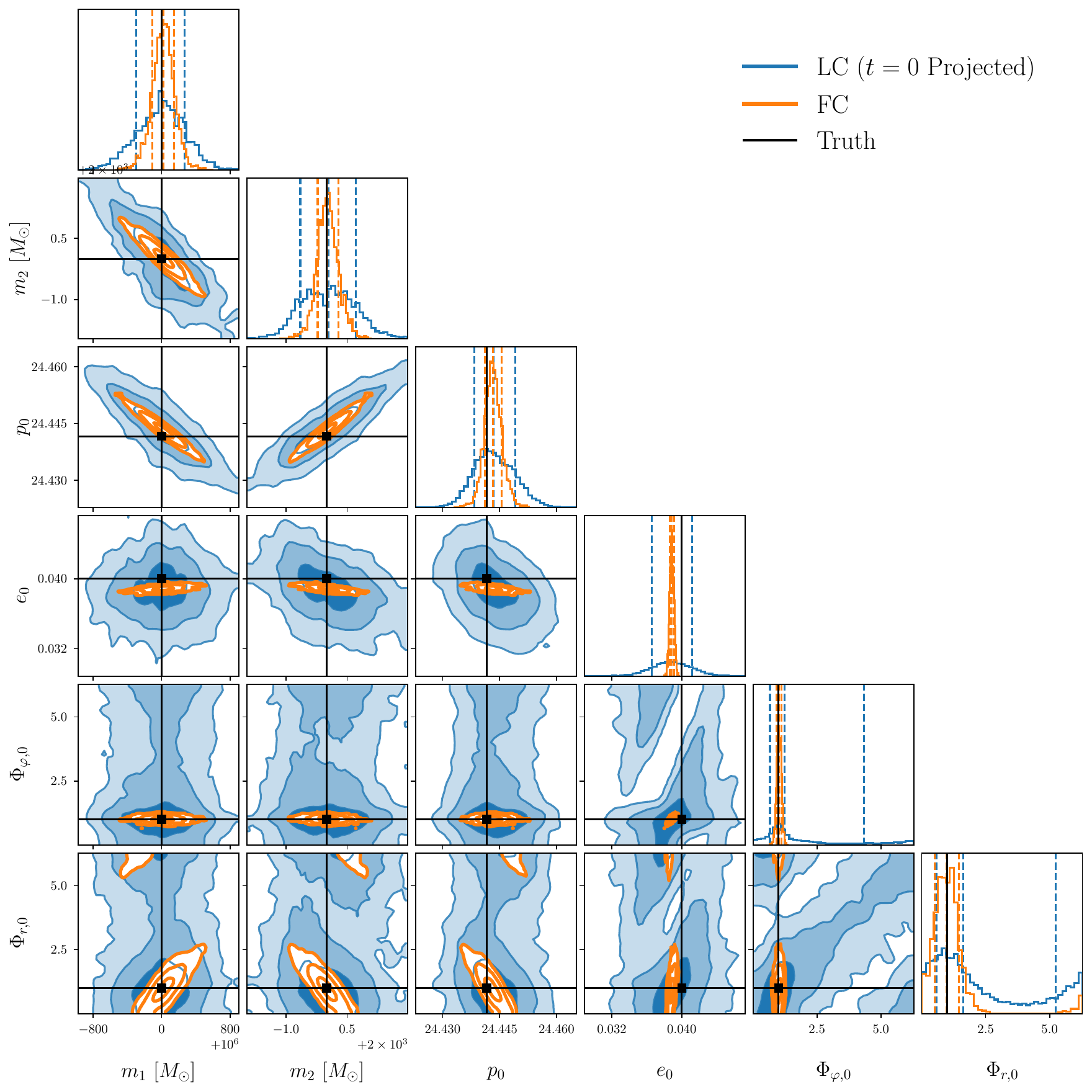}
    \caption{\justifying Full 6D triangle plot comparing the marginalized posterior distributions of the fully-coherent (FC, orange) and loosely-coherent (LC, blue) analyses for the environment-rich IMRI injection (Example \rom{3}, Sec.~\ref{sec:results_example3}). The LC posterior samples for the evolving parameters have been projected back to the initial time $t=0$ to align with the FC parameter space (see Sec.~\ref{sec:t0_backprop}). The exact injected parameter values are indicated by the solid black lines.}
    \label{fig:full_sc_vs_fc_example3}
\end{figure*}

\begin{figure*}[h]
    \centering
    \includegraphics[width=0.98\linewidth]{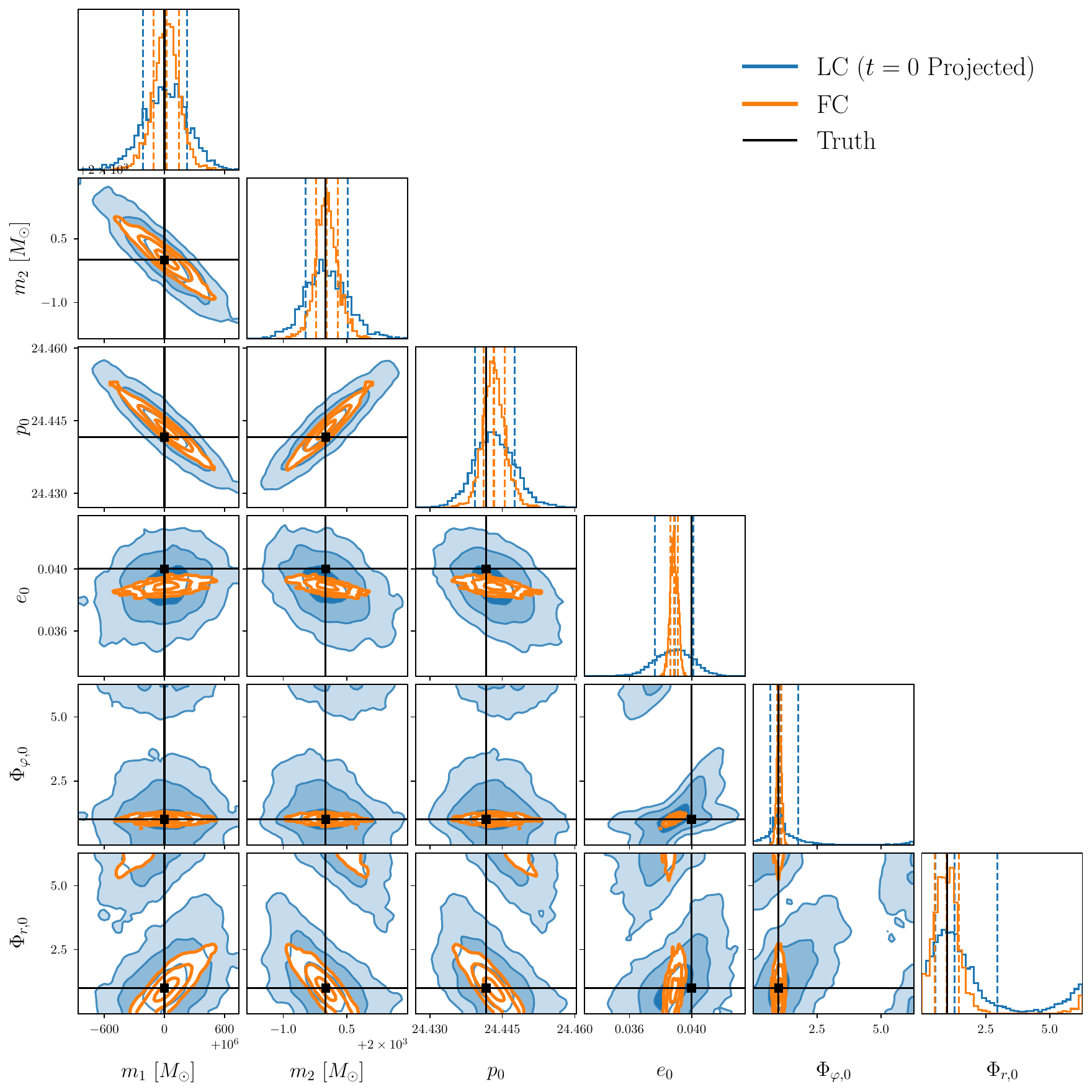}
    \caption{\justifying Full 6D triangle plot comparing the marginalized posterior distributions of the fully-coherent (FC, orange) and loosely-coherent (LC, blue) analyses for the environment-rich IMRI injection with $N_{\rm blocks} = 3$ (Sec.~\ref{sec:results_example3_Nblocks3}). The LC posterior samples for the evolving parameters have been projected back to the initial time $t=0$ to align with the FC parameter space (see Sec.~\ref{sec:t0_backprop}). The exact injected parameter values are indicated by the solid black lines. Note that the LC posteriors are tighter than in the $N_{\rm blocks} = 5$ case (Fig.~\ref{fig:full_sc_vs_fc_example3}).}
    \label{fig:full_sc_vs_fc_example3_Nblocks3}
\end{figure*}

\paragraph{Example~\rom{3} (Sec.~\ref{sec:results_example3}).} Finally, the 6-dimensional triangle plot for the environment-rich IMRI example at $t=0$ are presented in Fig.~\ref{fig:full_sc_vs_fc_example3}. The triangle plot corresponding to the reanalysis with $N_{\rm blocks} = 3$ in Sec.~\ref{sec:results_example3_Nblocks3} is presented in Fig.~\ref{fig:full_sc_vs_fc_example3_Nblocks3}.

\end{document}